\documentclass[preprint2]{aastex}
\usepackage{natbib}
\usepackage{lscape}
\usepackage{wasysym}
\usepackage{txfonts}
\usepackage{graphicx}
\usepackage{hyperref}
\bibpunct{(}{)}{;}{a}{}{,}
\usepackage{longtable}

\usepackage{afterpage}

\def\ms{\hbox{\,m\,s$^{-1}$}}         
\def\m2s2{\hbox{\,m$^{2}$\,s$^{-2}$}} 
\def\kms{\hbox{\,km\,s$^{-1}$}}       
\def\vsini{\hbox{$v$\,sin\,$i$ }}      

\newcommand{\titleast}{\ast}
\newcommand{\titlestar}{\star}

\shorttitle{SOAP 2.0}
\shortauthors{X. Dumusque et al.}

\begin{document}

\title{
SOAP 2.0: A tool to estimate the photometric and radial velocity variations induced by stellar spots and plages.\altaffilmark{\titleast}}

\author{X. Dumusque\altaffilmark{1}\altaffilmark{\titlestar}, 
		I.Boisse\altaffilmark{2},
		N.C. Santos\altaffilmark{3,4}} 

\altaffiltext{1}{Harvard-Smithsonian Center for Astrophysics, 60 Garden Street, Cambridge, Massachusetts 02138, USA}
\altaffiltext{2}{Laboratoire dÕAstrophysique de Marseille (UMR 6110),Technopole de Ch\^ateau-Gombert, 38 rue Fr\'ed\'eric Joliot-Curie,
13388 Marseille Cedex 13, France}
\altaffiltext{3}{Centro de Astrof\`isica, Universidade do Porto, Rua das Estrelas, 4150-762 Porto, Portugal}
\altaffiltext{4}{Departamento de F{\'\i}sica e Astronomia, Faculdade de Ci\^encias da Universidade do Porto, 4150-762 Porto, Portugal}

\altaffiltext{$\star$}
{Swiss National Science Foundation Fellow; xdumusque@cfa.harvard.edu}

\altaffiltext{$\ast$}
{The tool is available at \url{http://www.astro.up.pt/soap}. The work in this paper is based on observations made with the \emph{MOST} satellite, the HARPS instrument on the ESO 3.6-m telescope at La Silla Observatory (Chile), and the SOPHIE instrument at the Observatoire de Haute Provence (France).}


\begin{abstract}
This paper presents SOAP 2.0, a new version of the SOAP code that estimates in a simple way the photometric and radial velocity variations induced by active regions.
The inhibition of the convective blueshift inside active regions is considered, as well as the limb brightening effect of plages, a quadratic limb darkening law, and a realistic spot and plage contrast ratio.
SOAP 2.0 shows that the activity-induced variation of plages is dominated by the inhibition of the convective blueshift effect. For spots, this effect becomes significant only for slow rotators. 
In addition, in the case of a major active region dominating the activity-induced signal, the ratio between the full width at half maximum (FWHM) and the RV peak-to-peak amplitudes of the cross correlation function can be used to infer the type of active region responsible for the signal for stars with \vsini$\le8$\kms. A ratio smaller than three implies a spot, while a larger ratio implies a plage.
Using the observation of HD189733, we show that SOAP 2.0 manages to reproduce the activity variation as well as previous simulations when a spot is dominating the activity-induced variation. In addition, SOAP 2.0 also reproduces the activity variation induced by a plage on the slowly rotating star $\alpha$ Cen B, which is not possible using previous simulations. Following these results, SOAP 2.0 can be used to estimate the signal induced by spots and plages, but also to correct for it when a major active region is dominating the RV variation.
%
%
%
\end{abstract}


\section{Introduction} \label{sect:1}


The radial velocity (RV) technique is an indirect method that does not allow us to directly detect a planet: it measures the stellar wobble induced by a planet orbiting its host star. The technique is sensitive not only to possible companions, but also to signals induced by the host star. At the meter-per-second level, RV measurements are affected by solar-type oscillations \citep{Arentoft-2008,Kjeldsen-2005,Bouchy-2005b,Mayor-2003}, granulation phenomena \citep{Dumusque-2011a,Lindegren-2003,Dravins-1982}, and activity signals \citep[][]{Robertson-2014,Jeffers-2013,Meunier-2013,Lovis-2011b,Dumusque-2011b, Boisse-2011,Boisse-2009,Saar-2009,Huelamo-2008,Desort-2007, Queloz-2001,Santos-2000b,Saar-1997b}. The recent confirmation of Kepler-78 with HIRES@KECK and HARPS-N@TNG \citep{Sanchis-Ojeda-2013,Howard-2013b,Pepe-2013}, and the characterization of the the 17 Earth-mass planet Kepler-10c \citep[][]{Dumusque-2014} show that breakthrough results can be obtained with RV measurements. It is therefore extremely important to understand the different type of stellar signals affecting RV measurements to be able to correct them if we want the RV technique to be efficient in the future. 

Optimal observational strategies can be used to mitigate the effect of oscillations and granulation phenomena when observing solar-type stars \citep[][]{Dumusque-2011b}, which enables the detection of tiny planetary signals \citep{Dumusque-2012,Pepe-2011}. However, activity signals are more difficult to average out. Among the different types of activity signals affecting RV measurements, one should distinguish between short-term variations, with a timescale similar to the rotational period of the star \citep[][]{Saar-1997b}, and long-term perturbations induced by solar-like magnetic cycles \citep[][]{Meunier-2013,Gomes-da-Silva-2013,Lovis-2011b,Dumusque-2011c}.

This paper focuses on the short-term variation that is induced by stellar rotation in the presence of active regions, i.e regions like spots and plages that are the result of strong local magnetic fields. When the star is rotating, these active regions will induce a RV variation by two different physical processes. Because these regions will have temperatures that differ from the average surface temperature, their flux will be different, which shall be referred to as the flux effect throughout this paper. The second effect is induced by the inhibition of the convective blueshift inside active regions because of the strong local magnetic field \citep[][]{Cavallini-1985a, Dravins-1981}. An active region will therefore have a different RV than the average stellar surface, which will induce what is referred to as the convective blueshift effect throughout this paper.

A spot will induce a RV variation due to its small flux compared to the average stellar surface. Sunspots being $\sim 700\,K$ cooler than the effective temperature of the Sun \citep[][]{Meunier-2010a}, they have a much lower flux than the quiet solar photosphere regions\footnote{In the Plank function, the flux is proportional to the temperature elevated to the fourth power.}. A spot will therefore break the flux balance between the blueshifted approaching limb and the redshifted receding limb of a rotating star, and will induce a RV variation as it passes on the visible stellar disc. 
A plage at the disc center is only slightly hotter than the average effective temperature and will induce a small flux effect. A plage on the limb will be brighter due to a center-to-limb brightness dependence \citep[e.g][]{Meunier-2010a,Unruh-1999,Frazier-1971}, however at this location, the star emits less light due to limb darkening. Independently of its location, a plage will therefore induce a small flux effect compared to a spot, even if plages tend to be an order of magnitude more extended than spots \citep[][]{Chapman-2001}.

A plage or a spot are regions affected by strong local magnetic fields. These magnetic fields will inhibit locally the convection, which will suppress the convective blueshift effect inside active regions \citep[$\sim$ 300\ms for the Sun, e.g.][]{Dravins-1981}. These regions therefore appear redshifted in comparison to the quiet photosphere (see Figure 3 in \citet{Cavallini-1985a} for a plage and Figure 2 of this work for a spot), which induces a RV variation as active regions appear and disappear from the visible part of the stellar disk due to rotation. The convective blueshift effect for spots and plages of the same size will be similar.

Several simulations estimating the RV variation induced by active regions already exist. However most of them only consider the flux effect \citep[][]{Oshagh-2013a,Boisse-2012b,Barnes-2011,Desort-2007,Hatzes-2002,Saar-1997b}. The ones that include the convective blueshift effect do it in a simple way by assuming Gaussian spectral lines that are redshifted by a fixed amount to consider the convection inhibition inside active regions \citep[][]{Aigrain-2012,Lanza-2010,Meunier-2010a}. As we will see in this paper, considering observed spectral line shape is essential in reproducing the variation of the bisector span (BIS SPAN) and the Full Width at Half Maximum (FWHM)  of the Cross Correlation Function (CCF) used to derive precise RV measurements\footnote{The RV is measured on the CCF obtained by the cross correlation of the stellar spectrum with a synthetic stellar template \citep[][]{Pepe-2002a,Baranne-1996}. The contrast of the CCF, i.e its depth, is not used here because the variation of this observable is strongly correlated with the FWHM of the CCF, and more affected by instrumental noise than the FWHM in real data.}. \citet{Saar-2009,Saar-2003} estimated the convective blueshift effect using the observed bisector of a solar spectral line in and outside of an active region. However, the use of only a single spectral line makes it difficult to extrapolate this work to other stars than the Sun. For other stars, the signal-to-noise ratio required to detect activity signal in the spectrum can only be obtained on the CCF that is an average profile of all the spectral lines in the visible, and therefore only reflects partially the behavior of individual spectral lines.

Considering only the flux effect was justified a decade ago when the precision of the best RV instruments were only able to detect the flux effect induced by spots on rapid rotators. With the meter-per-second precision reached nowadays, the inhibition of the convective blueshift effect inside active regions is measurable and should be accounted for to reproduce the activity-related RV, BIS SPAN, and FWHM variations. 

This paper presents a new code to estimate the activity-induced RV variation. The convective blueshift effect is accounted for by using observed spectra of the quiet solar photosphere and of a sunspot, which is the natural way to include all the physics related to stellar atmosphere. After a presentation of the new code in Section \ref{sect:2} and a comparison between the different effects induced by spots, plages and stellar parameters in Sections \ref{sec:3} and \ref{sec:4}, the result of this activity simulation is confronted to observations of solar-type stars in Section \ref{sect:5}.

The SOAP 2.0 software is available for use along with a brief manual and some example data at the following url: \url{http://www.astro.up.pt/soap}. When using SOAP 2.0 in publications, it is appropriate to cite this paper.

\section{Simulating the effect induced by active regions} \label{sect:2}

In this section, we first describe briefly the code Spot Oscillation And Planet \citep[SOAP,][]{Boisse-2012b}, a code designed to estimate the flux effect of active regions on the photometric and spectroscopic measurements. We then discuss the improvements made to this code to include the convective blueshift effect, the limb brightening effect of plage, a quadratic limb darkening law, a realistic active region contrast and the resolution of the spectrograph used for the observations.

\subsection{The SOAP tool}  \label{sec:2-0}

Different tools to estimate the photometric and RV contribution of active regions have been published \citep[][]{Aigrain-2012,Lanza-2010,Meunier-2010a, Saar-1997b}, and we decided here to focus on SOAP \citep[][]{Boisse-2012b}. This code only considers the flux effect of active regions (see Introduction) to estimate the activity-induced variation on the photometry, RV, BIS SPAN and FWHM. In this section, only a general presentation of SOAP is done and defines the material that will be used in the next sections. For a complete and more detailed description of the tool, the reader is referred to the original SOAP paper \citep[][]{Boisse-2012b}.

In SOAP, and as a first step, the non-spotted emission of the star is computed. The visible stellar surface is divided in cells having the same projected area (whose number is defined by the user, generally more than 10000). Each cell presents a different radial velocity depending on the rotational period and radius of the star, and a different weight depending on a linear limb darkening law. In each cell, the emerging stellar spectrum is represented by a Gaussian line profile equivalent to the spectrum CCF of a solar-type star with zero-rotation. The velocity of this Gaussian line profile is shifted to the projected velocity of the cell (dependent on its position on the disk, the geometry of the system, and the rotational velocity of the star), and its contrast is weighted by a linear limb darkening law. Therefore, the integrated CCF of the quiet stellar disc, $\mathrm{CCF}_{\mathrm{tot,\,quiet}}$ is defined by:
\begin{equation} \label{eq:0}
\mathrm{CCF}_{\mathrm{tot,\,quiet}} = \sum_{x,y}^N I_{ld}(x,y) \, \mathrm{CCF}(x,y),
\end{equation}
where $x$ and $y$ scan the stellar grid that has a size $N\times N$, and $I_{ld}(x,y)$ represents the limb darkening law and CCF is in the case of SOAP a Gaussian line profile. The non-spotted emission of the star, $\mathrm{Flux}_{\mathrm{tot,\,quiet}}$, is simply the integration of the limb darkening law over the entire stellar disc:
\begin{equation} \label{eq:1}
\mathrm{Flux}_{\mathrm{tot,\,quiet}} = \sum_{x,y}^N I_{ld}(x,y).
\end{equation}

Then, an active region of a given size is added at a given longitude and latitude. The Gaussian CCF considered for this region is the same that the one for the stellar disk, but its weight depend on the region brightness: [0:1] for a dark spot, $>$1 for a plage. In each cell affected by the active region, defined by $(x_a,y_a)$, SOAP estimates the difference between the quiet CCF at this location and the CCF of the active region, $\mathrm{CCF_a}$, which gives:
\small
\begin{eqnarray} \label{eq:2}
\Delta \mathrm{CCF}(x_a,y_a) &=&\sum_{x_a,y_a} I_{ld}(x_a,y_a) \, \big[\mathrm{CCF}(x_a,y_a)- I_a \, \mathrm{CCF_a}(x_a,y_a)\big] \nonumber \\
\end{eqnarray}
\normalsize
where $I_a = \mathrm{Planck}(\lambda_0,T_{\mathrm{active\,region}})/\mathrm{Planck}(\lambda_0,T_{\mathrm{eff}})$ is the relative brightness of the active region, $\lambda_0$ being the wavelength at which the Planck function is estimated, $T_{\mathrm{active\,region}}$ the temperature of the active region and $T_{\mathrm{eff}}$ the effective temperature of the star.

\begin{figure}
\begin{center}
\includegraphics[width=8cm]{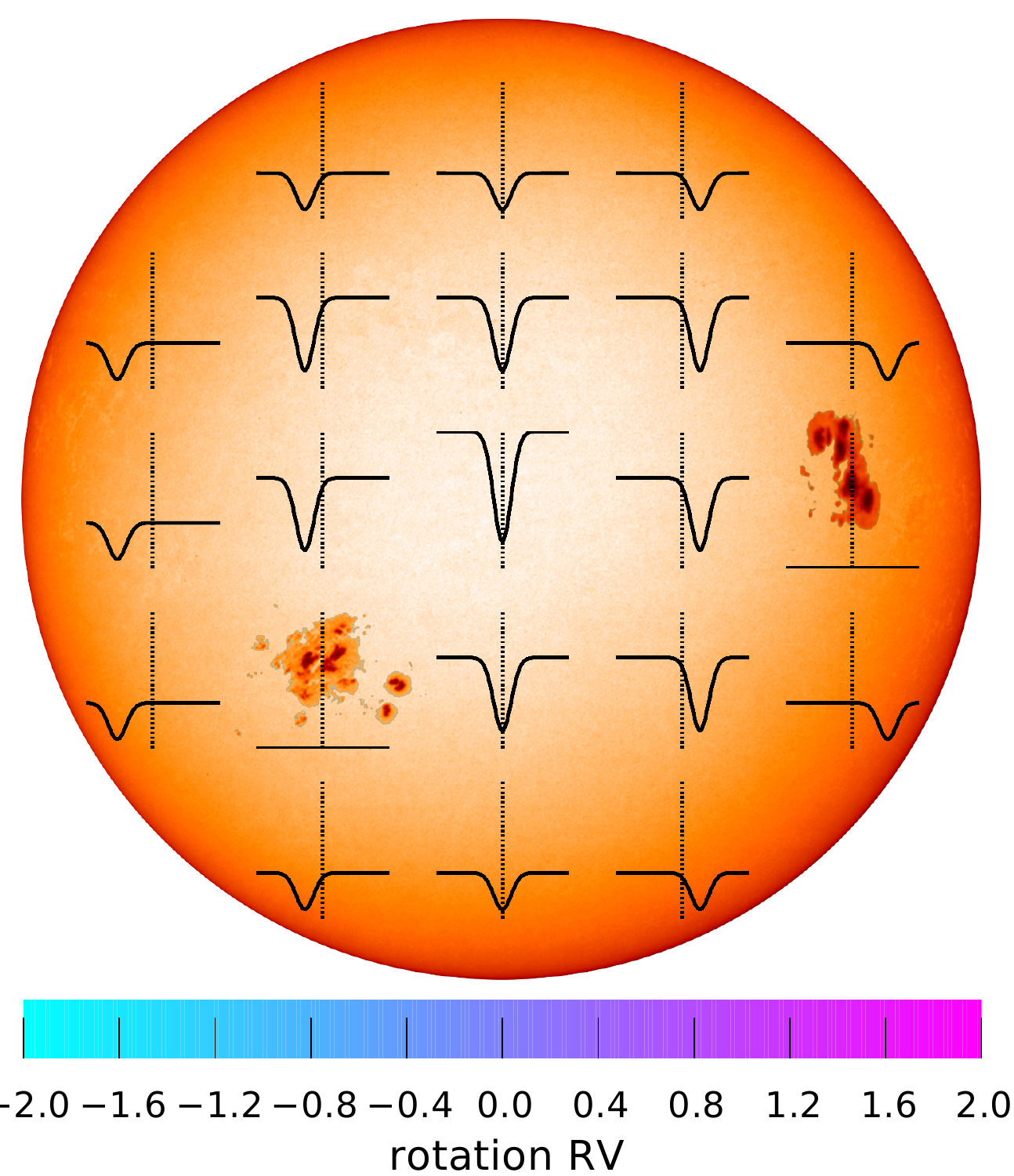}
\caption[How SOAP simulates the effect of stellar spots]
{Figure explaining how SOAP simulates the effect of active regions. The star is divided in a grid of N $\times$ N cells, each of them having its own Gaussian corresponding to the stellar CCF. Depending on the cell position, the Gaussian is Doppler shifted to account for rotation. The rotational speed \vsini in the case of the Sun is given as the horizontal bar at the bottom of the figure. The intensity of each cell is weighted by a limb darkening law, and the intensity is reduced (or increased) in presence of a dark spot (or a bright plage). Here dark spots with no emitting flux are considered. The integrated CCF over the stellar disc is obtained by summing up all the cells. Note that here all the Gaussian have the same depth. On the limb, the Gaussians appear shallower because they are weighted by the limb darkening.}
\label{fig:2-1-0}
\end{center}
\end{figure}

Finally, the difference between the quiet CCF at the active region location (if the active region would not exist) and the CCF of the active region is subtracted from the quiet integrated CCF which give us the integrated CCF taking into account the effect of the active region:
\small
\begin{eqnarray} \label{eq:3}
\mathrm{CCF}_{\mathrm{tot,\,active}} &=&\mathrm{CCF}_{\mathrm{tot,\,quiet}} - \Delta \mathrm{CCF}(x_a,y_a)  \nonumber \\
							   &=&\sum_{x,y}^N I_{ld} \, \mathrm{CCF} - \sum_{x_a,y_a}   I_{ld} \, \big[\mathrm{CCF} - I_a \, \mathrm{CCF_a}\big]  \nonumber \\
							   &=&\sum_{x,y \neq x_a,y_a}^N  I_{ld} \, \mathrm{CCF} - \sum_{x_a,y_a}   I_{ld} \, \big[\mathrm{CCF}  \left(1-1\right) - I_a \, \mathrm{CCF_a} \big]  \nonumber \\
							   &=&\sum_{x,y \neq x_a,y_a}^N  I_{ld} \, \mathrm{CCF} + \sum_{x_a,y_a}   I_{ld} \, I_a \, \mathrm{CCF_a}.
\end{eqnarray}
This approach of estimating first the integrated CCF of the quiet star and then assuming only the contribution of the active region is computationally efficient, because this prevents us from estimating the integrated CCF of the entire star at each step of the rotational phase when looking at the variation induced by an active region passing on the visible stellar disc. The integrated flux of the disc in the presence of an active region, $\mathrm{Flux}_{\mathrm{tot,\,active}}$, is equal to the non-spotted emission of the star minus the flux difference between the quiet photosphere and the active region (inside that region), which gives:
\begin{eqnarray} \label{eq:4}
\mathrm{Flux}_{\mathrm{tot,\,active}} &=&\sum_{x,y}^N I_{ld} - \sum_{x_a,y_a} I_{ld} (1-I_a) \nonumber\\
							   &=&\sum_{x,y \neq x_a,y_a} I_{ld} + \sum_{x_a,y_a} I_{ld} \, I_a.
\end{eqnarray}
\normalsize
An illustration of the SOAP simulation showing the variation of the CCF across the stellar disc can be found in Fig. \ref{fig:2-1-0}. \\

In the end SOAP returns the photometry, which is directly given by $\mathrm{Flux}_{\mathrm{tot,\,active}}$, the BIS SPAN, which is calculated \footnote{The BIS SPAN is defined here as the difference between the top of the bisector ranging from 10 to 40\% of the depth and the bottom of the bisector ranging from 60 to 90\% of the depth. This is the definition adopted for HARPS measurements.} directly on $\mathrm{CCF}_{\mathrm{tot,\,active}}$, and the RV and the FWHM, which are estimated by fitting a Gaussian to $\mathrm{CCF}_{\mathrm{tot,\,active}}$. Estimating these parameters for different rotational phase of the star give us the variation induced by an active region as it passes on the visible stellar disc.

The total integrated CCF, $\mathrm{CCF}_{\mathrm{tot,\,active}}$, can be modified in two ways. On the one hand, we can change the intensity of the active region, $I_a$, assuming that the CCF inside the quiet photosphere and inside the active region is the same ($\mathrm{CCF}$=$\mathrm{CCF}_a$). This is what is done in SOAP because the CCF of a solar-type stars inside the photosphere or inside an active region can be approximated at first order by the same Gaussian. Because in this case only the active region intensity is affecting the total integrated CCF, this corresponds to the flux (F) effect. On the other hand, we can fix the intensity of the active region and consider a different CCF inside the active region than inside the quiet photosphere ($\mathrm{CCF} \neq \mathrm{CCF}_a$). This is what will be done in the next sections to include the convective blueshift (CB) effect. These two ways of modifying the total integrated CCF can be written:
\begin{equation} \label{eq:5}
\mathrm{CCF}_{\mathrm{tot,\,active,\,F\,effect}} = \sum_{x,y \neq x_a,y_a}^N  I_{ld} \, \mathrm{CCF} + \sum_{x_a,y_a}   I_{ld} \, I_a \, \mathrm{CCF}.
\end{equation}
\begin{equation} \label{eq:6}
\mathrm{CCF}_{\mathrm{tot,\,active,\,CB\,effect}}  = \sum_{x,y \neq x_a,y_a}^N  I_{ld} \, \mathrm{CCF} + \sum_{x_a,y_a}   I_{ld} \, \mathrm{CCF_a}.
\end{equation}
%


\subsection{Limitation of SOAP and equivalent tools}  \label{sec:2-1}

At present, models like SOAP trying to reproduce the effect induced by active regions on RV measurements only consider the flux effect \citep[e.g.][]{Boisse-2012b, Barnes-2011, Desort-2007, Saar-1997b}. In these simulations, the spectrum or its CCF used in each cell dividing the stellar surface is assumed the same in both the quiet photosphere and the active regions (see Eq. \ref{eq:5}). This implies a negligible RV effect for plages, for which the flux effect is nearly zero due to the low contrast difference between a plage and the quiet photosphere.
However, inhibition of convection in plages (spots) due to the strong local magnetic field produces a $\sim300$\ms redshift of spectral lines formed inside a plage \citep[or spot,][]{Cavallini-1985a}, and therefore induces a non-negligible RV variation \citep[][]{Meunier-2010a, Saar-2009,Dravins-1981}. This inhibition of the convective blueshift effect will be weighted by the contrast difference between the active region and the quiet photosphere. Therefore the activity variation induced by the inhibition of the convective blueshift effect will be significant for plages, and nearly negligible for spots. Contrary to what simulations only considering the flux effect predict, the effect of plage cannot be neglected. Moreover, plages are an order of magnitude more extended than spots \citep[][]{Chapman-2001}, increasing even more their activity-induced RV effect.

In conclusion, simulations only considering the flux effect can reproduce the observed variation induced by active regions for stars that are spot-dominated \citep[][]{Shapiro-2014,Lockwood-2007}, which is the case for very active stars \citep[e.g. HD166435,][]{Boisse-2012b}. However, these models fail to explain the variations observed in more quiet stars like our Sun, which are dominated by plages.

\subsection{Including the convective blueshift effect and some additional stellar physics}  \label{sec:2-2}

\begin{figure}
\begin{center}
\includegraphics[width=8cm]{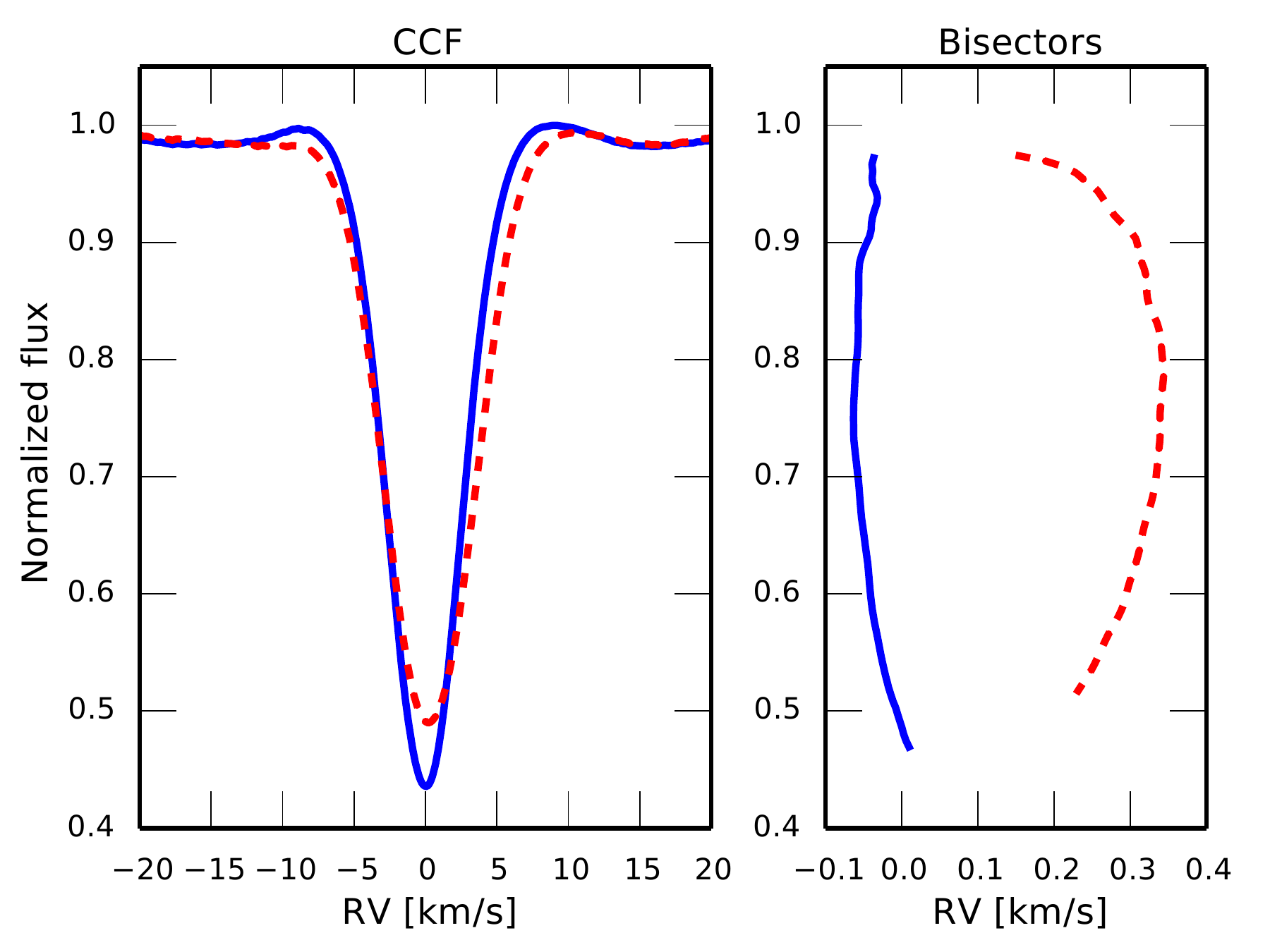}
\caption{\emph{Left:} CCFs of the quiet photosphere in blue (continuous line) and of a spot in red (dashed line) calculated from observed FTS spectra of the Sun. \emph{Right:} Bisectors of the same CCFs. A $350 m\,s^{-1}$ redshift of the spectrum inside a spot can be seen when comparing the two bisectors.}
\label{fig:2-2-0}
\end{center}
\end{figure}

Some authors have tried to include the convective blueshift effect and its inhibition inside active regions. \citet{Saar-2009,Saar-2003} derived the activity-related RV variation for a plage, using solar observations of an iron line measured inside the quiet photosphere and inside a plage. In solar spectra, the "C" shape of spectral lines formed inside a quiet photosphere region reflects the presence of the convective blueshift effect \citep[][]{Dravins-1981}. Therefore using solar observations of the quiet photosphere and of a plage naturally includes the convective blueshift effect and its inhibition in plages. However only one single spectral line has been used in these studies. Nowadays the most precise spectrographs use the information of the entire visible spectrum to reach a high precision in RV. This RV is measured on the CCF obtained by the cross correlation of the stellar spectrum with a synthetic stellar template \citep[][]{Pepe-2002a,Baranne-1996} . Because each spectral line is affected in a different way by convection, it is not straightforward from one spectral line to estimate the variation of the CCF, and therefore comparison with observations is difficult. \citet{Lanza-2011b} use a synthesized line profile, that they assume being equal to a CCF. However, for the same reason as above, this assumption is only correct at first order as spectral lines will have a different shape than the CCF. The inhibition of the convection is included by shifting the line profile by a fixed amount, which does not include the warping of the bisector that is observed on the Sun (see Fig. \ref{fig:2-2-0}). \citet{Meunier-2010b} use some simulated solar spectra comparable to those obtained with HARPS in terms of spectral coverage and resolution, however the inhibition of convective blueshift is also considered as a shift of the spectral lines. 

The most reliable way to include the effect of convective blueshift, as well as its inhibition inside active regions due to strong magnetic fields, is to use observed spectra of the quiet solar photosphere and of a solar active region. Because our goal is to simulate the RV effect of active regions as seen with a high-resolution instrument like HARPS, we require some high-resolution solar spectra covering the visible spectral range. The Fourier Transform Spectrograph (FTS) at the Kitt Peak Observatory has obtained such spectra. In the archive of the instrument, we were able to obtain a spectrum for the quiet photosphere\citep[][]{Wallace-1998} in the disc center, and one for a sunspot \citep[][]{Wallace-2005}. 
We then computed the CCFs of these two spectra using the G2 HARPS template \citep[][]{Pepe-2002a} and associated these CCFs to the non-active region and active region of the star, respectively. The same spectrum will be used for a spot and a plage, because no spectrum for a plage with the required properties could be found in the literature. It is clear that the differences in temperature for a spot and a plage will influence the emerging spectrum and modify the CCF bisector, however at first order, the inhibition of convection, shifting spectral lines by $\sim300\,m\,s^{-1}$, will be the dominant effect. 
The CCFs for the quiet photosphere and the active region that we use, with their respective bisector are displayed in Fig. \ref{fig:2-2-0}. As we can see, the inhibition of convective blueshift inside an active region induce a CCF redshift of 350\ms and changes considerably the shape of the bisector. 

The choice of using the CCFs here rather than the spectra is for computational efficiency. Indeed, with only a few hundred data points, the CCF carries the averaged information of the entire spectrum (500'000 points for the FTS spectra compared to 201 points in our case for the CCF). In appendix \ref{app:2}, we show analytically that the integrated CCF over the stellar disc obtained when assuming a CCF in each cell of the simulation and then integrating over the disc is the same that when assuming a spectrum in each cells, integrating over the disc and finally calculating the CCF of the integrated spectrum. This equivalence is also shown numerically using our simulation (Fig. \ref{fig:app2-0} of appendix \ref{app:2}).

In addition to using observed solar spectra, we also added in SOAP the observed contrast for sunspots and plages. For the Sun, \citet{Meunier-2010a} found an average temperature difference between a sunspot and a quiet region of $\Delta T_S=-663$\,K. Using a Planck function with the solar effective temperature (5778\,K), we arrive to a contrast of 0.54 at 5293 \AA\,(middle of the FTS spectral range). For a plage, the temperature and therefore the contrast is dependent on the position of the active region on the stellar disc. \citet{Meunier-2010a} show that the temperature difference between a plage and the quiet photosphere is given by the law:
\begin{equation}  \label{eq:2-2-0}
\Delta T_P=250.9-407.7\cos{\theta}+190.9\cos^2\theta,
\end{equation}
where $\theta$ is the angle between the normal to the stellar surface and the observer ($\theta=0$ at the stellar disc center and $\pi/2$ at the limb). A plage is therefore brighter on the solar limb than on the solar disc center, with a contrast ranging from 1.22 to 1.03 at 5293 \AA, respectively.

We modified as well the linear limb darkening law originally present in SOAP by a more realistic quadratic model \citep[][]{Mandel-2002}, as it was done in SOAP-T \citep[][]{Oshagh-2013a}: 
\begin{equation} \label{eq:2-2-1}
I_{ld}(\cos{\theta}) = 1 - \gamma_1(1-\cos{\theta}) - \gamma_2(1-\cos{\theta})^2 \quad \gamma_1+\gamma_2 < 1. 
\end{equation}
For a stellar effective temperature close to that of the Sun (5778\,K), we choose $\gamma_1=0.29$ and $\gamma_2=0.34$ \citep[][]{Oshagh-2013a,Claret-2011,Sing-2010}.
In the following sections, these values will be used unless specified otherwise.

Finally, to compare the results of this simulation to observations, the resolution of the spectrograph has to be considered as it might influence the estimation of the RV, BIS SPAN and FWHM \citep[][]{Boisse-2012b,Desort-2007}. The spectra obtained with the FTS have a resolution higher than 700000, and therefore the resolution of the disk integrated CCF has to be reduced to match observations made with HARPS or other instruments. This step is done by convolving to the CCF an instrumental profile that can be approximated by a Gaussian for fiber-fed spectrographs like HARPS, HARPS-N, SOPHIE, CORALIE, TRES. The FWHM of this Gaussian is equal to the velocity resolution $\Delta v$ given by the formulae $\Delta v = c/R$, where $R$ is the instrumental resolution and $c$ the speed of light in vacuum. For example, this gives us a FWHM of 2.6 and 5.5\kms for HARPS ($R = 115000$) and CORALIE ($R = 55000$), respectively.

\section{Modifications brought by the convective blueshift effect and additional stellar physics}  \label{sec:3}

In the previous section, we listed the different improvements implemented into SOAP to include more stellar atmospheric physics based on solar observations. Here, to investigate the effect brought by each improvement, we simulate an active region on a rotating star, and estimate the variations seen in photometry\footnote{The photometric effect is only dependent on the active region contrast compared to the quiet photosphere in our simulation (see Equation \ref{eq:4}). This contrast is estimated at 5293 \AA, therefore the photometric effect that will be derived in this paper is for Sun-like active regions observed in a narrow filter centered on 5293 \AA}, RV, BIS SPAN and FWHM. Note that we will use the differential value for the variation of all these observables, which explains why the FWHM variation is sometimes equals to zero or smaller. We will first study the variation brought by different limb darkening laws and by different instrumental resolutions. We will then study the effect induced by the use of observed solar spectra. In our simulation, the star has a radius and a projected rotational velocity fixed to the solar value (\vsini$=2$\kms), and is
seen equator on. The active region is set on the equator and has a size of $S=1$\%. This size is defined as the fraction of the surface of the visible hemisphere covered by the active region\footnote{$S=\pi R_{sp}^2 / 2\pi R_{\star}^2$, where $R_{sp}$ is the radius of the spot and $R_{\star}$ the radius of the star}. The results of this section are obtained considering observed CCFs in each of the 300$\times$300 cells (except for Section \ref{sec:3-3} and Figure \ref{fig:3-3} were we compare the use of Gaussian and observed CCFs).

\subsection{Modifications brought by the limb darkening}  \label{sec:3-1}

\begin{figure*}
\begin{center}
\includegraphics[width=8cm]{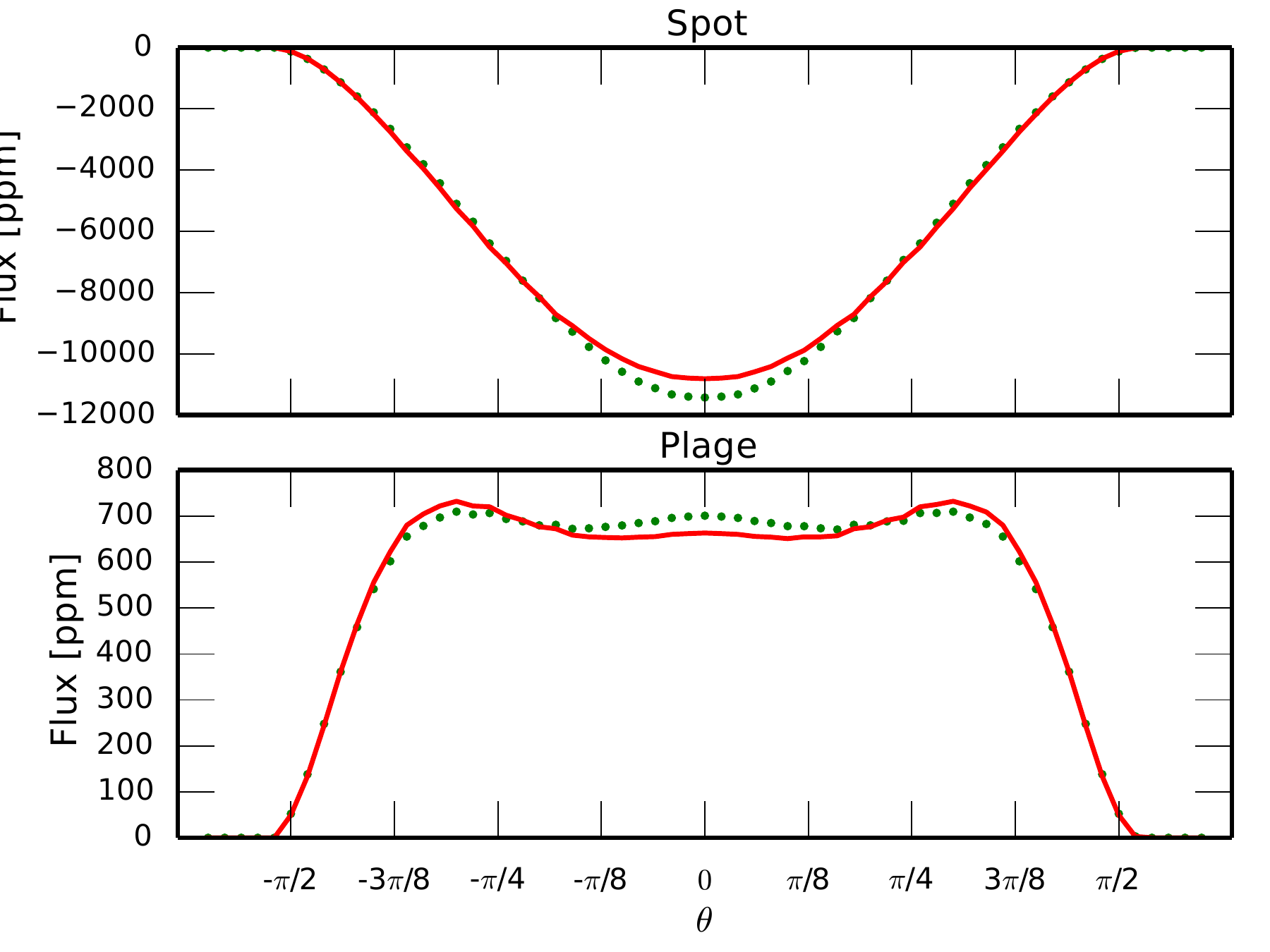}
\includegraphics[width=8cm]{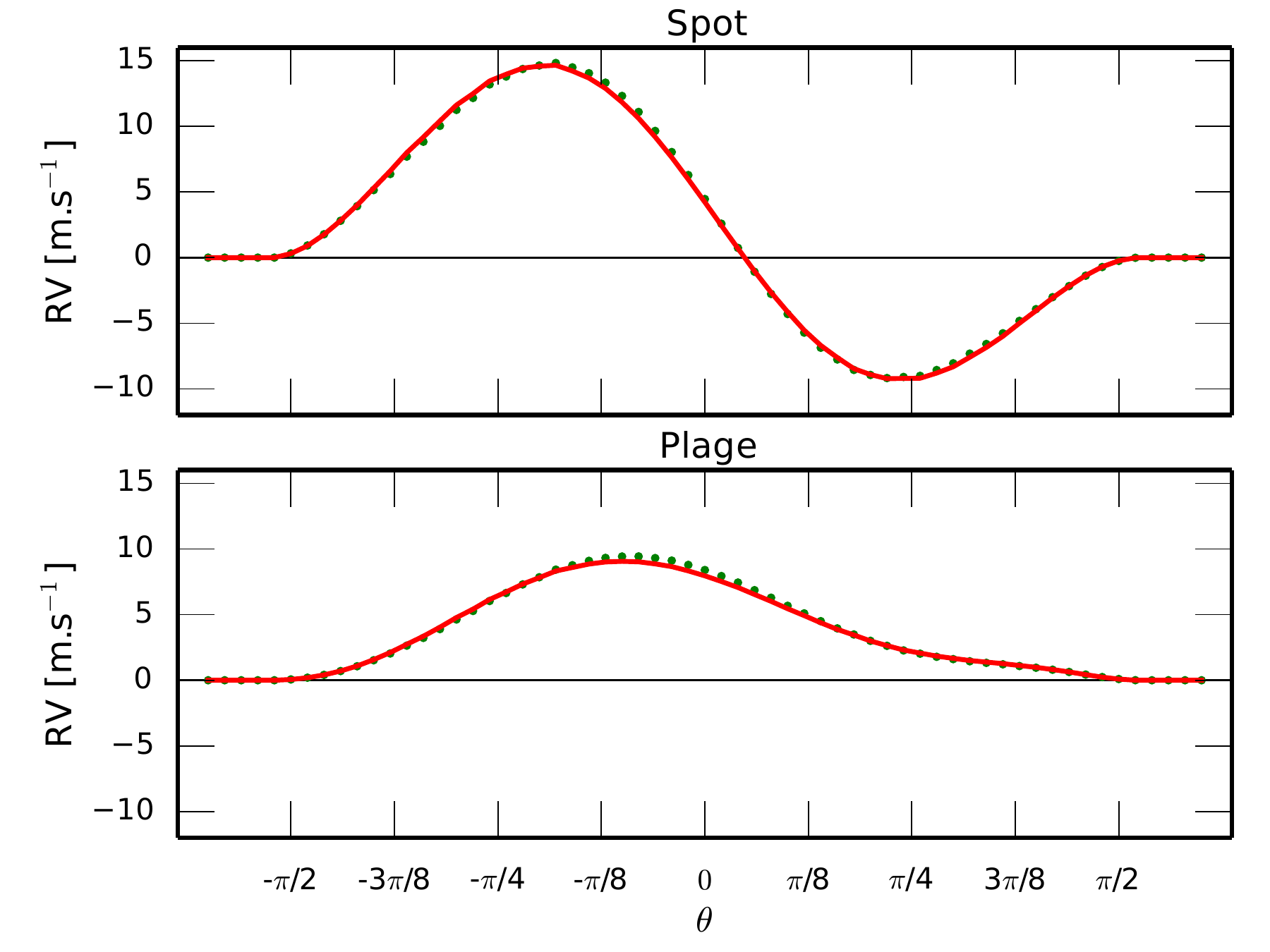}
\includegraphics[width=8cm]{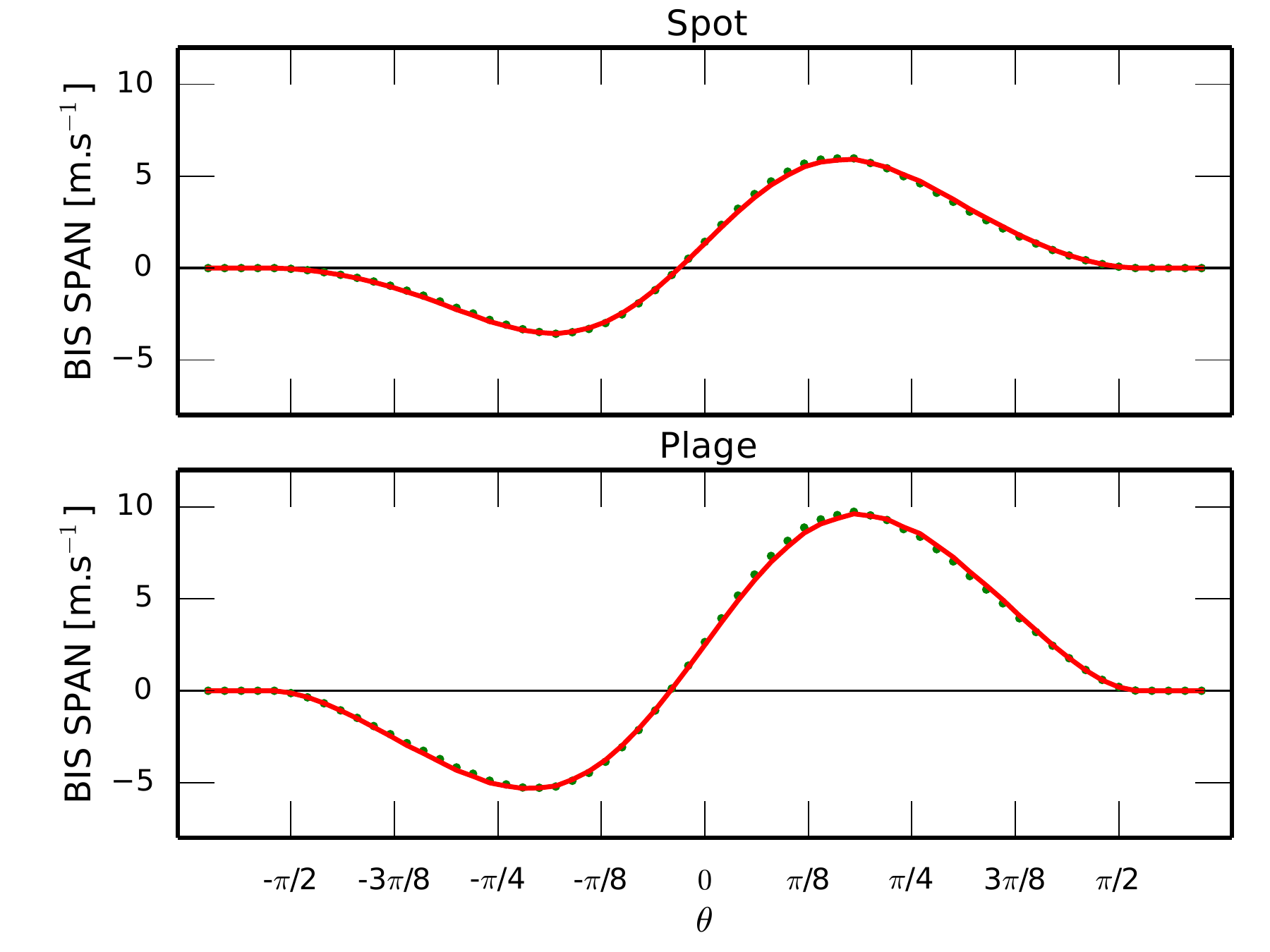}
\includegraphics[width=8cm]{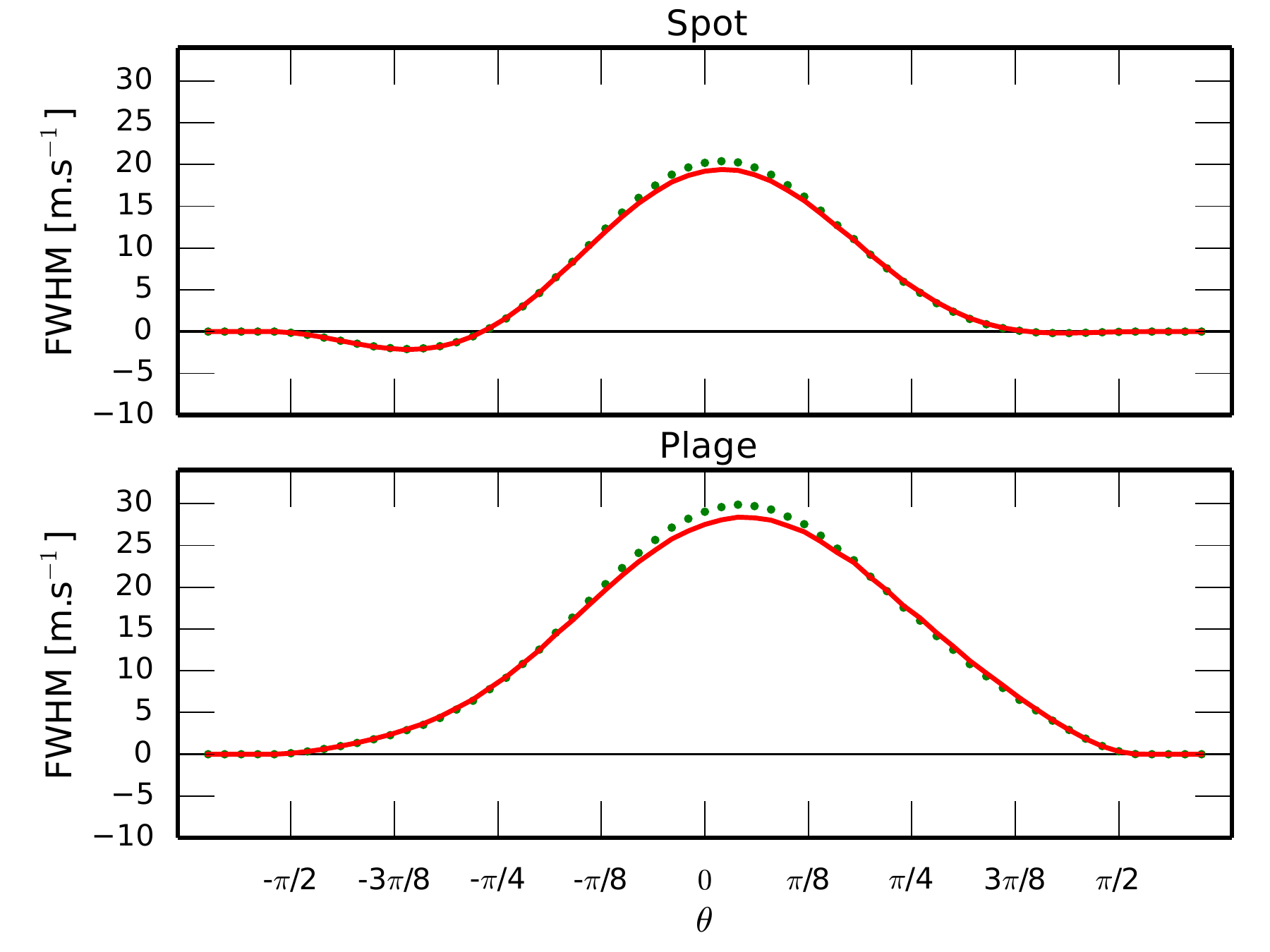}
\caption{Photometric, RV BIS SPAN, and FWHM variations induced by an equatorial spot or plage on an equator on star when assuming a linear limb darkening law (green dotted line) and a quadratic one (red continuous line). The size of the active region is 1\%. The contrast of the active region is 0.54 in the case of a spot ($663 K$ cooler than the effective temperature of the Sun, at 5293\,\AA.), and is given by Eq. \ref{eq:2-2-0} in the case of a plage. The active region is on the stellar disc center when $\theta =$ 0 and on the limb when $\theta = \pm\,\pi/2$.}
\label{fig:3-1}
\end{center}
\end{figure*}

In SOAP, as we can see in Eq. \ref{eq:3} and \ref{eq:4}, the limb darkening law affects quiet regions, as well as active ones. Using different limb darkening laws will therefore affect all regions of the star. 
Fig. \ref{fig:3-1} displays the variations induced by a spot or a plage, assuming a linear limb darkening law with with parameter value $0.6$ and a quadratic one with $\gamma_1=0.29$ and $\gamma_2=0.34$ (see Equation \ref{eq:2-2-1}). As we can see in Table \ref{tab:1}, the RV and BIS SPAN peak-to-peak differences for the variation induced by an active region when comparing a linear and quadratic law is at the detection limit of future spectrographs designed to reach the 0.1\ms precision level (e.g. ESPRESSO@VLT, G-CLEF@GMT). The impact of the limb-darkening law on the RV and BIS SPAN measurements is therefore very small, which implies that limb-darkening laws more precise than quadratic are not useful for estimating the variation of these observables. The photometry is however significantly affected when assuming a quadratic rather than a linear limb darkening law for a spot. An effect of 600 ppm is easily detected by a space-based instruments like \emph{Kepler} and therefore a quadratic limb darkening law should be used to properly estimate the photometric effect of spots. Finally, the FWHM is also significantly influenced.
\begin{deluxetable}{ccccc}
	\tabletypesize{\scriptsize}
	\tablewidth{8cm} 
	\tablecaption{Peak-to-peak differences in photometry and in the CCF parameters when comparing different limb darkening laws. \label{tab:1}}
		\startdata 
			\tableline
			\tableline  
			Spot & Flux & RV & BIS SPAN & FWHM \\
			  & [ppm] &  [m\,s$^{-1}$] &  [m\,s$^{-1}$] &  [m\,s$^{-1}$]\\
			\tableline
Lin - quad & 613 (5\%) & 0.1 (0\%) & $<$0.1 (0\%) & 0.9 (4\%) \\
			\tableline
			\tableline
			Plage & Flux & RV & BIS SPAN & FWHM\\
			  & [\%] &  [m\,s$^{-1}$] &  [m\,s$^{-1}$] &  [m\,s$^{-1}$]\\
			\tableline
Lin - quad & 22 (3\%) & 0.4 (4\%) & $<$0.1 (0\%) & 1.5 (5\%) \\
			\tableline
		\enddata
	\tablecomments{In brackets, the fraction of the difference is shown in percent. Values higher than 10\% are highlighted in bold face. The spot and the plage have the same properties than in Fig. \ref{fig:3-1}.}
\end{deluxetable}
%


\subsection{Modifications brought by the spectrograph resolution}  \label{sec:3-2}

\begin{figure*}
\begin{center}
\includegraphics[width=8cm]{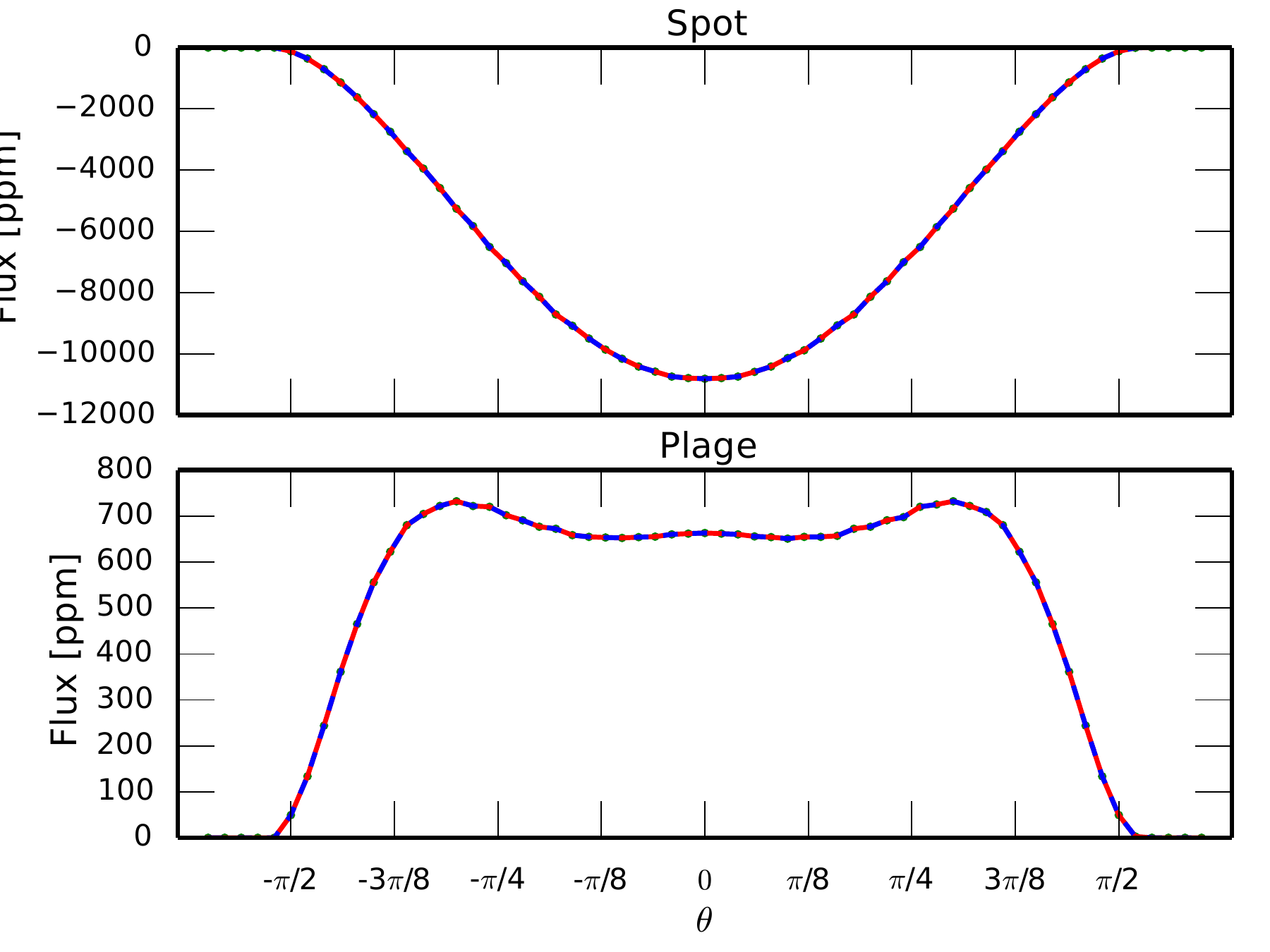}
\includegraphics[width=8cm]{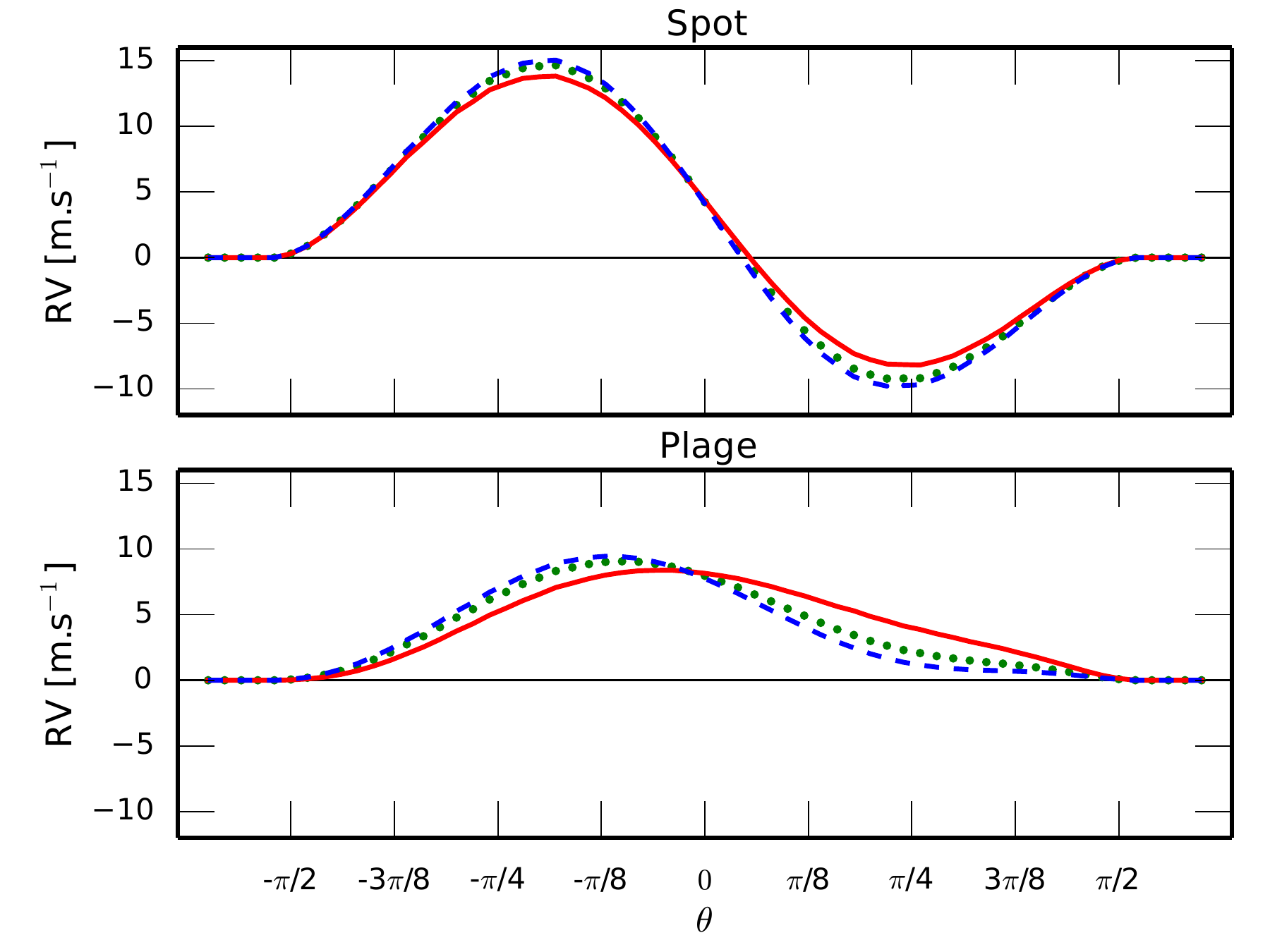}
\includegraphics[width=8cm]{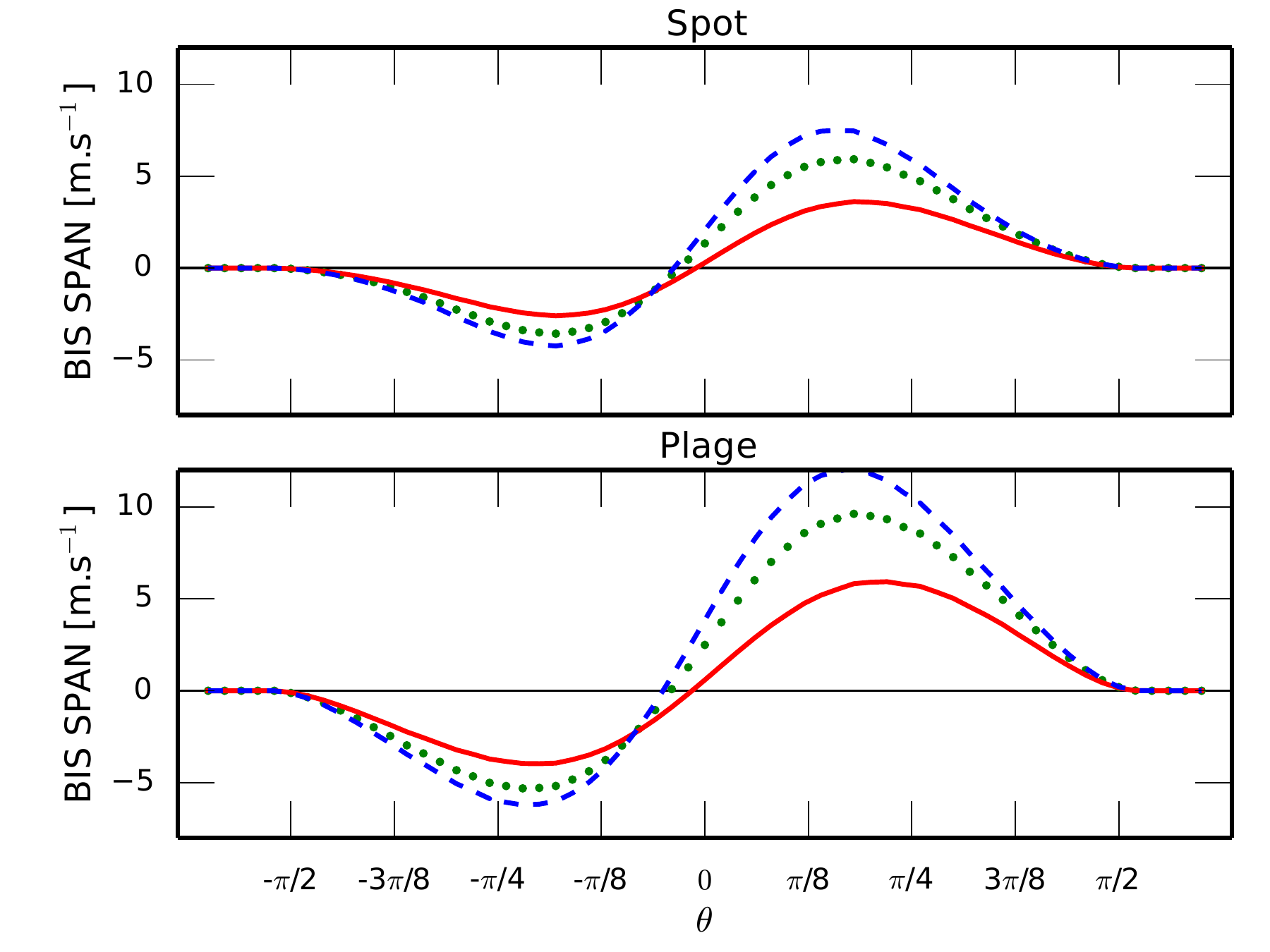}
\includegraphics[width=8cm]{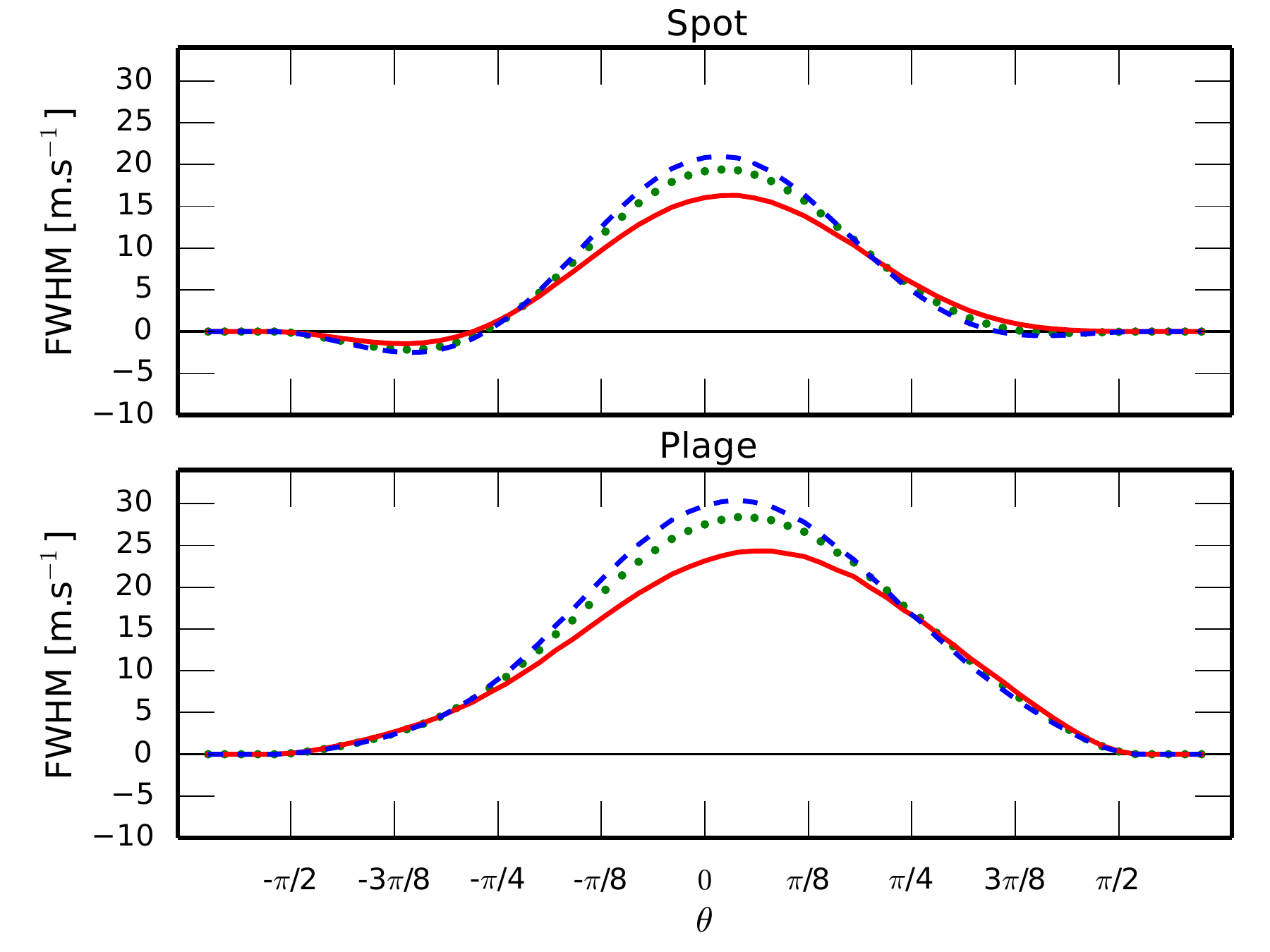}
\caption{Photometric, RV BIS SPAN, and FWHM variations induced by an equatorial spot or plage on an equator on star when assuming different instrument resolution: $R>700000$ (FTS, blue dashed line), $R=115000$ (HARPS, green dotted line) and $R=55000$ (CORALIE, red continuous line). The size of the active region is 1\%. The contrast of the active region is 0.54 in the case of a spot ($663 K$ cooler than the effective temperature of the Sun, at 5293\,\AA.), and is given by Eq. \ref{eq:2-2-0} in the case of a plage. The active region is on the stellar disc center when $\theta =$ 0 and on the limb when $\theta = \pm\,\pi/2$.}
\label{fig:3-2}
\end{center}
\end{figure*}

When observing with different instrumental resolutions, the integrated CCF over the stellar disc will be modified because the star is a point source, and thus the integrated light over the stellar disc is fed into the spectrograph. This integrated CCF will be convolved with a Gaussian instrumental profile that has a FWHM that depends on the instrumental resolution. As we can see in the last paragraph of Sec. \ref{sec:2-2}, the lower the resolution, the larger will be the FWHM of the instrumental profile. Fig. \ref{fig:3-2} displays the photometric, RV, BIS SPAN and FWHM variations when assuming the instrumental resolutions of the FTS ($R>700000$), of HARPS ($R=115000$) and of CORALIE ($R=55000$). As we can see in Fig. \ref{fig:3-2} and Table \ref{tab:2}, the photometry is not affected because this observable only depends on the intensity of the active region and the limb darkening law used (Equation \ref{eq:4}), which are not modified here. The difference in RV variation is small, $\sim10$\%, because convolving with a Gaussian will not modify significantly the center of the CCF. However, the peak-to-peak amplitudes seen in the BIS SPAN and the FWHM will be strongly reduced by lower resolutions because convolving with wider instrument profiles will average out any asymmetry or intrinsic width of the CCF. In Table \ref{tab:2}, we can see that the BIS SPAN peak-to-peak amplitude obtained with the FTS resolution is nearly twice that obtained with the CORALIE resolution. The differences between the FTS and HARPS resolutions are smaller but still approximatively 20\% of the BIS SPAN peak-to-peak amplitude (see Table \ref{tab:2}). As already demonstrated by other authors \citep[][]{Boisse-2012b,Desort-2007}, a high resolution will therefore strengthen the correlation between the RV and the BIS SPAN or the FWHM variations. These correlations are often a powerful diagnostic to decide if the observed RV signal is of activity or planetary origin \citep[e.g.][]{Boisse-2009,Bonfils-2007,Queloz-2001}.
\begin{deluxetable}{ccccc}
	\tabletypesize{\scriptsize}
	\renewcommand{\arraystretch}{0.7}    
	\tablewidth{8cm} 
	\tablecaption{Peak-to-peak differences in photometry and in the CCF parameters when comparing different instrumental resolutions. \label{tab:2}}
		\startdata 
			\tableline
			\tableline  
			Spot & Flux & RV & BIS SPAN & FWHM \\
			  & [ppm] &  [m\,s$^{-1}$] &  [m\,s$^{-1}$] &  [m\,s$^{-1}$]\\
			\tableline
			R$_{FTS}$ - R$_{HARPS}$ & 0 (0\%) & 1.0 (4\%) & 2.2 ({\bf 23\%}) & 1.9 (9\%) \\
			R$_{FTS}$ - R$_{CORALIE}$ & 0 (0\%) & 2.8 ({\bf 12\%}) & 5.5 ({\bf 88\%}) & 5.7 ({\bf 32\%}) \\
			R$_{HARPS}$ - R$_{CORALIE}$ & 0 (0\%) & 1.9 (8\%) & 3.3 ({\bf 52\%}) & 3.8 ({\bf 21\%}) \\
			\tableline
			\tableline
			Plage & Flux & RV & BIS SPAN & FWHM\\
			  & [\%] &  [m\,s$^{-1}$] &  [m\,s$^{-1}$] &  [m\,s$^{-1}$]\\
			\tableline
			R$_{FTS}$ - R$_{HARPS}$ & 0 (0\%) & 0.4 (4\%) & 3.4 ({\bf 22\%}) & 2.0 (7\%) \\
			R$_{FTS}$ - R$_{CORALIE}$ & 0 (0\%) & 1.1 ({\bf 12\%}) & 8.4 ({\bf 85\%}) & 6.1 ({\bf 25\%}) \\
			R$_{HARPS}$ - R$_{CORALIE}$ & 0 (0\%) & 0.7 (8\%) & 5.0 ({\bf 50\%}) & 4.0 ({\bf 16\%}) \\
			\tableline		
		\enddata
	\tablecomments{Different resolution: R$_{FTS}>700000$, R$_{HARPS}=115000$ and R$_{CORALIE}=55000$. In brackets, the fraction of the difference is shown in percent. Values higher than 10\% are highlighted in bold face. The spot and the plage have the same properties as in Fig. \ref{fig:3-2}.}
\end{deluxetable}

\subsection{Modifications brought by the convective blueshift}  \label{sec:3-3}

\begin{figure*}
\begin{center}
\includegraphics[width=8cm]{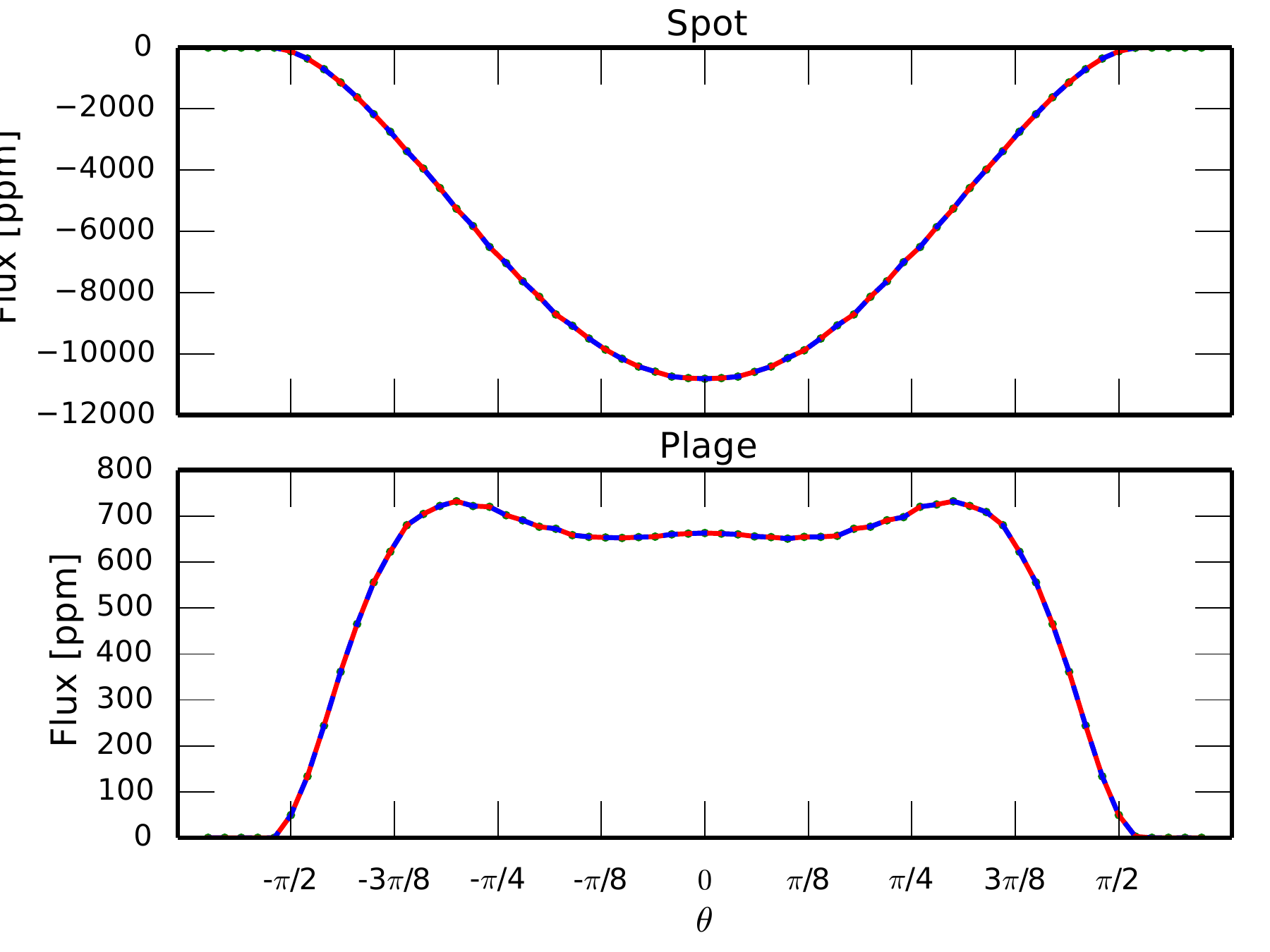}
\includegraphics[width=8cm]{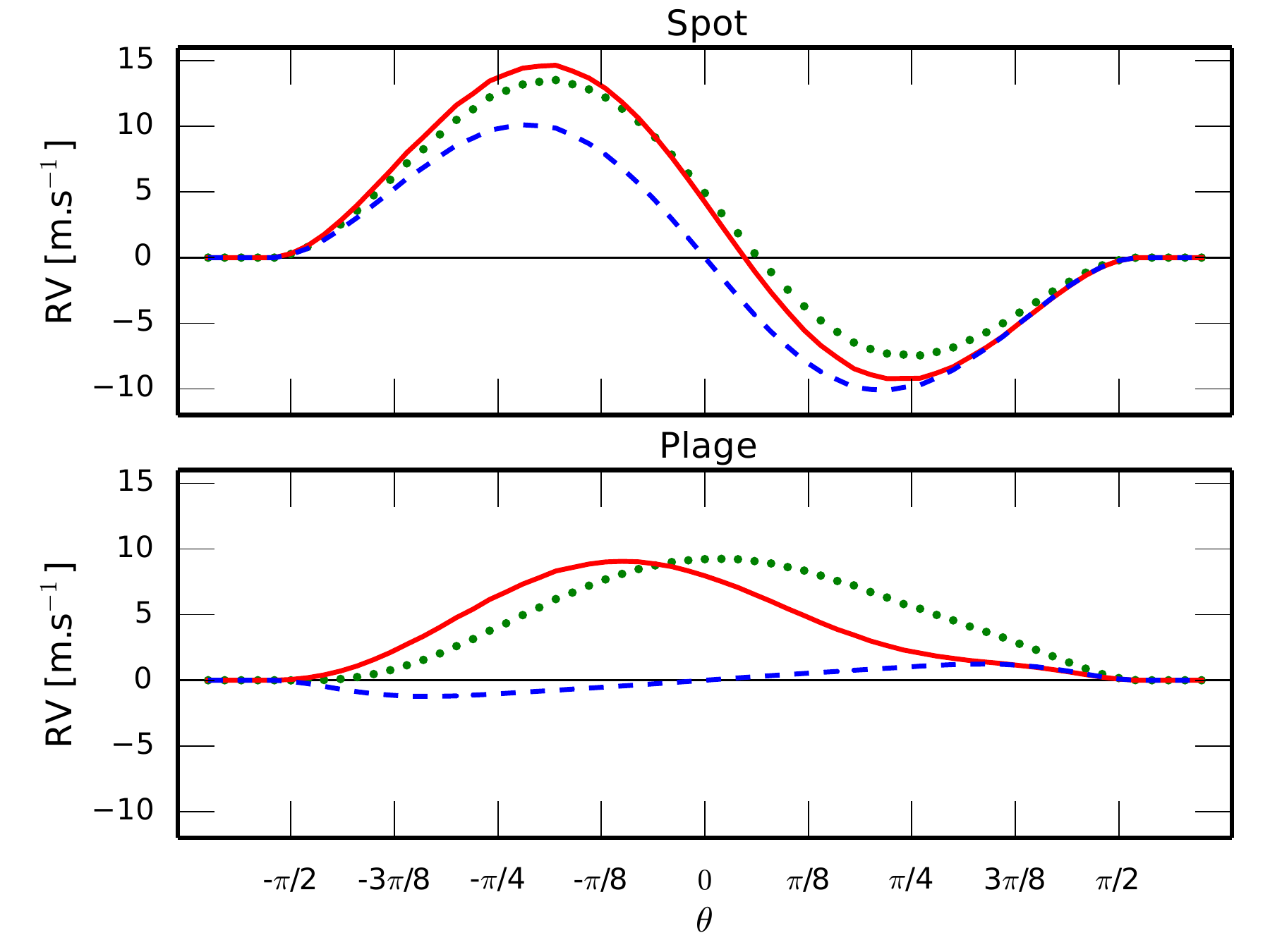}
\includegraphics[width=8cm]{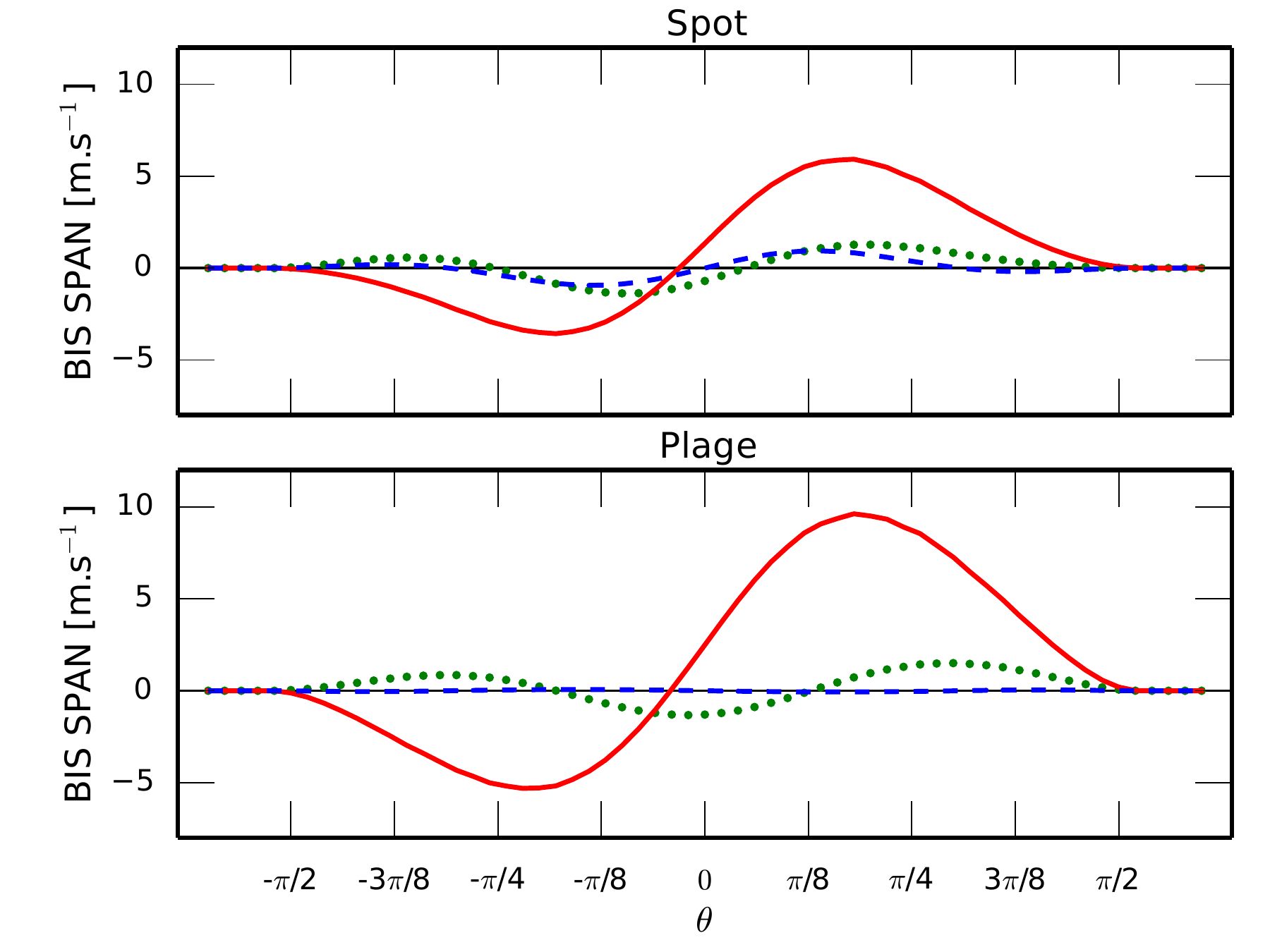}
\includegraphics[width=8cm]{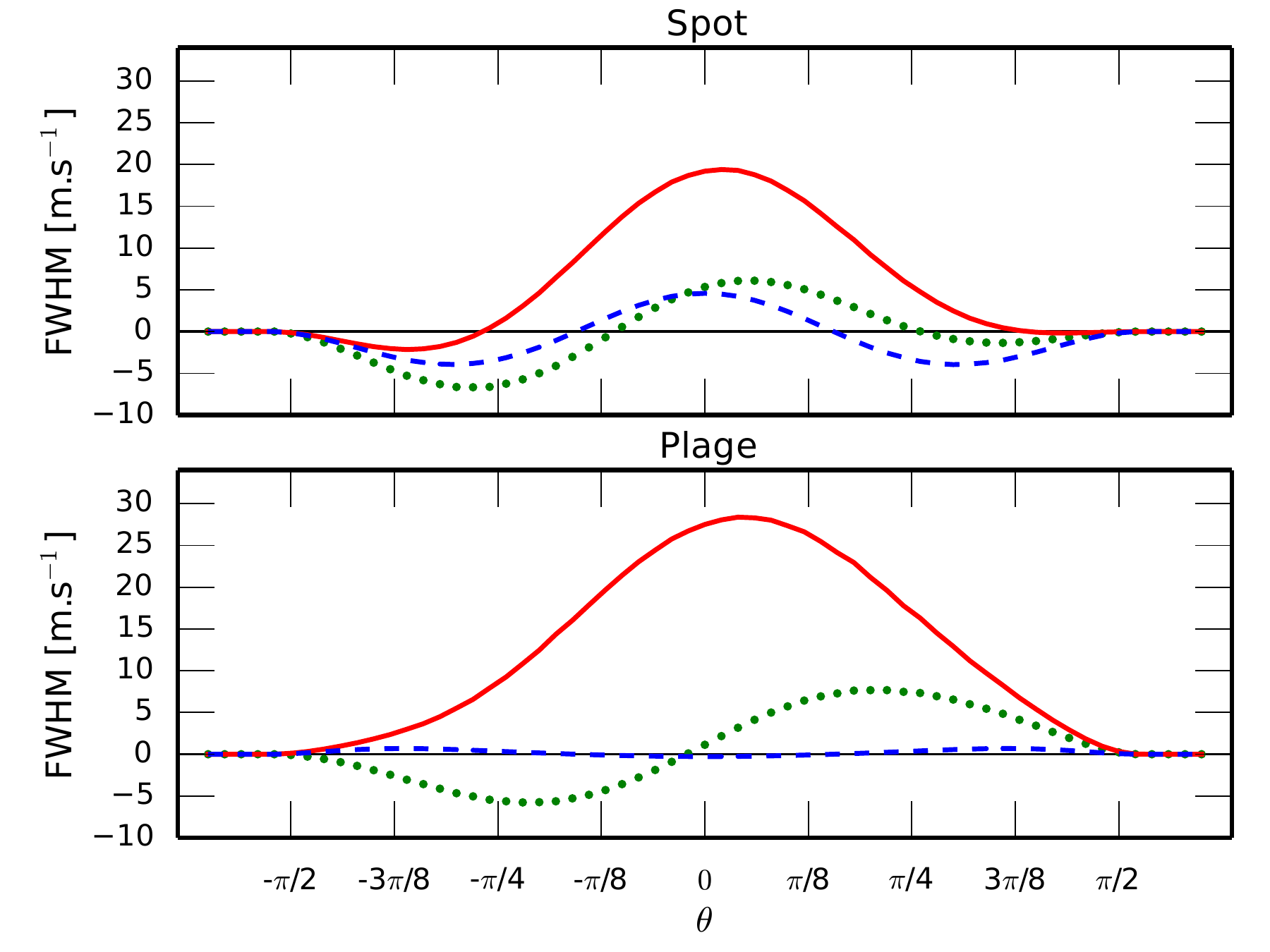}
\caption{Photometric, RV BIS SPAN, and FWHM variations induced by an equatorial spot or plage on an equator on star when assuming different CCFs in SOAP: same Gaussian CCF in the quiet photosphere and in the active region (blue dashed line), same Gaussian CCF but shifted by 350\ms in the active region (green dotted line) and observed solar CCFs (red continuous line). The size of the active region is 1\%. The contrast of the active region is 0.54 in the case of a spot ($663 K$ cooler than the effective temperature of the Sun, at 5293\,\AA.), and is given by Eq. \ref{eq:2-2-0} in the case of a plage. The active region is on the stellar disc center when $\theta =$ 0 and on the limb when $\theta = \pm\,\pi/2$.}
\label{fig:3-3}
\end{center}
\end{figure*}

\begin{figure*}
\begin{center}
\includegraphics[width=8cm]{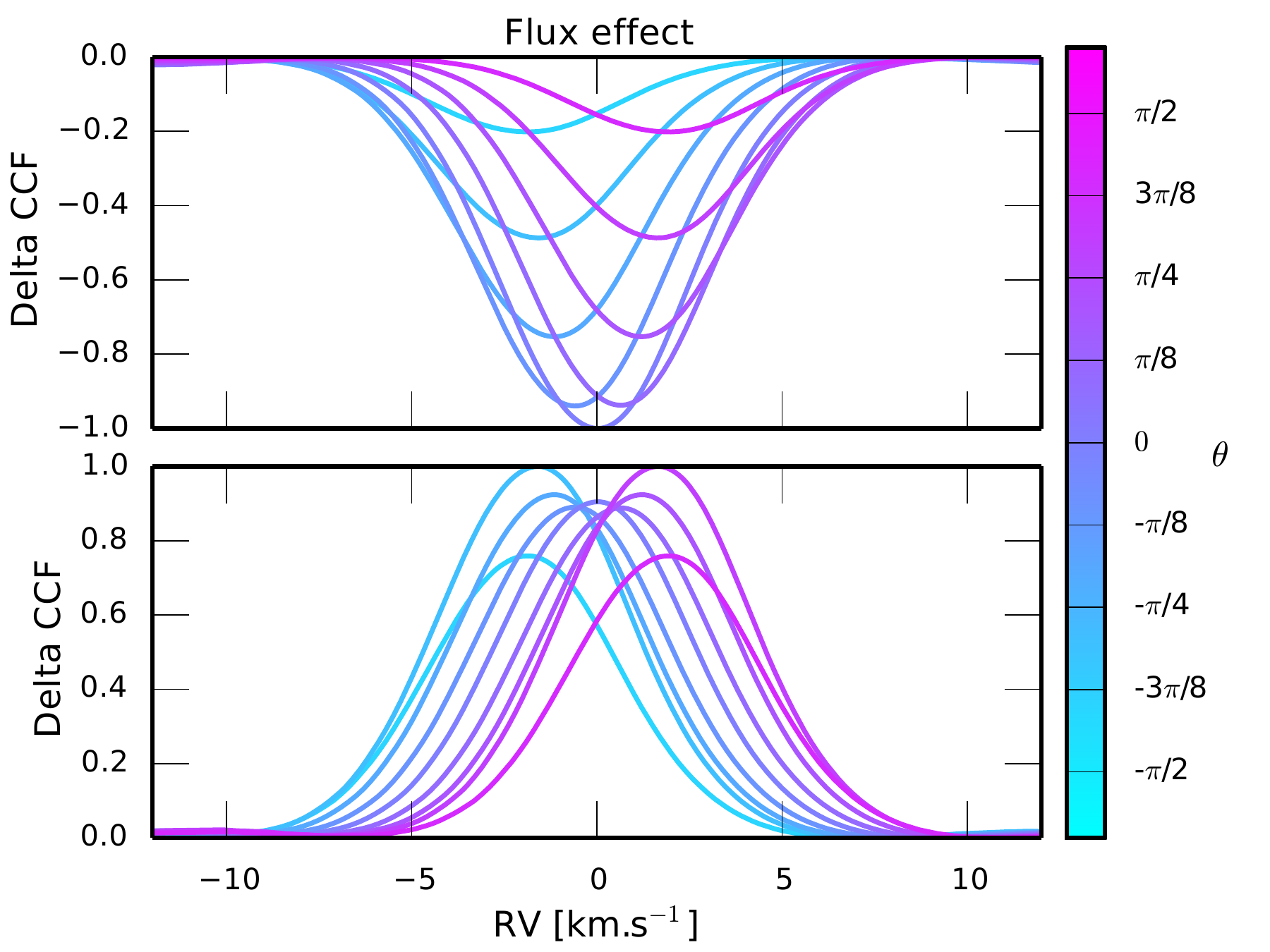}
\includegraphics[width=8cm]{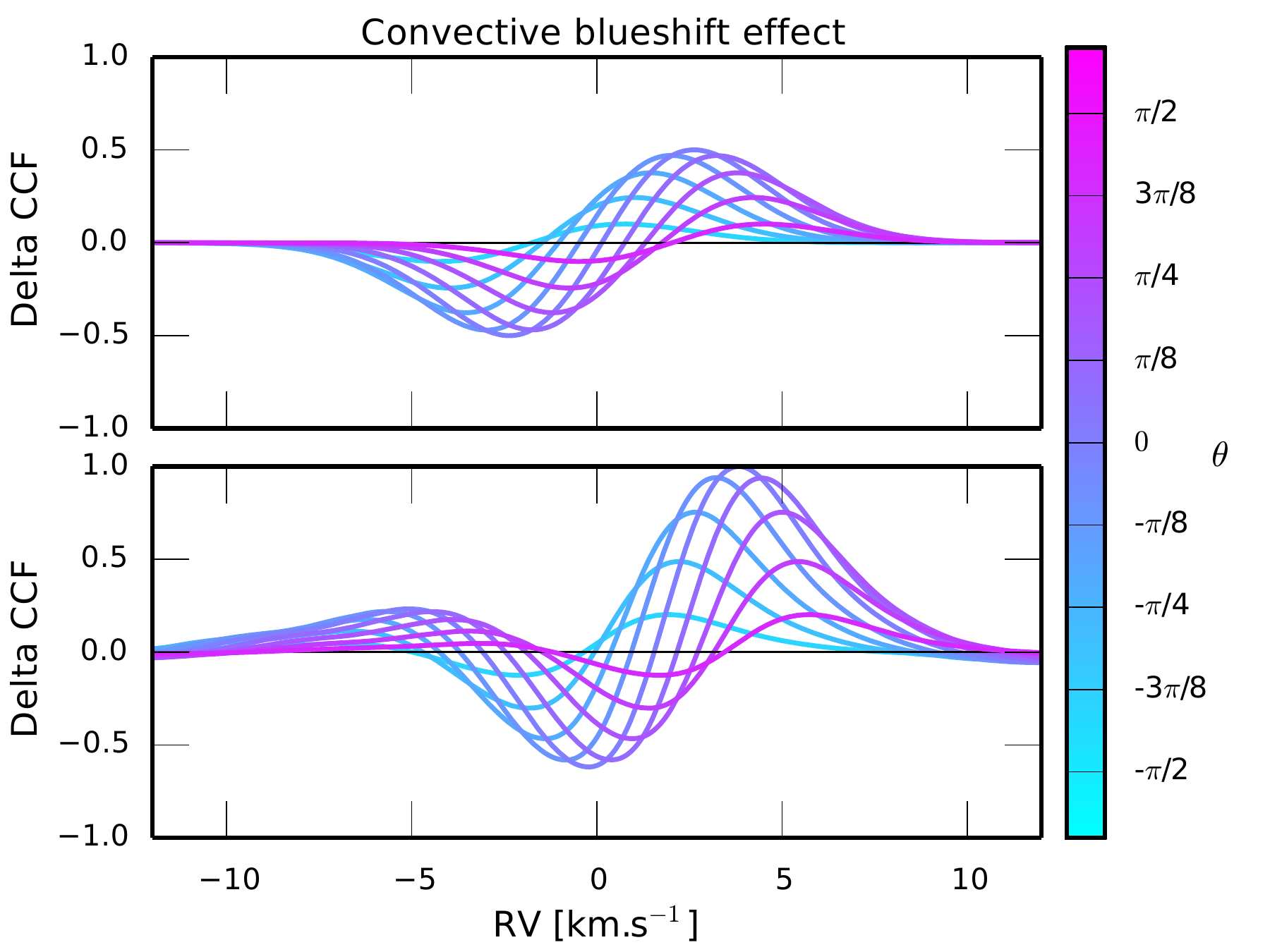}
\caption{Correction $\Delta \mathrm{CCF}$ to apply to the quiet Sun integrated CCF, $\mathrm{CCF}_{\mathrm{tot,\,quiet}}$, to estimate the effect induced by an equatorial spot or plage of size 1\% on a star seen equator on.
		\emph{Left: } Flux correction $\Delta \mathrm{CCF}_{\mathrm{flux}}$ for an equatorial spot (\emph{top graph}) or plage (\emph{bottom graph}).
	        \emph{Right: } Convective blueshift correction $\Delta \mathrm{CCF}_{\mathrm{conv\,blue}}$ of an active region when assuming Gaussian CCFs shifted by 350\ms (\emph{top graph}), and when assuming observed CCFs (\emph{bottom graph}). The use of observed (non-symmetric) CCFs, breaks the symmetry of the convective blueshift correction to apply to the quiet Sun integrated CCF.}
\label{fig:3-4}
\end{center}
\end{figure*}

We now consider the effect on the photometry, the RV, the BIS SPAN and the FWHM induced by the use of different CCFs. We will compute the differences between the following three cases: i) assuming the same Gaussian CCF in the quiet photosphere and in the active region, ii) assuming the same Gaussian CCF but shifted by 350\ms inside an active region, and iii) assuming CCFs of observed solar spectra of the quiet photosphere and of an active region. We use the quadratic limb darkening law and the HARPS resolution ($R=115000$) to derive the results of these three different assumptions that can be seen in Fig. \ref{fig:3-3} and Table \ref{tab:3}. As in the previous case when comparing different resolutions, the photometry is not affected because this observable only depends on the intensity of the active region and the limb darkening law used (Equation \ref{eq:4}), which are not modified here. The RV induced effect of a spot is not very different when assuming Gaussian CCFs or observed CCFs. This is because the spot intensity is 54\% of the quiet photosphere brightness ($663 K$ cooler than the effective temperature of the Sun, at 5293\,\AA), and therefore the flux effect of the spot is more important than the effect induced by the convective blueshift and its inhibition in active regions. A plage is however only 3\% to 22\% brighter than the quiet photosphere and in this case the convective blueshift effect becomes $\sim$10 times more important than the flux effect. We see in Fig. \ref{fig:3-3} that there is only a small difference between the RV peak-to-peak amplitudes when using the same Gaussian CCF but shifted by 350\ms inside the active region, or when using observed CCFs (compare the green and red curves). Therefore, the main parameter influencing the RV effect of a plage is the $\sim350$\ms shift difference in velocity between the CCF in the quiet photosphere and the one in the active region, due to the inhibition of the convective blueshift (compare the green and red curves to the blue curve). Another important point is that the flux effect on RVs is anti-symmetric compared to the center of the star (see the RV variation of the spot), while the convective blueshift RV effect is symmetric\footnote{ The effect is not totally symmetric because the bisector of the CCF inside the quiet photosphere is different from the one inside the active region (see Figure \ref{fig:3-4} and related text).} (see the RV variation of the plage). This can be explained because the spot-induced RV variation, dominated by the flux effect, is sensitive to the stellar projected rotational velocity that is anti-symmetric being blueshifted on one half of the star and redshifted on the other half. The plage-induced RV variation, dominated by the convective blueshift effect, will be sensitive to the limb darkening and the plage intensity variation, which are both symmetric effects with respect to the disc center.

Looking at the variations of the BIS SPAN and the FWHM, we see that using observed CCFs, which are not symmetric, induces a much higher peak-to-peak amplitude for both the spot and the plage than when assuming symmetric CCFs (compare the red curve to the blue and green curves). Therefore, the use of observed CCFs modify significantly the BIS SPAN and FWHM amplitude, without influencing the RV amplitude. By studying the RV data of a few active stars, we could test if using observed CCFs rather than Gaussian CCFs gives a better description of the induced effect of active regions. This will be the topic of Section \ref{sect:5}

We can also study the correction $\Delta \mathrm{CCF}$ (see Eq. \ref{eq:2}) to apply to the quiet Sun integrated CCF, $\mathrm{CCF}_{\mathrm{tot,\,quiet}}$, to estimate the effect induced by a spot and a plage. On the left plot of Fig. \ref{fig:3-4}, we can see the flux effect (see Eq. \ref{eq:5}) for a spot and a plage. The contribution of a spot is negative and increases close to the disc center, due to limb darkening. The contribution of a plage is always positive and is maximum in a region between the limb and the center of the disc. This can be explained by a balance between two opposite effects: the limb darkening of the star and the limb brightening of plages. On the right plot of Fig. \ref{fig:3-4}, we compare the convective blueshift effect (see Eq. \ref{eq:6}) when assuming the use of the same Gaussian CCF for all the star but shifted by 350\ms inside the active region (\emph{top plot}), and when assuming the use of observed CCFs (\emph{bottom plot}). Because the CCF that we use for the spot and the plage are the same, we will have the same convective blueshift effect for both type of active region. When assuming a Gaussian CCF, the effect is anti-symmetric with respect to the disc center. This is no longer the case when assuming observed CCFs because the bisector in the quiet photosphere is different from the one in the active region. This explains the important differences in the BIS SPAN and the FWHM variations that we see when using Gaussian CCFs or observed CCFs in our simulation (see Fig. \ref{fig:3-3}).

\begin{deluxetable}{ccccc}
	\tabletypesize{\scriptsize}
	\tablewidth{8cm} 
	\tablecaption{Peak-to-peak differences in photometry and in the CCF parameters when comparing different CCFs injected in each cell. \label{tab:3}}
		\startdata 
			\tableline
			\tableline  
			Spot & Flux & RV & BIS SPAN & FWHM \\
			  & [ppm] &  [m\,s$^{-1}$] &  [m\,s$^{-1}$] &  [m\,s$^{-1}$]\\
			\tableline
G - G shift & 0 (0\%) & 0.7 (3\%) & 0.8 ({\bf 29\%}) & 4.2 ({\bf 33\%}) \\
G - CCF & 0 (0\%) & 3.6 ({\bf 15\%}) & 7.6 ({\bf 80\%}) & 13.0 ({\bf 60\%}) \\
G shift - CCF & 0 (0\%) & 2.9 ({\bf 12\%}) & 6.8 ({\bf 72\%}) & 8.8 ({\bf 40\%}) \\
			\tableline
			\tableline
			Plage & Flux & RV & BIS SPAN & FWHM\\
			  & [\%] &  [m\,s$^{-1}$] &  [m\,s$^{-1}$] &  [m\,s$^{-1}$]\\
			\tableline
G - G shift & 0 (0\%) & 6.8 ({\bf 73\%}) & 2.7 ({\bf 95\%}) & 12.5 ({\bf 92\%}) \\
G - CCF & 0 (0\%) & 6.6 ({\bf 72\%}) & 14.8 ({\bf 99\%}) & 27.4 ({\bf 96\%}) \\
G shift - CCF & 0 (0\%) & 0.2 (2\%) & 12.1 ({\bf 81\%}) & 14.9 ({\bf 52\%}) \\
			\tableline
		\enddata
	\tablecomments{Three cases are considered: the same Gaussian CCF is used for the quiet photosphere and for the active region (G), a Gaussian CCF is used for the quiet photosphere and the same Gaussian CCF, however shifted by 350\ms, is used for the active region (G shift), and CCFs of observed solar spectra of the quiet photosphere and of a spot are used. In brackets, the fraction of the difference is shown in percent. Values higher than 10\% are highlighted in bold face. The spot and the plage have the same properties as in Fig. \ref{fig:3-3}.}
\end{deluxetable}

\section{Testing different stellar and spot parameters}  \label{sec:4}

In this section, we use a quadratic limb darkening law and observed CCFs of the quiet solar photosphere and of a solar active region. From all the tested configurations in the preceding section, this one, that we call from hereon SOAP 2.0, is the most similar to what is observed on the Sun. In the present section, after fixing the instrumental resolution to that of HARPS ($R=115000$), we investigate the effect of different projected stellar rotational velocities, different active region sizes, different stellar inclinations compared to the line of sight, and different latitudes of active regions. If not specified in the text, the star will be seen equator on and its radius and projected rotational velocity will be fixed to the solar value (\vsini$=2$\kms). The active region will be on the equator and will have a size of $S=1$\%. The results of this section are obtained for a star divided in 300$\times$300 cells.

\begin{figure*}
\begin{center}
\includegraphics[width=8cm]{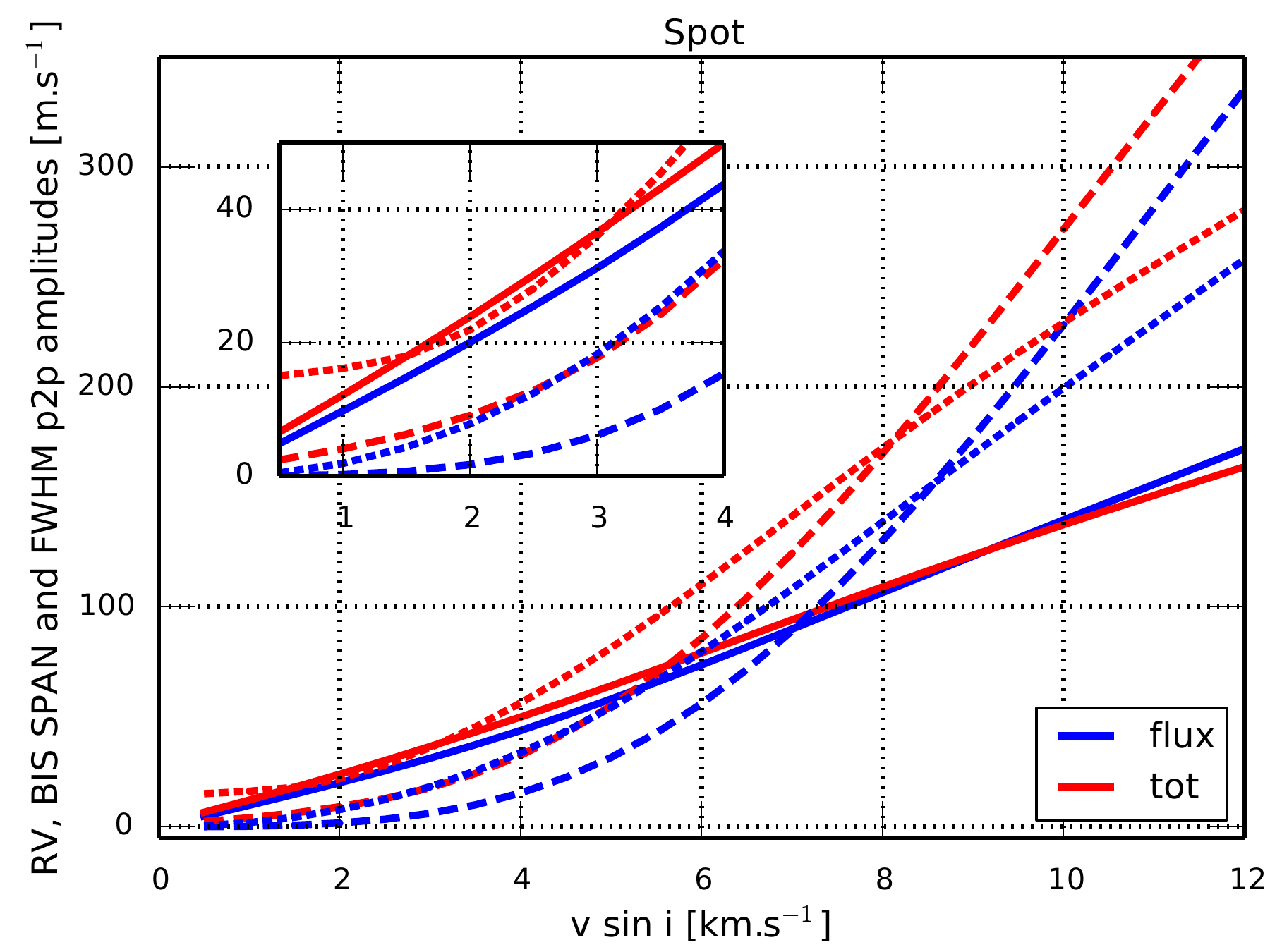}
\includegraphics[width=8cm]{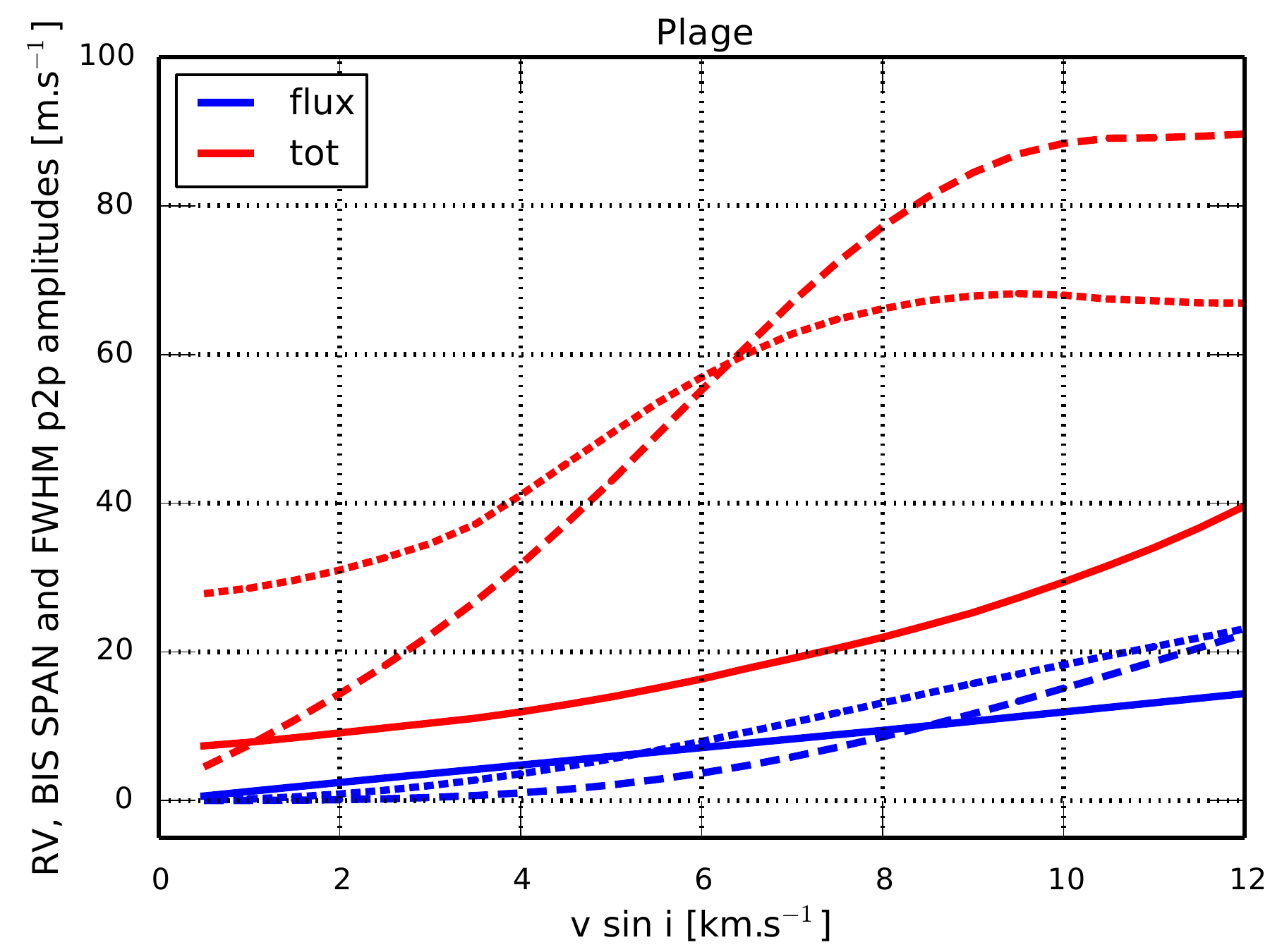}
\caption{Peak-to-peak (p2p) variations of the RV, BIS SPAN and FWHM induced by a spot (\emph{left}) and a plage (\emph{right}) as a function of the stellar projected rotational velocity, \vsini. The variations when assuming the flux effect only (flux, blue lines), and when assuming the flux and the convective blueshift effects (tot, red lines) are shown. The RV, BIS SPAN and FWHM are represented by continuous, dashed and dotted lines, respectively.}
\label{fig:4-0}
\end{center}
\end{figure*}

In Fig. \ref{fig:4-0}, we show the RV, BIS SPAN and FWHM peak-to-peak amplitudes induced by an active region as a function of the projected rotational velocity. We compare the variation induced by the flux effect only (see Eq. \ref{eq:5}), with the variation induced by the total effect, which includes the flux and the convective blueshift effects (see Eq. \ref{eq:3}). When comparing our results for the flux effect only with the ones obtained with SOAP \citep[see Figure 5 in][]{Boisse-2012b}, we notice that our estimation of the peak-to-peak amplitudes for the RV and the BIS SPAN are twice as small\footnote{We arrive to similar values than Figure 5 in \citet{Boisse-2012b}, however we show the peak-to peak amplitudes, while \citet{Boisse-2012b} show the semi-amplitudes.} In \citet{Boisse-2012b}, the spot inducing the RV and the BIS SPAN variations is totally black, while in our case the spot emits 54\% of the flux of a non-active region, explaining this factor of two difference.

For a spot, the RV peak-to-peak amplitude that we obtain from the flux effect and from the total effect are very similar, independently of the projected stellar rotational velocity. This implies that the flux effect dominates the RV variation for any \vsini. For the BIS SPAN and the FWHM peak-to-peak amplitudes, the flux effect dominates for rapid rotators. However, for small \vsini similar to the solar value, the flux effect vanishes and the convective blueshift effect becomes dominant (see the zoom in the left plot of Fig. \ref{fig:4-0}). Therefore, simulations like the original SOAP only including the flux effect manage well to reproduce the variations induced by a spot on rapid rotators. For slow rotators like the Sun, SOAP 2.0 predicts that the BIS SPAN and the FWHM peak-to-peak variations are larger than when considering the flux effect only. 

For a plage, the RV, BIS SPAN and FWHM peak-to-peak amplitudes are dominated by the convective blueshift effect for any projected stellar rotational velocity. This can be explained by the small contrast difference between a plage and the quiet photosphere, which implies a small flux effect. For a plage, SOAP 2.0 yields a much higher RV, BIS SPAN and FWHM peak-to-peak amplitudes than simulations only considering the flux effect predict. 

The introduction of the inhibition of the convective blueshift effect causes the FWHM peak-to-peak amplitude of a plage to be always three times greater than the RV peak-to-peak amplitude, for every stellar rotational velocities smaller than 8\kms. For spots, on the contrary, this is never the case. Therefore the ratio between the FWHM peak-to-peak amplitude and the RV peak-to-peak amplitude informs us on the type of active regions at the origin of the activity-induced variation: spot or plage. For \vsini$\le8$\kms, a ratio smaller than three implies the presence of a spot, while a ratio larger than three indicates the presence of a plage.

\begin{figure*}
\begin{center}
\includegraphics[width=8cm]{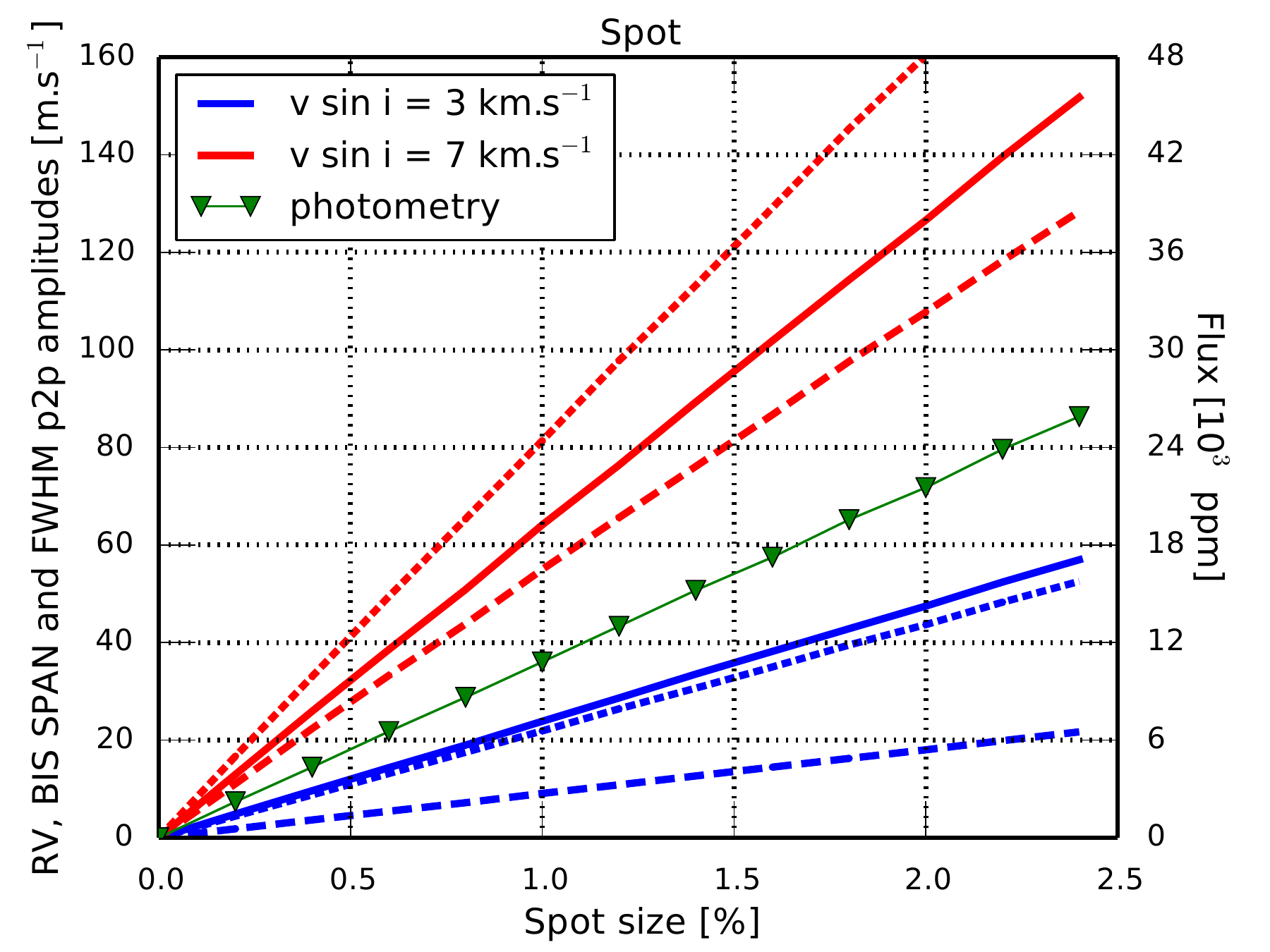}
\includegraphics[width=8cm]{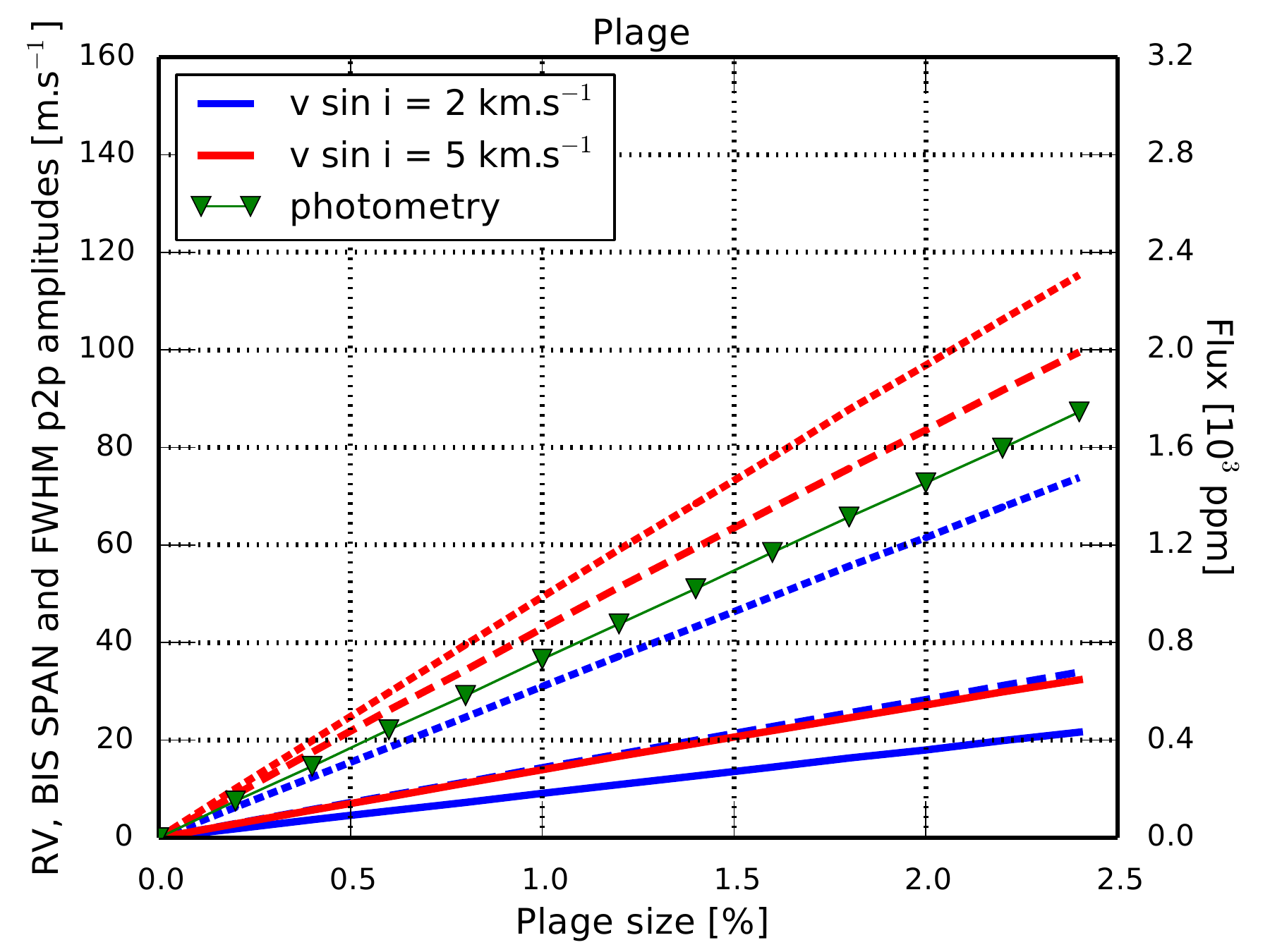}
\caption{Peak-to-peak (p2p) amplitudes of the photometry, RV, BIS SPAN and FWHM induced by a spot (\emph{left}) and a plage (\emph{right}) as a function of the active region size. Two different values for \vsini are considered: 2 and 5\kms (blue lines and red lines, respectively). The RV, BIS SPAN and FWHM are represented by continuous, dashed and dotted lines, respectively. As we can see, the photometric effect is independent of the stellar projected rotational velocity. Note that in the right plot, the red continuous line corresponding to the RV effect for a \vsini=5\kms is overploted on top of the blue dashed line corresponding to the BIS SPAN effect for a \vsini=2\kms.}
\label{fig:4-2}
\end{center}
\end{figure*}

In Fig. \ref{fig:4-2}, we show the RV, BIS SPAN and FWHM peak-to-peak amplitudes as a function of active region size. As we can see, these parameters are proportional to the active region size, which is the case because we stay in the small active region regime where the geometry of these regions does not change. As discussed in the preceding paragraphs, the faster a star rotates, the higher will be the activity-induced effect on the RV, the BIS SPAN and the FWHM. The photometry is the only observable independent of the stellar projected rotational velocity.
%

Using the results of Figs. \ref{fig:4-0} and \ref{fig:4-2}, we can estimate the photometric, RV, BIS SPAN and FWHM peak-to-peak amplitudes as a function of \vsini and the active region size. These results are obtained assuming an equatorial spot on a star seen equator on, which is the configuration leading to the maximal effect. The following equations gives us thus the maximal photometric, RV, BIS SPAN and FWHM peak-to-peak amplitudes that a spot Sp or a plage Pl with a size $S$ (in percent) can induce considering a stellar projected rotational velocity in the range $0.5 \le v\sin{i} \le 12$\kms:
\small
\begin{eqnarray} \label{eq:4-0}
\Delta_{\mathrm{FLUX,Sp}} &=& S\cdot10819\nonumber\\
\Delta_{\mathrm{RV,Sp}} &=& S\cdot \big[1.83\,  + 9.52\, \vsini + 0.79\, \vsini^2 - 0.04\, \vsini^3 \big]\nonumber\\
\Delta_{\mathrm{BIS,Sp}} &=& S\cdot \big[1.84\,  + 2.21\, \vsini -0.46\, \vsini^2 + 0.57\, \vsini^3\nonumber\\
						& & \,\,\,\,\,\,\,\,\,\,\,\, - 0.03\, \vsini^4 \big]\nonumber\\
\Delta_{\mathrm{FWHM,Sp}} &=& S\cdot \big[16.66\,  - 6.06\, \vsini + 4.88\, \vsini^2 -0.21\, \vsini^3 \big]\nonumber\\
\nonumber\\
\tableline
\nonumber\\
\Delta_{\mathrm{FLUX,Pl}} &=& S\cdot728\nonumber\\
\Delta_{\mathrm{RV,Pl}} &=& S\cdot \big[7.66\,  + 0.25\, \vsini + 0.20\, \vsini^2 \big]\nonumber\\
\Delta_{\mathrm{BIS,Pl}} &=& S\cdot \big[4.38\,  + 1.10\, \vsini + 1.98\, \vsini^2 -0.12\, \vsini^3 \big]\nonumber\\
\Delta_{\mathrm{FWHM,Pl}} &=& S\cdot \big[31.93\,  -7.35\, \vsini + 3.92\, \vsini^2 -0.41\, \vsini^3\nonumber\\
						& & \,\,\,\,\,\,\,\,\,\,\,\, + \,0.01\, \vsini^4 \big].					
\end{eqnarray}
\normalsize
%
The flux peak-to-peak amplitude is expressed in ppm and the RV, BIS SPAN and FWHM peak-to-peak amplitudes in\ms. The parameters or these polynomials have been estimated by fitting the results of the simulation.

\begin{figure*}
\begin{center}
\includegraphics[width=8cm]{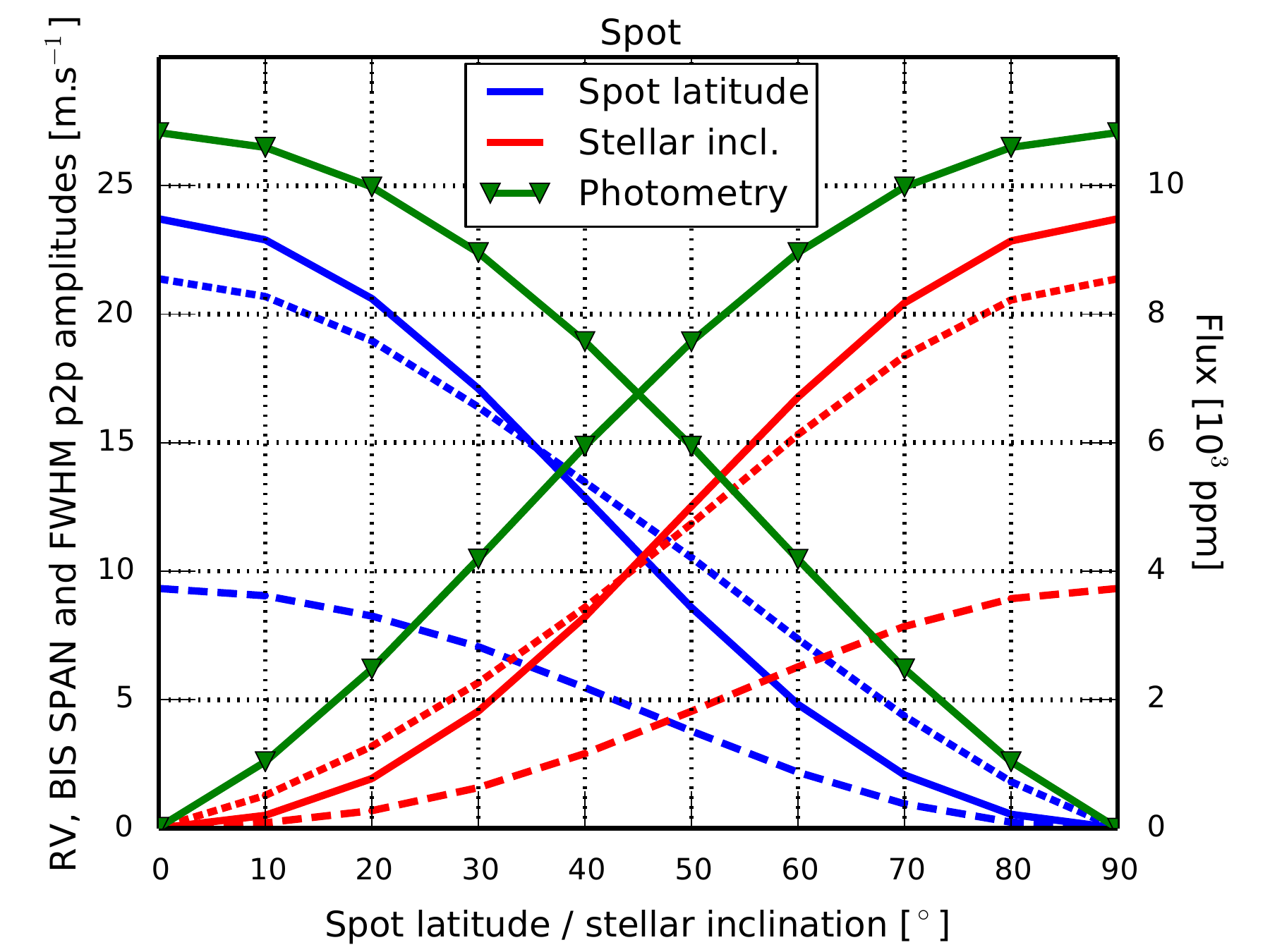}
\includegraphics[width=8cm]{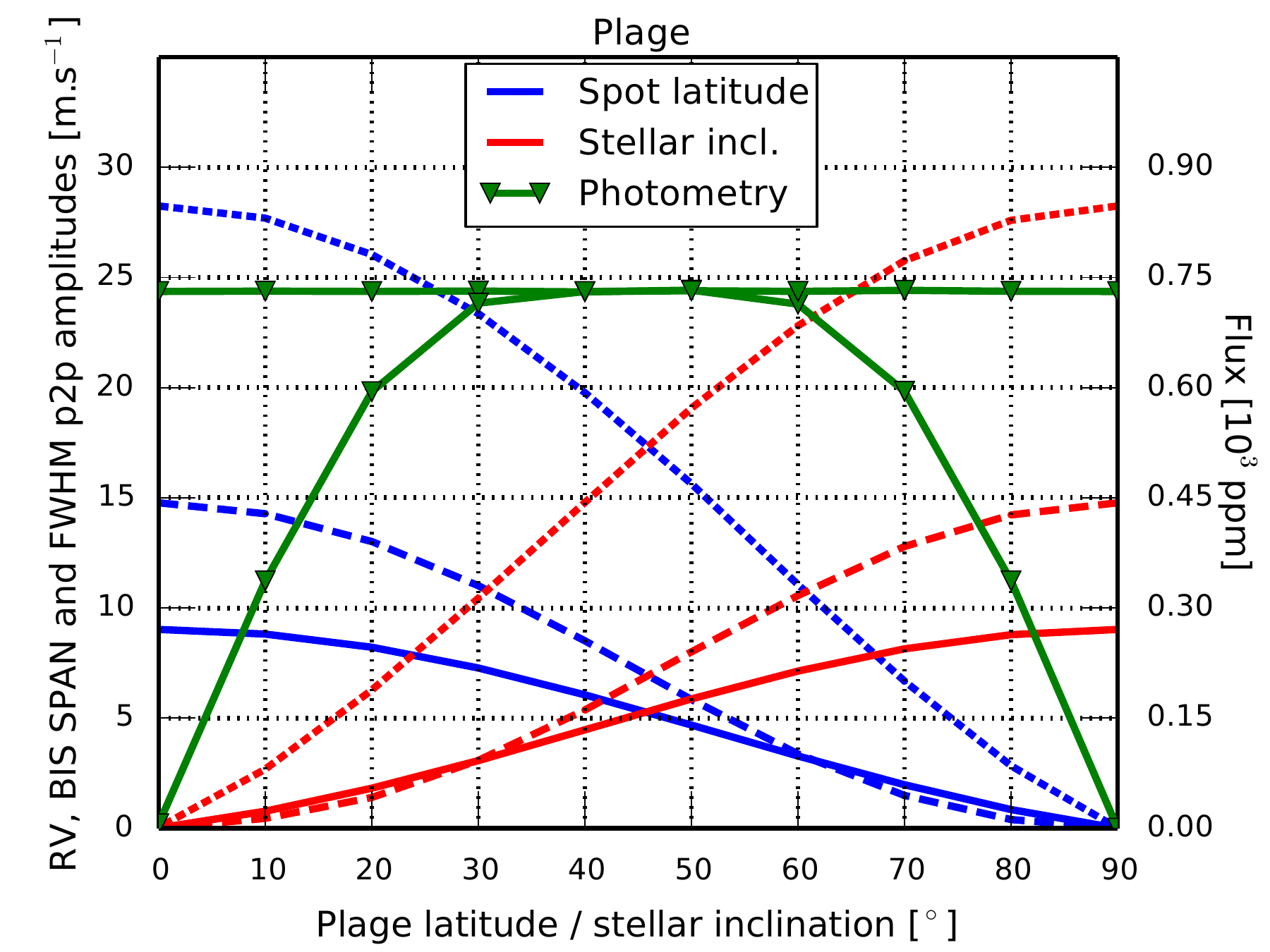}
\caption{Peak-to-peak (p2p) amplitudes of the photometry, RV, BIS SPAN and FWHM induced by a spot (\emph{left}) and a plage (\emph{right}) as a function of the stellar inclination (the active region being at the equator, red lines) and of the active region latitude (the star being equator on, blue lines). The RV, BIS SPAN and FWHM are represented by continuous, dashed and dotted lines, respectively. The photometric variation is shown as green lines with triangle markers. Note that the photometric variation (green line with triangle markers) is the same when the inclination or the latitude is modified.}
\label{fig:4-3}
\end{center}
\end{figure*}

In Fig. \ref{fig:4-3}, we display the photometric, RV, BIS SPAN and FWHM peak-to-peak amplitudes as a function of stellar inclination $i$ (the active region being at the equator), and as a function of active region latitude $\phi$ (the star being equator on, i.e. $i=90$ degrees). The maximum peak-to-peak amplitude for all the different observables are obtained for an equatorial active region on a star observed equator on. For both the spot and the plage, the photometric, RV, BIS SPAN and FWHM peak-to-peak amplitudes vary in a similar way when modifying the stellar inclination or the active region latitude. The amplitudes of the induced variations vary roughly as $\sin{i}^2$ or $\cos{\phi}^2$. These formulae give us an indication of how the peak-to-peak variations change but should not be used to get precise values. Looking at the photometric effect, it is interesting to see that for the plage, the limb darkening is competing with the increase of the active region intensity when approaching the limb, which induce a nearly constant flux from $i=90$ to 30 or from $\phi=0$ to 60 degrees.

\section{Observational test}  \label{sect:5}

In the preceding sections, we presented SOAP 2.0, a new simulation of stellar activity that estimates the impact of spots and plages  on the photometry, RV, BIS SPAN and FWHM. In this section, we compare the result of this simulation with observations of HD189733 and $\alpha$ Cen B. The full description of the method using SOAP 2.0 to fit the data of both stars is described in another paper (Dumusque 2014, submitted). We however show the results here to demonstrate that the results of SOAP 2.0 reproduce the stellar activity observed on solar-type stars.

\subsection{HD189733}\label{sect:5-1}

HD189733 will be used as a first example. This star is rather active because its photometric variability in the visible reaches the percent level \citep[][]{Winn-2007,Croll-2007}. In addition, signs of stellar activity have also been detected in the X-ray \citep[][]{Poppenhaeger-2013a} and in the calcium H \& K activity index \citep[][]{Boisse-2009,Moutou-2007}, which is compatible with a star rotating moderately fast, with a \vsini of $\sim$3\kms \citep[e.g.][]{Triaud-2009} and a rotational period of 11.95 days \citep[][]{Henry-2008}. All these indicators favor a star for which the activity should be dominated by the effect of spots on the stellar surface \citep[][]{Shapiro-2014,Lockwood-2007}. In that case, the flux effect should explain the major part of the photometric, RV, BIS SPAN and FWHM activity-induced variations.

HD189733 has been observed in July 2007 simultaneously in spectroscopy with SOPHIE at the Observatoire de Haute Provence in France and in photometry with the \emph{MOST} satellite. These data have been used to test other activity simulations \citep[][]{Boisse-2012b,Aigrain-2012,Lanza-2011b} that could reproduce the photometric and RV activity-induced variations fairly well. Here we will use the same data set to check if the SOAP 2.0 code manages to also reproduce the variations seen in photometry and RV, in addition to the variation observed in BIS SPAN. The FWHM of HD189733 for the same period exhibits a peak-to-peak amplitude of 135\ms, which is unlikely to be due to activity variations. We therefore decided not to include the FWHM in our fitting procedure.

\begin{figure*}
\begin{center}
\includegraphics[width=16cm]{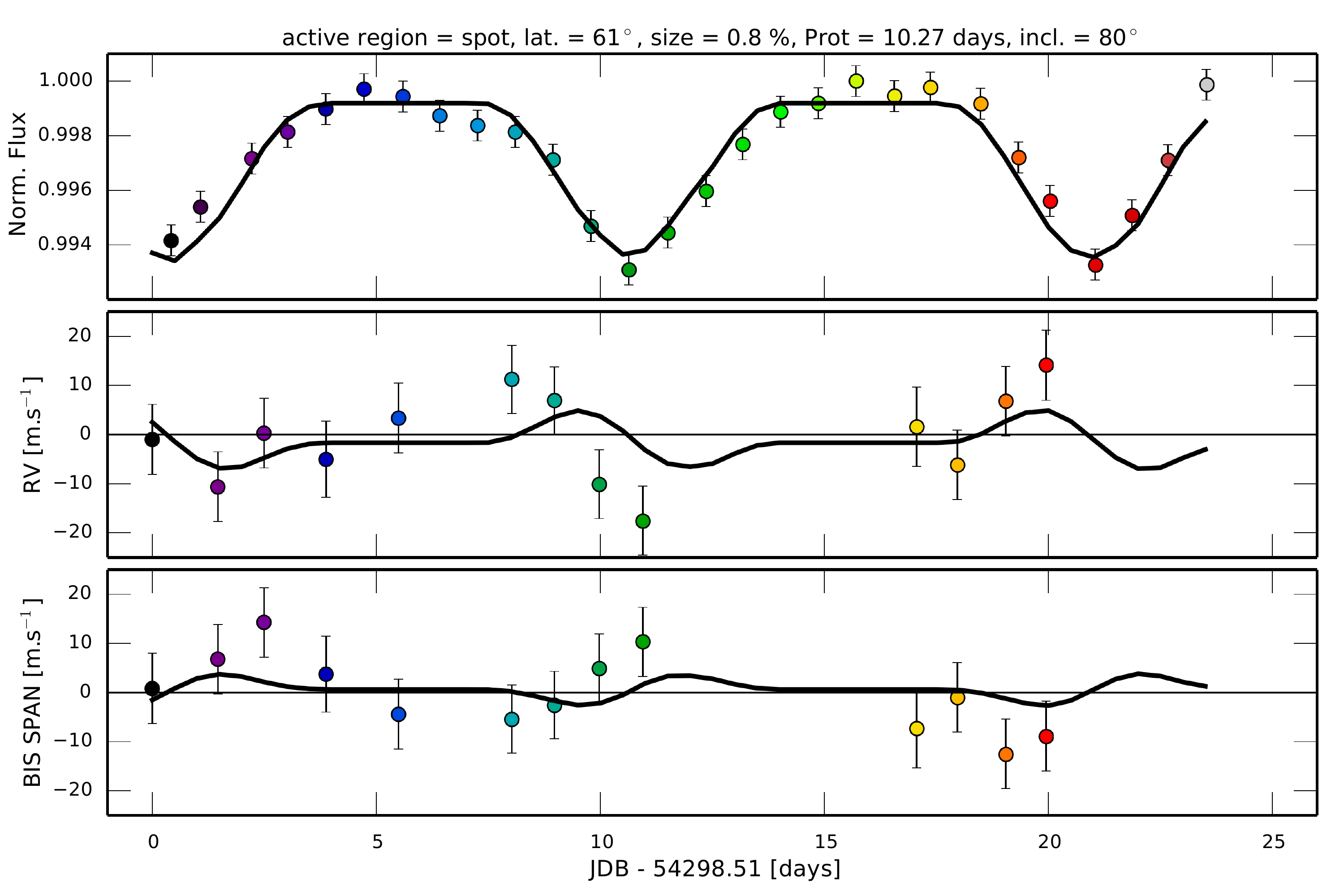}
\caption{Photometric, RV, and BIS SPAN variations and best fit (black continuous line) of the activity-induced signal observed on HD189733. Our best fit to the data corresponds to a 0.8\% spot that can be found at a latitude of 61 degrees. The star rotates in 10.27 days and is seen nearly equator-on with an inclination of 80 degrees. Note that the planetary signal has been removed using the planetary solution of \citet{Boisse-2009} and that the residual RVs and the BIS SPAN have been centered on zero using a weighted mean.}
\label{fig:5-0}
\end{center}
\end{figure*}

Given the regular photometric variation over two rotational periods (see Fig. \ref{fig:5-0}), it is justified to consider that one main active region is present on the stellar surface\footnote{A linear trend was fitted to the \emph{MOST} photometric data to account for an instrumental drift or a long-term activity variation not related to rotational modulation.}. We therefore try to reproduce the activity-induced variation using only one active region. If a plage was at the origin of the photometric variation, the RV and BIS SPAN variations would be much larger in amplitude, and thus we decided to use a spot to fit the data.
%

After binning the \emph{MOST} and SOPHIE measurements over one day and running a Markov Chain Monte Carlo (MCMC) to fit the results of SOAP 2.0 to the photometric, RV and BIS SPAN data (see Dumusque 2014, submitted), we obtain our best fitted solution represented by the black curve in Figure \ref{fig:5-0}. The reduced $\chi^2$ of this model is 1.18 compared to 20.31 for a flat model.

The fit does not match the RV data around BJD = 2454308.5 (10 days in the abscissa of Figure \ref{fig:5-0}).
In the studies by \citet{Aigrain-2012} and \citet{Lanza-2011b}, the same anomaly was reported using a spot model taking into account the flux effect and in some way the convective blueshift effect. The results of SOAP 2.0 show that the convective blueshift effect is not dominating the activity-induced RV variation when spots are present on the stellar surface (see Section \ref{sec:4}). Therefore the flux effect is dominating in this case, and because this effect is estimated in SOAP 2.0 in a similar way than in the works published by \citet{Aigrain-2012} and \citet{Lanza-2011b}, it is not surprising that we find the same anomaly. 
The RV data around BJD = 2454308.5 were obtained near the full moon (BJD = 2454311.5), which can contaminate some spectra in case of clouds. \citet{Boisse-2009} removed strongly contaminated spectra from the observations, however, without simultaneous observation of the sky\footnote{the second fiber was illuminated by a thorium lamp for cross calibration, and not by the nearby sky.}, it is possible that some of the remaining spectra are slightly contaminated. Note that this contamination could be at the origin of the large peak-to-peak amplitude observed in the FWHM. In addition, as already discussed by \citet{Lanza-2011b}, flares could also be the cause of this anomaly, because the calcium activity index of HD189733 can sometimes vary on a very short timescale \citep[][]{Fares-2010,Moutou-2007}.

Removing the two bad points of the anomaly and considering only the spectroscopic data (RV and BIS SPAN), the reduced $\chi^2$ of the fit is 1.17 compared to 1.26 for a flat model, and the standard deviation of the RV residuals is 5.57\ms compared to 7.53\ms, which is an improvement of 5.06\ms. Although the improvement in $\chi^2$ only considering the spectroscopy is not very significant comparing our best fit model to a flat model, we have to note that photometry and spectroscopy are both fitted together and that photometry his much more constraining the fit than spectroscopy in this case. With this slight improvement in $\chi^2$ and the improvement in standard deviation, we are confident that our best fit reproduces better the data than a flat model.

To fit the data of HD189733, we used observed CCF in our simulation. The result would have been similar using the same Gaussian CCFs in the quiet photosphere and the active region, or Gaussian CCFs shifted by 350\ms, because the three different prescriptions predict a similar RV effect (see Figure \ref{fig:3-3}). The prescription using observed CCF predicts though a larger BIS SPAN amplitude, however the precision on the BIS SPAN for HD189733 is not good enough to be able to differentiate between one or the other prescription.

The stellar inclination obtained from the marginalized posterior of our MCMC, $i=84^{+6}_{-20}$ degrees, as well as the latitude of the spot found, $67^{+12}_{-36}$ degrees, are compatible with previous measurements of the spin-orbit angle close to zero degree\footnote{Because the spin-orbit alignment of HD189733b is close to zero degree, the stellar inclination can be different from 90 degrees only in the plane perpendicular to the planetary orbit that contains the line of sight. The probability of the stellar spin being in this plane is very small compared to all the possible orientations and therefore there is a high probability that the stellar inclination is close to 90 degrees.} \citep[][]{Collier-Cameron-2010,Triaud-2009,Winn-2006}, and HST observation of the transiting planet HD189733b occulting stellar spots at $\sim30$ degrees in latitude \citep[][]{Pont-2007}. More information about confronting our results to previous measurements can be found in Dumusque (2014, submitted).
This compatibility with previous observations brings us confidence that the results of SOAP 2.0 manages to reproduce the activity-induce variation of stars that are spot-dominated and rotate moderately fast, like HD189733.

\subsection{$\alpha$ Cen B}\label{sect:5-2}

\begin{figure*}
\begin{center}
\includegraphics[width=16cm]{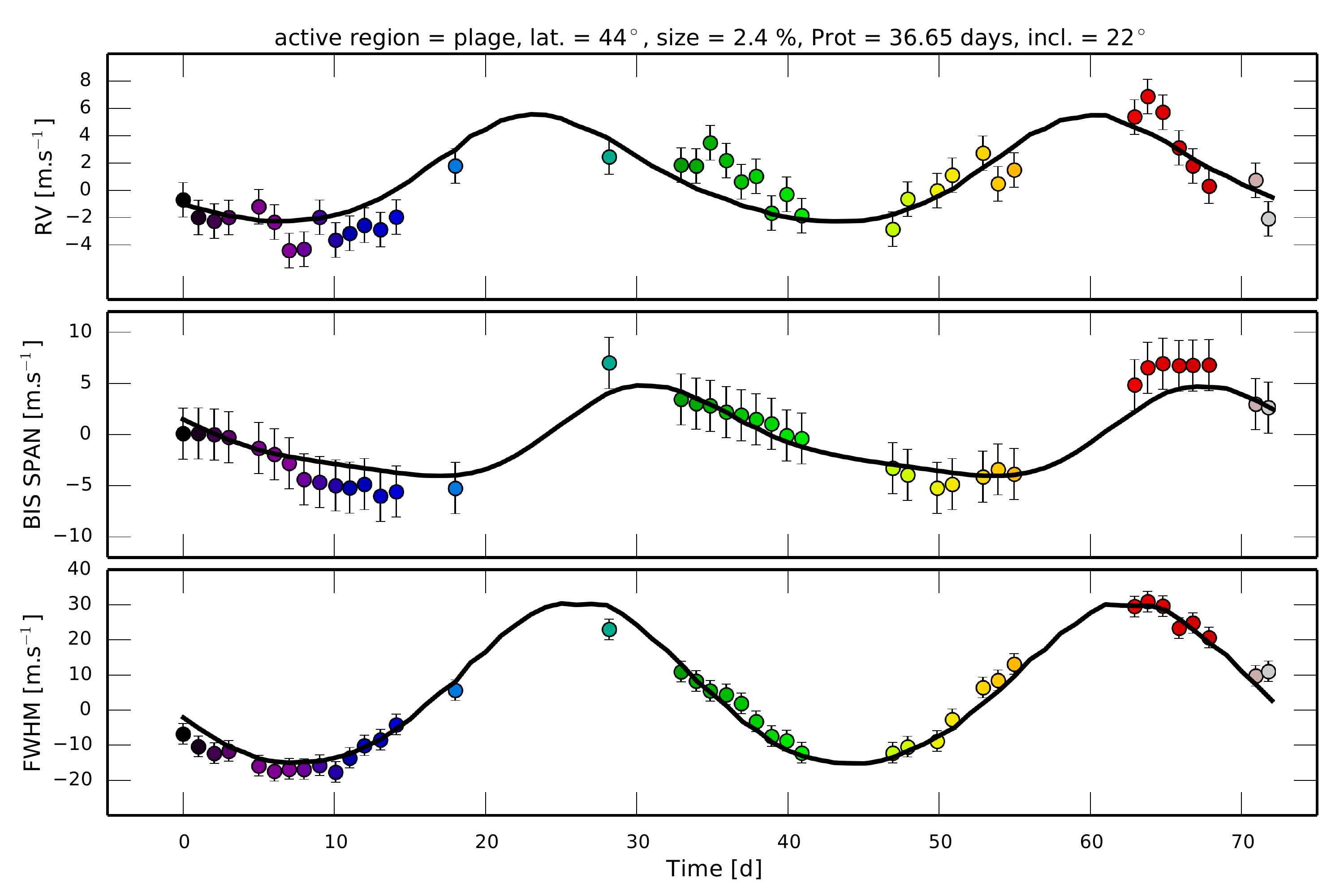}
\caption{RV, BIS SPAN and FWHM variations and best fit (black continuous line) of the activity-induced signal observed on $\alpha$ Cen B. The best-fitted solution corresponds to a plage of size 2.4\% that can be found at a latitude of 44 degrees. The star rotates in 36.65 days and is seen with an inclination of  22 degrees. Note that the binary contribution of $\alpha$ Cen A has been removed from the raw RVs published in \citet{Dumusque-2012} and that the residual RVs, the BIS SPAN and the FWHM have been centered on zero using a weighted mean.}
\label{fig:5-1}
\end{center}
\end{figure*}

As a second example, we want to see if SOAP 2.0 can reproduce the activity induced variation of slow rotators, for which the effect of plages should dominate the activity-induced variation \citep[][]{Shapiro-2014,Lockwood-2007}.
Many slow rotators have been observed with HARPS, HARPS-N and HIRES to search for planets with the sufficient RV precision and cadence to study stellar activity. However, RV surveys are biased towards non-active stars, or stars at the minimum of their activity cycle, for which the RV activity-induced signature is at the level of the instrumental precision. Nevertheless, a few RV observations of slow rotators during their high-activity phase exist, and the best RV measurements to study stellar activity are probably the ones used to detect the closest planet to our Solar System orbiting $\alpha$ Cen B \citep[][]{Dumusque-2012}. The data for 2010 exhibit an important and extremely regular activity index variation \citep[in Ca II H and K,][]{Dumusque-2012} that can be modeled by a single major active region present on the stellar surface. Looking at the spectroscopic measurements of $\alpha$ Cen B in Figure \ref{fig:5-1}, we can notice that the FWHM peak-to-peak amplitude is nearly four times larger than the RV peak-to-peak amplitude. Using the results of Section \ref{sec:4}, this ratio can be explained if a plage is responsible for the activity-induced variation. 

The results of SOAP 2.0 can be used to predict the projected rotational velocity of $\alpha$ Cen B. The ratio between the RV, the BIS SPAN, and the FWHM peak-to-peak amplitudes implies a \vsini of $\sim1$\kms (see Figure \ref{fig:4-0}). This is compatible with the projected rotational velocity calculated using the rotational period of $\alpha$ Cen B, i.e. \vsini$\le 1.15$\kms \citep[rotational period of 37.8 days,][]{Dumusque-2012}.

As for HD189733, we used the results of SOAP 2.0 and run an MCMC to fit the data of $\alpha$ Cen B. The details of the fit can be found in Dumusque (2014, submitted).
Our best fit to the data, shown by the black curve in Figure \ref{fig:5-1}, matches well the observed variations. The reduced $\chi^2$ of this fit is 1.00 compared to 11.17 for a flat model. Only considering the RVs, the reduced $\chi^2$ of the fit is 1.85 compared to 4.90 for a flat model, and the standard deviation of the RV residuals is 1.58\ms compared to 2.73\ms, which is an improvement of 2.22\ms. Our best fitted model is therefore a better representation of the observed RV variations than a flat model and it can be used to correct the RVs from activity variations.

To fit the data of  $\alpha$ Cen B, we used observed CCF in our simulation. Looking at Figure \ref{fig:3-3} and comparing with the observation in Figure \ref{fig:5-1}, it seems clear that to reproduce a significant variation in BIS SPAN and in FWHM, we need to use same Gaussian CCFs but shifted by 350\ms or observed CCFs in our simulation. These two different prescriptions predict a similar RV amplitude, but not the same BIS SPAN and FWHM amplitudes. In the observations, the peak-to-peak amplitude in FWHM is about four times larger than the RV peak-to-peak amplitude, which can only be explained when using observed CCF in the simulation.

\section{Discussion and Conclusion}  \label{sect:6}

This paper presents SOAP 2.0, a new version of SOAP \citep[][]{Boisse-2012b} that estimates the activity-induced variations seen in photometry and spectroscopy. The convective blueshift effect and its inhibition inside active regions is included in the simulation by using spectra of the Sun taken in the quiet photosphere and inside an active region. This inhibition is one of the major cause of the activity-induced variation for slow rotators \citep[][]{Meunier-2010a}. Limb-brightening of plages, i.e. that plages are brighter on the limb that on the disc center \citep[][]{Meunier-2010a}, is also taken into account, as well as a quadratic limb darkening law. Finally, SOAP 2.0 is optimized to estimate the activity-induced variation as seen by high-resolution fiber-fed spectrographs like HARPS, HARPS-N, SOPHIE, CORALIE.

An important result obtained with SOAP 2.0 is that for slow rotators, the convective blueshift effect and its inhibition in active regions plays an important role in the activity-induced variation. For stars with \vsini smaller than 8\kms, the convective blueshift effect dominates the activity-induced FWHM variation, regardless of the type of active region considered, spot or plage. This is different for the RV variation that is dominated by the flux effect in presence of a spot, or by the convective blueshift effect in presence of a plage. This difference has a direct impact on the ratio between the FWHM and the RV peak-to-peak variations, which can be used to characterize the type of the active region responsible for the activity-induced signal. When a major active region is dominating the activity-induced signal, this active region is a spot if this ratio is smaller than three, while it is associated to a plage for larger ratios. Note that this result should be the same when several active regions are present on the star because the total activity variation will just be the sum of the variation of each individual active region.

For fast rotators, the results of SOAP 2.0 show that the flux effect dominates the activity-induced RV, BIS SPAN and FWHM variations when spots are at the origin of the activity-induced signal, while it is the convective blueshift effect that dominates when plages are present on the stellar surface.

When the Sun is at its maximum activity level, it is not uncommon to see one long-lived main active region on the stellar surface. The data presented here for HD189733 and $\alpha$ Cen B show a similar behavior, that therefore seems to be something common among solar-type stars. In the case where only one active region dominates the activity-induced variation and if this active region evolves slowly in comparison with the stellar rotation period, SOAP 2.0 can be used to fit the observed data. Accounting for the convective blueshift effect in SOAP 2.0 allows to better reproduce the variations seen in photometry, RV, BIS SPAN and FWHM. Fitting all these observables simultaneously allows to lift the degeneracy between active region size, active region latitude, and stellar inclination. For more information, readers are referred to the paper from Dumusque (2014, submitted), where the author shows how the stellar inclination can be obtained using the results of SOAP 2.0, even in the case of stellar projected rotational velocity smaller than 2\kms.

When several active regions are present on the stellar surface, it is more difficult to fit the activity signal due to degeneracy between the active region latitudes and sizes, and the stellar inclination. However, if the stellar inclination can be measured on a small subset of the data that shows only one dominant active region, the posterior of the stellar inclination estimated on that subset can then be used as a prior to fit the other parts of the data. Once the stellar inclination is fixed, fitting several active regions becomes less degenerated and computationally more efficient. This will be tested in forthcoming papers. In the case of HD189733 and $\alpha$ Cen B, fitting the stellar activity signal and correcting the RV measurements from it reduces the standard deviation by 5.06 and 2.22\ms, respectively.
 
\begin{figure}
\begin{center}
\includegraphics[width=8cm]{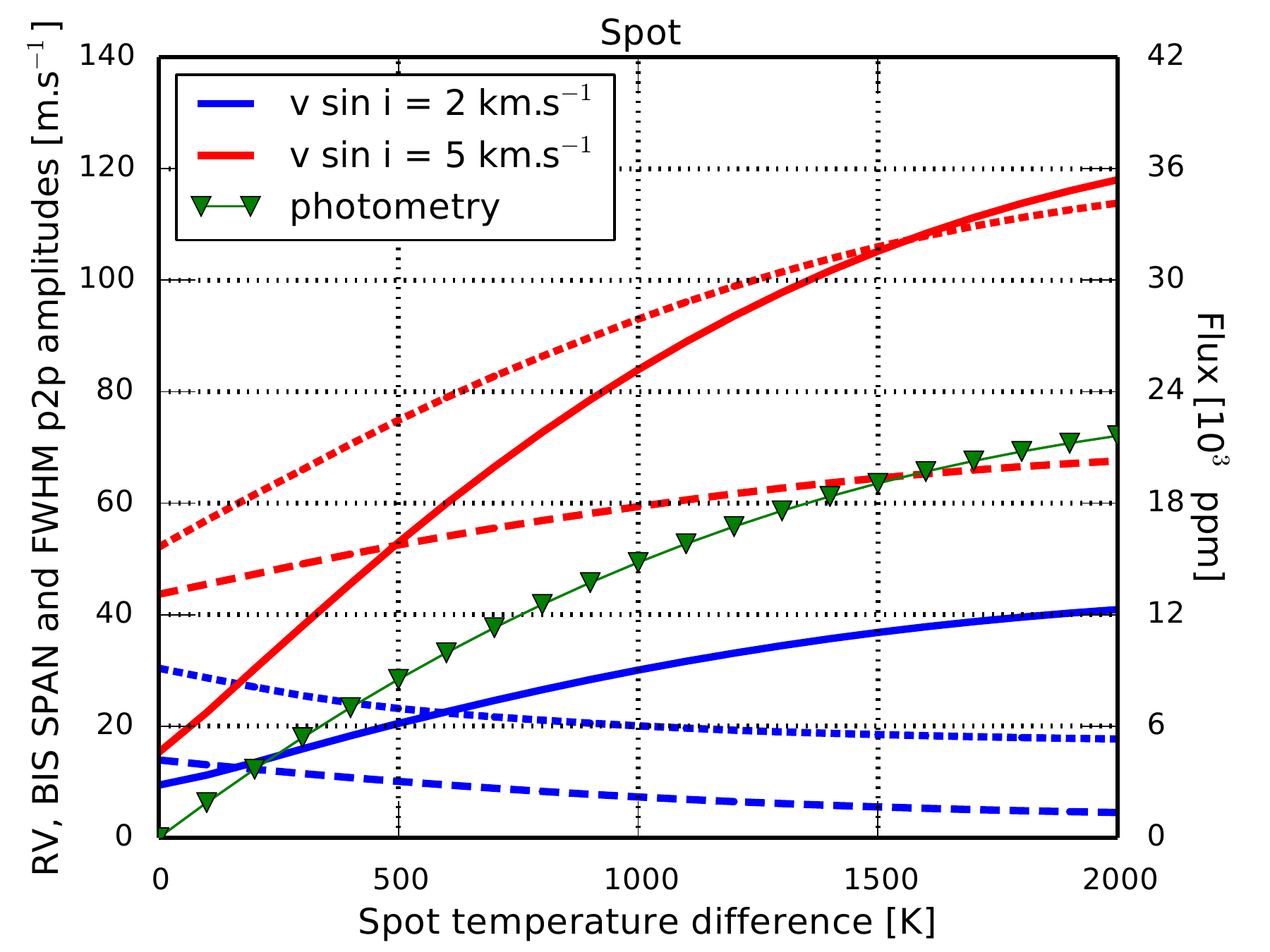}
\caption{Peak-to-peak (p2p) amplitudes of the photometry, RV, BIS SPAN and FWHM induced by a spot as a function of the spot temperature difference with the photosphere. Note that in our simulation, the temperature difference is used to estimate the flux ratio between the active region and the photosphere, and this flux ratio is used to weight the contribution of the active region (see Section \ref{sec:2-0}). Two different values for \vsini are considered: 2 and 5\kms (blue lines and red lines, respectively). The RV, BIS SPAN and FWHM are represented by continuous, dashed and dotted lines, respectively. As we can see, the photometric effect (green line with triangle markers) is independent of the stellar projected rotational velocity.}
\label{fig:6-0}
\end{center}
\end{figure}

The results of this paper are obtained for active regions with a temperature fixed to the solar value. When fitting stellar activity on other stars than the Sun, like the K1 dwarfs HD189733 and $\alpha$ Cen B, the temperature difference between active region and photosphere could be different. The active region temperature is always degenerate with the active region size to first order, because the signal of a big active region with a small contrast can be reproduced by a smaller active region with a higher contrast. However, if we study in detail the activity-induced signal of a spot as a function of the spot temperature difference (see Fig. \ref{fig:6-0}), the ratios between the photometric, RV, BIS SPAN and FWHM peak-to-peak variations change as a function of spot temperature. It is therefore possible that precise spectroscopic measurements can lift some degeneracy between spot temperature and spot size. Another possibility to measure the spot temperature would be multi-band photometry, which has been obtained in the case of HD189733. In \citet{Pont-2013}, the temperature difference of a spot occulted during the transit of HD189733b is estimated to be $-750\pm250\,K$, compatible with our value of $-663\,K$ adopted in SOAP 2.0.
%

Although SOAP 2.0 includes some additional solar physics compared to previous activity codes, the simulation is still simple compared to the complexity of stellar atmosphere and further improvements can be made. In SOAP 2.0, the spectrum of a plage is considered the same as the one of a spot. At first order, this can be done because the most important effect is the inhibition of convection inside a spot or a plage due to strong magnetic fields. This inhibition shifts towards the red the bisectors of the spectral lines, which has an influence on the RV, BIS SPAN and FWHM. The temperature inside a plage is however hundreds of Kelvin higher than inside a spot, which will modify the emerging spectrum of the active region and thus modify the bisector shape. Therefore an observed spectrum of a solar plage should be used in order to include the correct bisector shape of a plage in SOAP 2.0. Furthermore, the convection seen inside a quiet photosphere region depends on the position of this region on the stellar disc, as different depths inside the star are probed. This difference induces a decrease of the granulation contrast when going towards the limb \citep[][]{Sanchez-Cuberes-2003,Beckers-1980,Beckers-1978}, and consequently the amount of convective blueshift decreases as well. This limb-shift effect influences the shape of spectral lines, and the observed "\emph{C}" shape of the bisector in the disc center disappears and is transformed into a "\emph{D}" shape on the stellar limb, as it is observed for the Sun \citep[][]{Cavallini-1985b}. We tried to account for this effect in our simulation by warping the CCF depending on the position on the stellar disc. However, as seen in Appendix \ref{app:3}, we were not able to reproduce the CCF bisector shape observed on solar twins by including this limb-shift variation. More work is therefore needed to take into account this effect in the correct way. This could be done by either using high-resolution spectroscopic observation of the Sun at different center-to-limb angles, or by using MHD stellar atmosphere models. The effect induced by this limb-shift on the shape of the stellar integrated CCF and therefore on the RV, BIS SPAN and FWHM variations is however of the second order (see Figure \ref{fig:app3-1}) because this limb-shift effect affects mostly the limb of the star, which are much fainter than the stellar disc center because of limb darkening.

To finish, we would like to make it clear that the results of SOAP 2.0 are based on one solar observation of the quiet photosphere and one of a spot. We make the assumption that these two spectra are representative of all quiet photosphere regions and active regions. These spectra will be different if we consider different localization on the stellar surface, different magnetic field strengths and configuration, or a star with a different spectral type. However, our test of including the limb shift effect show that if these differences are small, what we think, they will induce a second order effect because the inhibition of the convection inside active region will dominate and because limb effects will be averaged out by limb-darkening. The fact that SOAP 2.0 manages to reproduce the activity-induced variation of the early-K dwarfs HD189733 and $\alpha$ Cen B shows that we have some margins, however we have to be cautious when using SOAP 2.0 on stars with a different spectral type than the Sun.

%

\acknowledgments
We thanks the anonymous referee for his valuable comments that improved the first version of the paper and made the results more significant. The authors are very grateful to the \emph{MOST} team for making available the observations analyzed in this work and for useful discussions. In addition, interesting discussions with R. Anderson, F. Pepe, S. Saar, D. S\'egransan and A. Triaud helped in improving the paper. We thank R. Anderson for a careful read of the paper and very helpful comments in return. X. Dumusque thanks the Swiss National Science Foundation for its financial support through an \emph{Early PostDoc Mobility} fellowship. We also acknowledge the support from the European Research Council/European Community under the FP7 through Starting Grant agreement number 239953.
NCS was supported by FCT through the Investigador FCT contract reference IF/00169/2012
and POPH/FSE (EC) by FEDER funding through the program "Programa Operacional de Factores de Competitividade - COMPETE.
NSO/Kitt Peak FTS data used here were produced by NSF/NOAO.


\bibliographystyle{apj}
\bibliography{dumusque_bibliography}

\begin{thebibliography}{}
\expandafter\ifx\csname natexlab\endcsname\relax\def\natexlab#1{#1}\fi

\bibitem[{{Aigrain} {et~al.}(2012){Aigrain}, {Pont}, \&
  {Zucker}}]{Aigrain-2012}
{Aigrain}, S., {Pont}, F., \& {Zucker}, S. 2012, \mnras, 419, 3147

\bibitem[{{Arentoft} {et~al.}(2008){Arentoft}, {Kjeldsen}, {Bedding}, {Bazot},
  {Christensen-Dalsgaard}, {Dall}, {Karoff}, {Carrier}, {Eggenberger},
  {Sosnowska}, {Wittenmyer}, {Endl}, {Metcalfe}, {Hekker}, {Reffert}, {Butler},
  {Bruntt}, {Kiss}, {O'Toole}, {Kambe}, {Ando}, {Izumiura}, {Sato}, {Hartmann},
  {Hatzes}, {Bouchy}, {Mosser}, {Appourchaux}, {Barban}, {Berthomieu},
  {Garcia}, {Michel}, {Provost}, {Turck-Chi{\`e}ze}, {Marti{\'c}}, {Lebrun},
  {Schmitt}, {Bertaux}, {Bonanno}, {Benatti}, {Claudi}, {Cosentino}, {Leccia},
  {Frandsen}, {Brogaard}, {Glowienka}, {Grundahl}, \&
  {Stempels}}]{Arentoft-2008}
{Arentoft}, T., {Kjeldsen}, H., {Bedding}, T.~R., {et~al.} 2008, \apj, 687,
  1180

\bibitem[{{Baranne} {et~al.}(1996){Baranne}, {Queloz}, {Mayor}, {Adrianzyk},
  {Knispel}, {Kohler}, {Lacroix}, {Meunier}, {Rimbaud}, \&
  {Vin}}]{Baranne-1996}
{Baranne}, A., {Queloz}, D., {Mayor}, M., {et~al.} 1996, \aaps, 119, 373

\bibitem[{{Barnes} {et~al.}(2011){Barnes}, {Linscott}, \&
  {Shporer}}]{Barnes-2011}
{Barnes}, J.~W., {Linscott}, E., \& {Shporer}, A. 2011, \apjs, 197, 10

\bibitem[{{Beckers} \& {Nelson}(1978)}]{Beckers-1978}
{Beckers}, J.~M., \& {Nelson}, G.~D. 1978, \solphys, 58, 243

\bibitem[{{Beckers} \& {Taylor}(1980)}]{Beckers-1980}
{Beckers}, J.~M., \& {Taylor}, W.~R. 1980, \solphys, 68, 41

\bibitem[{{Boisse} {et~al.}(2012){Boisse}, {Bonfils}, \&
  {Santos}}]{Boisse-2012b}
{Boisse}, I., {Bonfils}, X., \& {Santos}, N.~C. 2012, \aap, 545, 109

\bibitem[{{Boisse} {et~al.}(2011){Boisse}, {Bouchy}, {H{\'e}brard}, {Bonfils},
  {Santos}, \& {Vauclair}}]{Boisse-2011}
{Boisse}, I., {Bouchy}, F., {H{\'e}brard}, G., {et~al.} 2011, \aap, 528, A4

\bibitem[{{Boisse} {et~al.}(2009){Boisse}, {Moutou}, {Vidal-Madjar}, {Bouchy},
  {Pont}, {H{\'e}brard}, {Bonfils}, {Croll}, {Delfosse}, {Desort}, {Forveille},
  {Lagrange}, {Loeillet}, {Lovis}, {Matthews}, {Mayor}, {Pepe}, {Perrier},
  {Queloz}, {Rowe}, {Santos}, {S{\'e}gransan}, \& {Udry}}]{Boisse-2009}
{Boisse}, I., {Moutou}, C., {Vidal-Madjar}, A., {et~al.} 2009, \aap, 495, 959

\bibitem[{{Bonfils} {et~al.}(2007){Bonfils}, {Mayor}, {Delfosse}, {Forveille},
  {Gillon}, {Perrier}, {Udry}, {Bouchy}, {Lovis}, {Pepe}, {Queloz}, {Santos},
  \& {Bertaux}}]{Bonfils-2007}
{Bonfils}, X., {Mayor}, M., {Delfosse}, X., {et~al.} 2007, \aap, 474, 293

\bibitem[{{Bouchy} {et~al.}(2005){Bouchy}, {Bazot}, {Santos}, {Vauclair}, \&
  {Sosnowska}}]{Bouchy-2005b}
{Bouchy}, F., {Bazot}, M., {Santos}, N.~C., {Vauclair}, S., \& {Sosnowska}, D.
  2005, \aap, 440, 609

\bibitem[{Cavallini(1985{\natexlab{a}})}]{Cavallini-1985a}
Cavallini, F.;~Ceppatelli, G. R.~A. 1985{\natexlab{a}}, \aap, 143, 116

\bibitem[{Cavallini(1985{\natexlab{b}})}]{Cavallini-1985b}
---. 1985{\natexlab{b}}, \aap, 150, 256

\bibitem[{{Chapman} {et~al.}(2001){Chapman}, {Cookson}, {Dobias}, \&
  {Walton}}]{Chapman-2001}
{Chapman}, G.~A., {Cookson}, A.~M., {Dobias}, J.~J., \& {Walton}, S.~R. 2001,
  \apj, 555, 462

\bibitem[{{Claret} \& {Bloemen}(2011)}]{Claret-2011}
{Claret}, A., \& {Bloemen}, S. 2011, \aap, 529, A75

\bibitem[{{Collier Cameron} {et~al.}(2010){Collier Cameron}, {Bruce}, {Miller},
  {Triaud}, \& {Queloz}}]{Collier-Cameron-2010}
{Collier Cameron}, A., {Bruce}, V.~A., {Miller}, G.~R.~M., {Triaud},
  A.~H.~M.~J., \& {Queloz}, D. 2010, \mnras, 403, 151

\bibitem[{{Croll} {et~al.}(2007){Croll}, {Matthews}, {Rowe}, {Gladman},
  {Miller-Ricci}, {Sasselov}, {Walker}, {Kuschnig}, {Lin}, {Guenther},
  {Moffat}, {Rucinski}, \& {Weiss}}]{Croll-2007}
{Croll}, B., {Matthews}, J.~M., {Rowe}, J.~F., {et~al.} 2007, \apj, 671, 2129

\bibitem[{{Desort} {et~al.}(2007){Desort}, {Lagrange}, {Galland}, {Udry}, \&
  {Mayor}}]{Desort-2007}
{Desort}, M., {Lagrange}, A.-M., {Galland}, F., {Udry}, S., \& {Mayor}, M.
  2007, \aap, 473, 983

\bibitem[{{Dravins}(1982)}]{Dravins-1982}
{Dravins}, D. 1982, \araa, 20, 61

\bibitem[{Dravins(1981)}]{Dravins-1981}
Dravins, D.;~Lindegren, L. N.~A. 1981, \aap, 96, 345

\bibitem[{{Dumusque} {et~al.}(2011{\natexlab{a}}){Dumusque}, {Santos}, {Udry},
  {Lovis}, \& {Bonfils}}]{Dumusque-2011b}
{Dumusque}, X., {Santos}, N.~C., {Udry}, S., {Lovis}, C., \& {Bonfils}, X.
  2011{\natexlab{a}}, \aap, 527, A82

\bibitem[{{Dumusque} {et~al.}(2011{\natexlab{b}}){Dumusque}, {Udry}, {Lovis},
  {Santos}, \& {Monteiro}}]{Dumusque-2011a}
{Dumusque}, X., {Udry}, S., {Lovis}, C., {Santos}, N.~C., \& {Monteiro},
  M.~J.~P.~F.~G. 2011{\natexlab{b}}, \aap, 525, A140

\bibitem[{{Dumusque} {et~al.}(2011{\natexlab{c}}){Dumusque}, {Lovis},
  {S{\'e}gransan}, {Mayor}, {Udry}, {Benz}, {Bouchy}, {Lo Curto}, {Mordasini},
  {Pepe}, {Queloz}, {Santos}, \& {Naef}}]{Dumusque-2011c}
{Dumusque}, X., {Lovis}, C., {S{\'e}gransan}, D., {et~al.} 2011{\natexlab{c}},
  \aap, 535, A55

\bibitem[{{Dumusque} {et~al.}(2012){Dumusque}, {Pepe}, {Lovis}, {Segransan},
  {Sahlmann}, {Benz}, {Bouchy}, {Mayor}, {Queloz}, {Santos}, \&
  {Udry}}]{Dumusque-2012}
{Dumusque}, X., {Pepe}, F., {Lovis}, C., {et~al.} 2012, \nat, 491, 207

\bibitem[{{Dumusque} {et~al.}(2014){Dumusque}, {Bonomo}, {Haywood},
  {Malavolta}, {S{\'e}gransan}, {Buchhave}, {Collier Cameron}, {Latham},
  {Molinari}, {Pepe}, {Udry}, {Charbonneau}, {Cosentino}, {Dressing},
  {Figueira}, {Fiorenzano}, {Gettel}, {Harutyunyan}, {Horne}, {Lopez-Morales},
  {Lovis}, {Mayor}, {Micela}, {Motalebi}, {Nascimbeni}, {Phillips}, {Piotto},
  {Pollacco}, {Queloz}, {Rice}, {Sasselov}, {Sozzetti}, {Szentgyorgyi}, \&
  {Watson}}]{Dumusque-2014}
{Dumusque}, X., {Bonomo}, A.~S., {Haywood}, R.~D., {et~al.} 2014, \apj, 789,
  154

\bibitem[{{Fares} {et~al.}(2010){Fares}, {Donati}, {Moutou}, {Jardine},
  {Grie{\ss}meier}, {Zarka}, {Shkolnik}, {Bohlender}, {Catala}, \& {Collier
  Cameron}}]{Fares-2010}
{Fares}, R., {Donati}, J.-F., {Moutou}, C., {et~al.} 2010, MNRAS, 406, 409

\bibitem[{{Frazier}(1971)}]{Frazier-1971}
{Frazier}, E.~N. 1971, \solphys, 21, 42

\bibitem[{{Gomes da Silva} {et~al.}(2013){Gomes da Silva}, {Santos}, {Boisse},
  {Dumusque}, \& {Lovis}}]{Gomes-da-Silva-2013}
{Gomes da Silva}, J., {Santos}, N.~C., {Boisse}, I., {Dumusque}, X., \&
  {Lovis}, C. 2013, ArXiv e-prints, arXiv:1311.6642

\bibitem[{{Hatzes}(2002)}]{Hatzes-2002}
{Hatzes}, A.~P. 2002, Astronomische Nachrichten, 323, 392

\bibitem[{{Henry} \& {Winn}(2008)}]{Henry-2008}
{Henry}, G.~W., \& {Winn}, J.~N. 2008, \aj, 135, 68

\bibitem[{{Howard} {et~al.}(2013){Howard}, {Sanchis-Ojeda}, {Marcy}, {Johnson},
  {Winn}, {Isaacson}, {Fischer}, {Fulton}, {Sinukoff}, \&
  {Fortney}}]{Howard-2013b}
{Howard}, A.~W., {Sanchis-Ojeda}, R., {Marcy}, G.~W., {et~al.} 2013, Nature,
  503, 381

\bibitem[{{Hu{\'e}lamo} {et~al.}(2008){Hu{\'e}lamo}, {Figueira}, {Bonfils},
  {Santos}, {Pepe}, {Gillon}, {Azevedo}, {Barman}, {Fern{\'a}ndez}, {di Folco},
  {Guenther}, {Lovis}, {Melo}, {Queloz}, \& {Udry}}]{Huelamo-2008}
{Hu{\'e}lamo}, N., {Figueira}, P., {Bonfils}, X., {et~al.} 2008, \aap, 489, L9

\bibitem[{{Jeffers} {et~al.}(2013){Jeffers}, {Barnes}, {Jones}, {Reiners},
  {Pinfield}, \& {Marsden}}]{Jeffers-2013}
{Jeffers}, S.~V., {Barnes}, J.~R., {Jones}, H.~R.~A., {et~al.} 2013, ArXiv
  e-prints, arXiv:1311.3617

\bibitem[{{Kjeldsen} {et~al.}(2005){Kjeldsen}, {Bedding}, {Butler},
  {Christensen-Dalsgaard}, {Kiss}, {McCarthy}, {Marcy}, {Tinney}, \&
  {Wright}}]{Kjeldsen-2005}
{Kjeldsen}, H., {Bedding}, T.~R., {Butler}, R.~P., {et~al.} 2005, \apj, 635,
  1281

\bibitem[{{Lanza} {et~al.}(2011){Lanza}, {Boisse}, {Bouchy}, {Bonomo}, \&
  {Moutou}}]{Lanza-2011b}
{Lanza}, A.~F., {Boisse}, I., {Bouchy}, F., {Bonomo}, A.~S., \& {Moutou}, C.
  2011, \aap, 533, A44

\bibitem[{{Lanza} {et~al.}(2010){Lanza}, {Bonomo}, {Moutou}, {Pagano},
  {Messina}, {Leto}, {Cutispoto}, {Aigrain}, {Alonso}, {Barge}, {Deleuil},
  {Auvergne}, {Baglin}, \& {Collier Cameron}}]{Lanza-2010}
{Lanza}, A.~F., {Bonomo}, A.~S., {Moutou}, C., {et~al.} 2010, \aap, 520, A53

\bibitem[{{Lindegren} \& {Dravins}(2003)}]{Lindegren-2003}
{Lindegren}, L., \& {Dravins}, D. 2003, \aap, 401, 1185

\bibitem[{{Lockwood} {et~al.}(2007){Lockwood}, {Skiff}, {Henry}, {Henry},
  {Radick}, {Baliunas}, {Donahue}, \& {Soon}}]{Lockwood-2007}
{Lockwood}, G.~W., {Skiff}, B.~A., {Henry}, G.~W., {et~al.} 2007, \apjs, 171,
  260

\bibitem[{{Lovis} {et~al.}(2011){Lovis}, {Dumusque}, {Santos}, {Bouchy},
  {Mayor}, {Pepe}, {Queloz}, {S{\'e}gransan}, \& {Udry}}]{Lovis-2011b}
{Lovis}, C., {Dumusque}, X., {Santos}, N.~C., {et~al.} 2011, ArXiv e-prints,
  arXiv:1107.5325

\bibitem[{{Mandel} \& {Agol}(2002)}]{Mandel-2002}
{Mandel}, K., \& {Agol}, E. 2002, \apjl, 580, L171

\bibitem[{{Mayor} {et~al.}(2003){Mayor}, {Pepe}, {Queloz}, {Bouchy},
  {Rupprecht}, {Lo Curto}, {Avila}, {Benz}, {Bertaux}, {Bonfils}, {Dall},
  {Dekker}, {Delabre}, {Eckert}, {Fleury}, {Gilliotte}, {Gojak}, {Guzman},
  {Kohler}, {Lizon}, {Longinotti}, {Lovis}, {Megevand}, {Pasquini}, {Reyes},
  {Sivan}, {Sosnowska}, {Soto}, {Udry}, {van Kesteren}, {Weber}, \&
  {Weilenmann}}]{Mayor-2003}
{Mayor}, M., {Pepe}, F., {Queloz}, D., {et~al.} 2003, The Messenger, 114, 20

\bibitem[{{Meunier} {et~al.}(2010{\natexlab{a}}){Meunier}, {Desort}, \&
  {Lagrange}}]{Meunier-2010a}
{Meunier}, N., {Desort}, M., \& {Lagrange}, A.-M. 2010{\natexlab{a}}, \aap,
  512, A39

\bibitem[{{Meunier} {et~al.}(2010{\natexlab{b}}){Meunier}, {Lagrange}, \&
  {Desort}}]{Meunier-2010b}
{Meunier}, N., {Lagrange}, A., \& {Desort}, M. 2010{\natexlab{b}}, \aap, 519,
  A66+

\bibitem[{{Meunier} \& {Lagrange}(2013)}]{Meunier-2013}
{Meunier}, N., \& {Lagrange}, A.-M. 2013, \aap, 551, A101

\bibitem[{{Moutou} {et~al.}(2007){Moutou}, {Donati}, {Savalle}, {Hussain},
  {Alecian}, {Bouchy}, {Catala}, {Collier Cameron}, {Udry}, \&
  {Vidal-Madjar}}]{Moutou-2007}
{Moutou}, C., {Donati}, J.-F., {Savalle}, R., {et~al.} 2007, \aap, 473, 651

\bibitem[{{Oshagh} {et~al.}(2013){Oshagh}, {Boisse}, {Bou{\'e}}, {Montalto},
  {Santos}, {Bonfils}, \& {Haghighipour}}]{Oshagh-2013a}
{Oshagh}, M., {Boisse}, I., {Bou{\'e}}, G., {et~al.} 2013, \aap, 549, A35

\bibitem[{Pepe {et~al.}(2002)Pepe, Mayor, Galland, Naef, Queloz, Santos, Udry,
  \& Burnet}]{Pepe-2002a}
Pepe, F., Mayor, M., Galland, F., {et~al.} 2002, \aap, 388, 632

\bibitem[{{Pepe} {et~al.}(2011){Pepe}, {Lovis}, {S{\'e}gransan}, {Benz},
  {Bouchy}, {Dumusque}, {Mayor}, {Queloz}, {Santos}, \& {Udry}}]{Pepe-2011}
{Pepe}, F., {Lovis}, C., {S{\'e}gransan}, D., {et~al.} 2011, \aap, 534, A58

\bibitem[{{Pepe} {et~al.}(2013){Pepe}, {Cameron}, {Latham}, {Molinari}, {Udry},
  {Bonomo}, {Buchhave}, {Charbonneau}, {Cosentino}, {Dressing}, {Dumusque},
  {Figueira}, {Fiorenzano}, {Gettel}, {Harutyunyan}, {Haywood}, {Horne},
  {Lopez-Morales}, {Lovis}, {Malavolta}, {Mayor}, {Micela}, {Motalebi},
  {Nascimbeni}, {Phillips}, {Piotto}, {Pollacco}, {Queloz}, {Rice}, {Sasselov},
  {S{\'e}gransan}, {Sozzetti}, {Szentgyorgyi}, \& {Watson}}]{Pepe-2013}
{Pepe}, F., {Cameron}, A.~C., {Latham}, D.~W., {et~al.} 2013, \nat, 503, 377

\bibitem[{{Pont} {et~al.}(2013){Pont}, {Sing}, {Gibson}, {Aigrain}, {Henry}, \&
  {Husnoo}}]{Pont-2013}
{Pont}, F., {Sing}, D.~K., {Gibson}, N.~P., {et~al.} 2013, \mnras, 432, 2917

\bibitem[{{Pont} {et~al.}(2007){Pont}, {Gilliland}, {Moutou}, {Charbonneau},
  {Bouchy}, {Brown}, {Mayor}, {Queloz}, {Santos}, \& {Udry}}]{Pont-2007}
{Pont}, F., {Gilliland}, R.~L., {Moutou}, C., {et~al.} 2007, \aap, 476, 1347

\bibitem[{{Poppenhaeger} {et~al.}(2013){Poppenhaeger}, {Schmitt}, \&
  {Wolk}}]{Poppenhaeger-2013a}
{Poppenhaeger}, K., {Schmitt}, J.~H.~M.~M., \& {Wolk}, S.~J. 2013, ArXiv
  e-prints, arXiv:1306.2311

\bibitem[{{Queloz} {et~al.}(2001){Queloz}, {Henry}, {Sivan}, {Baliunas},
  {Beuzit}, {Donahue}, {Mayor}, {Naef}, {Perrier}, \& {Udry}}]{Queloz-2001}
{Queloz}, D., {Henry}, G.~W., {Sivan}, J.~P., {et~al.} 2001, \aap, 379, 279

\bibitem[{{Robertson} {et~al.}(2014){Robertson}, {Mahadevan}, {Endl}, \&
  {Roy}}]{Robertson-2014}
{Robertson}, P., {Mahadevan}, S., {Endl}, M., \& {Roy}, A. 2014, ArXiv
  e-prints, arXiv:1407.1049

\bibitem[{{Saar}(2003)}]{Saar-2003}
{Saar}, S.~H. 2003, in Astronomical Society of the Pacific Conference Series,
  Vol. 294, Scientific Frontiers in Research on Extrasolar Planets, ed.
  D.~{Deming} \& S.~{Seager}, 65--70

\bibitem[{{Saar}(2009)}]{Saar-2009}
{Saar}, S.~H. 2009, in American Institute of Physics Conference Series, Vol.
  1094, 15th Cambridge Workshop on Cool Stars, Stellar Systems, and the Sun,
  ed. E.~{Stempels}, 152--161

\bibitem[{{Saar} \& {Donahue}(1997)}]{Saar-1997b}
{Saar}, S.~H., \& {Donahue}, R.~A. 1997, \apj, 485, 319

\bibitem[{{S{\'a}nchez Cuberes} {et~al.}(2003){S{\'a}nchez Cuberes},
  {V{\'a}zquez}, {Bonet}, \& {Sobotka}}]{Sanchez-Cuberes-2003}
{S{\'a}nchez Cuberes}, M., {V{\'a}zquez}, M., {Bonet}, J.~A., \& {Sobotka}, M.
  2003, \aap, 397, 1075

\bibitem[{{Sanchis-Ojeda} {et~al.}(2013){Sanchis-Ojeda}, {Rappaport}, {Winn},
  {Levine}, {Kotson}, {Latham}, \& {Buchhave}}]{Sanchis-Ojeda-2013}
{Sanchis-Ojeda}, R., {Rappaport}, S., {Winn}, J.~N., {et~al.} 2013, \apj, 774,
  54

\bibitem[{{Santos} {et~al.}(2000){Santos}, {Mayor}, {Naef}, {Pepe}, {Queloz},
  {Udry}, \& {Blecha}}]{Santos-2000b}
{Santos}, N.~C., {Mayor}, M., {Naef}, D., {et~al.} 2000, \aap, 361, 265

\bibitem[{{Shapiro} {et~al.}(2014){Shapiro}, {Solanki}, {Krivova}, {Schmutz},
  {Ball}, {Knaack}, {Rozanov}, \& {Unruh}}]{Shapiro-2014}
{Shapiro}, A.~I., {Solanki}, S.~K., {Krivova}, N.~A., {et~al.} 2014, ArXiv
  e-prints, arXiv:1406.2383

\bibitem[{{Sing}(2010)}]{Sing-2010}
{Sing}, D.~K. 2010, \aap, 510, A21

\bibitem[{{Triaud} {et~al.}(2009){Triaud}, {Queloz}, {Bouchy}, {Moutou},
  {Collier Cameron}, {Claret}, {Barge}, {Benz}, {Deleuil}, {Guillot},
  {H{\'e}brard}, {Lecavelier Des {\'E}tangs}, {Lovis}, {Mayor}, {Pepe}, \&
  {Udry}}]{Triaud-2009}
{Triaud}, A.~H.~M.~J., {Queloz}, D., {Bouchy}, F., {et~al.} 2009, \aap, 506,
  377

\bibitem[{{Unruh} {et~al.}(1999){Unruh}, {Solanki}, \& {Fligge}}]{Unruh-1999}
{Unruh}, Y.~C., {Solanki}, S.~K., \& {Fligge}, M. 1999, \aap, 345, 635

\bibitem[{{Wallace} {et~al.}(1998){Wallace}, {Hinkle}, \&
  {Livingston}}]{Wallace-1998}
{Wallace}, L., {Hinkle}, K., \& {Livingston}, W. 1998, {An atlas of the
  spectrum of the solar photosphere from 13,500 to 28,000 cm-1 (3570 to 7405
  A)}

\bibitem[{{Wallace} {et~al.}(2005){Wallace}, {Hinkle}, \&
  {Livingston}}]{Wallace-2005}
{Wallace}, L., {Hinkle}, K., \& {Livingston}, W.~C. 2005, {An atlas of sunspot
  umbral spectra in the visible from 15,000 to 25,500 cm-1 (3920 to 6664
  {\AA})}

\bibitem[{{Winn} {et~al.}(2006){Winn}, {Johnson}, {Marcy}, {Butler}, {Vogt},
  {Henry}, {Roussanova}, {Holman}, {Enya}, {Narita}, {Suto}, \&
  {Turner}}]{Winn-2006}
{Winn}, J.~N., {Johnson}, J.~A., {Marcy}, G.~W., {et~al.} 2006, \apjl, 653, L69

\bibitem[{{Winn} {et~al.}(2007){Winn}, {Holman}, {Henry}, {Roussanova}, {Enya},
  {Yoshii}, {Shporer}, {Mazeh}, {Johnson}, {Narita}, \& {Suto}}]{Winn-2007}
{Winn}, J.~N., {Holman}, M.~J., {Henry}, G.~W., {et~al.} 2007, \aj, 133, 1828

\end{thebibliography}

\appendix

%

\section{Equivalence in integrated CCF when using a spectrum or its CCF in each cell of the SOAP simulation} \label{app:2}

To obtain the quiet Sun CCF of the integrated disc, several linear processes are used: 
\begin{itemize}
\item calculation of the CCF of the quiet Sun spectrum. According to \citet{Pepe-2002a}, the value of the CCF at velocity $v_R$ is given by:
	\begin{equation} \label{eq:app2-0}
	\mathrm{CCF}(v_R) = \int \mathrm{S}(\lambda) \mathrm{M}(\lambda_{v_R})d\lambda ,
	\end{equation}
where $\lambda_{v_R} = \lambda \sqrt{\frac{1+v_R/c}{1-v_R/c}}$, S is the spectrum and $\mathrm{M}(\lambda_{v_R})$ represents the Doppler-shifted numerical mask used for the correlation
\item shift of the CCF according to the projected stellar velocity $v_j$ of cell j:
	\begin{equation}  \label{eq:app2-1}
	\mathrm{CCF}(v_R)_j = CCF(v_R + v_j).
	\end{equation}
\item summation of the CCFs present in every cell to obtain the integrated CCF, $\mathrm{CCF}(v_R)_{\mathrm{tot,\,quiet}}$:
	\begin{equation}  \label{eq:app2-2}
	\mathrm{CCF}(v_R)_{\mathrm{tot,\,quiet}} = \sum_j \mathrm{CCF}(v_R)_j.
	\end{equation}
\end{itemize}
Therefore, the integrated CCF over the stellar disc is given by:
\begin{equation}  \label{eq:app2-3}
\mathrm{CCF}(v_R)_{\mathrm{tot,\,quiet}} = \sum_j \int \mathrm{S}(\lambda)M(\lambda_{v_R+v_j})d\lambda.
\end{equation} 
One can also consider the quiet Sun spectrum in each cell and shift it to the corresponding projected stellar velocity $v_j$ of each cell. The integrated spectrum over the entire disc is obtained by summing all the spectra in the cells and then the CCF can be calculated. In this case:
\begin{equation}  \label{eq:app2-4}
\mathrm{CCF}(v_R)_{\mathrm{tot,\,quiet}} = \int \sum_j S(\lambda_{v_j})\mathrm{M}(\lambda_{v_R})d\lambda ,
\end{equation} 
where $\lambda_{v_j} = \lambda \sqrt{\frac{1+v_j/c}{1-v_j/c}}$.
By doing the change of variable $\lambda \rightarrow \lambda' = \lambda_{v_j}$, we have $d\lambda \rightarrow \sqrt{\frac{1-v_j/c}{1+v_j/c}}d\lambda'$, and therefore Eq. \ref{eq:app2-4} can be rewritten:
\begin{equation}  \label{eq:app2-5}
\mathrm{CCF}(v_R)_{\mathrm{tot,\,quiet}} = \sum_j \sqrt{\frac{1-v_j/c}{1+v_j/c}} \int S(\lambda')\mathrm{M}(\lambda_{v_R+v_j})d\lambda',
\end{equation}
where we used that:
\begin{eqnarray}
\mathrm{M}(\lambda_{v_R}) =  \lambda \sqrt{\frac{1+v_R/c}{1-v_R/c}} &\rightarrow& \lambda' \sqrt{\frac{1+v_R/c}{1-v_R/c}} \nonumber\\
													     &\rightarrow& \lambda \sqrt{\frac{(1+v_j/c)(1+v_R/c)}{(1-v_j/c)(1-v_R/c)}} \nonumber\\
													     &\rightarrow& \lambda \sqrt{\frac{(1+(v_R+v_j)/c+v_R\cdot v_j/c^2}{(1-(v_R+v_j)/c+v_R\cdot v_j/c^2}} \nonumber\\
													     &\rightarrow& \lambda \sqrt{\frac{1+(v_R+v_j)/c}{1-(v_R+v_j)/c}},\nonumber\\
													     &\rightarrow& \mathrm{M}(\lambda_{v_R+v_j})
\end{eqnarray}
where the last equivalence can be obtained for small velocities compared to the speed of light, $v_R \cdot v_j \ll  c^2$, which is the case here.

\begin{figure}
\begin{center}
\includegraphics[width=8cm]{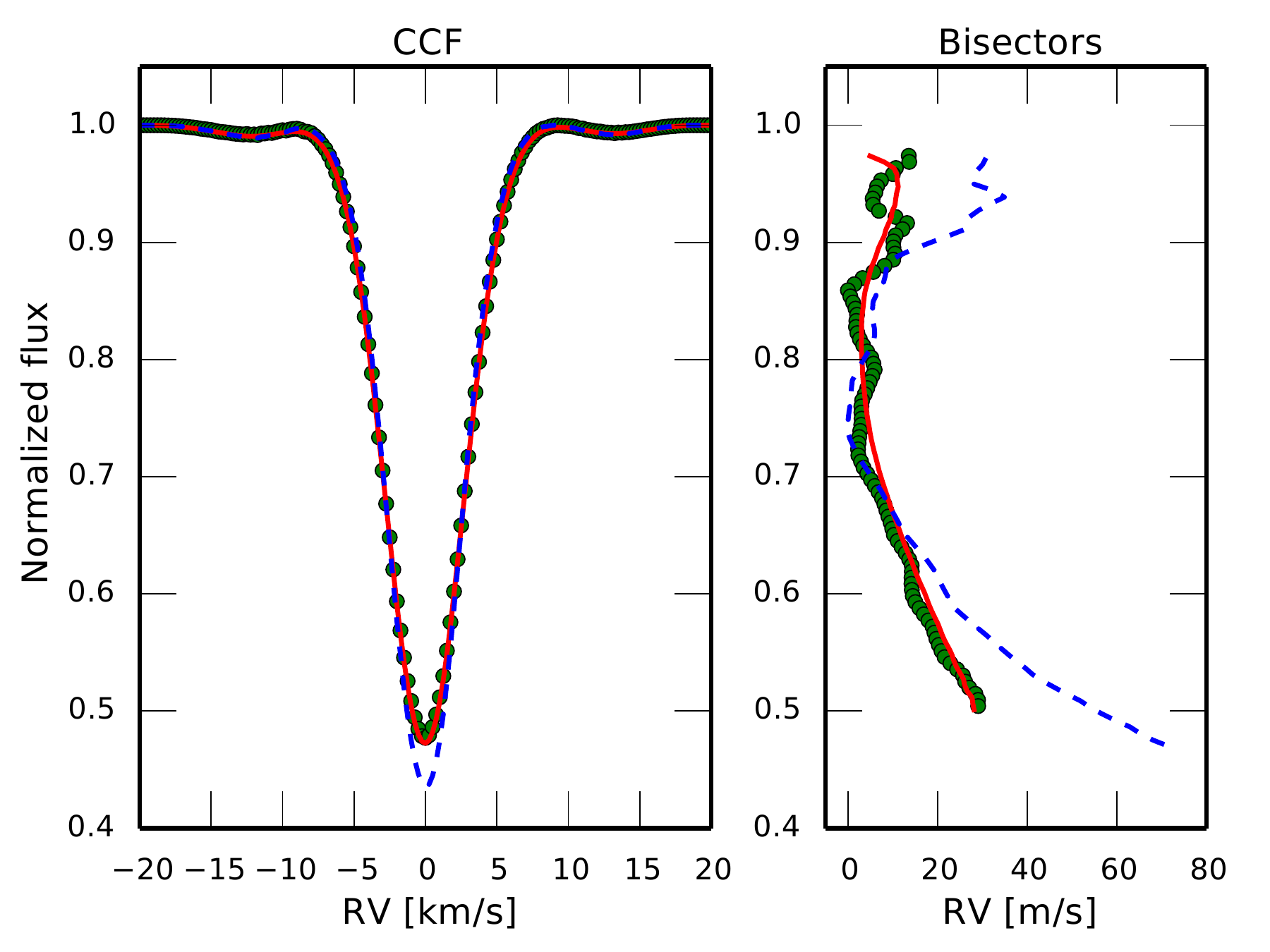}
\caption{Full-disk integrated CCFs of a quiet star, i.e without active regions, and its bisectors. The green CCF and bisector (large dots) obtained when using the FTS spectrum of the quiet Sun in each of the simulation cell  match the red CCF and bisector (continuous line) obtained when using the CCF of the FTS spectrum of the quiet Sun in each cell. To illustrate that the integration over the entire disc modify the CCF and the bisector, we show in blue (dashed line) the CCF and the bisectors of the FTS quiet Sun spectrum taken in the stellar disc center. Note that the result obtained when using the full spectrum in each cell is noisier. This is an artifact of the way each spectrum was Doppler shifted to account for stellar rotation.}
\label{fig:app2-0}
\end{center}
\end{figure}

Considering a CCF in each cell and obtaining the integrated CCF by summing all the cells together is therefore equivalent to considering a spectrum in each cell, obtaining the integrated spectrum and finally calculating the CCF of this integrated spectrum. Figure \ref{fig:app2-0} illustrates this analytical result using our simulation. By injecting spectra or CCFs in each cell of SOAP 2.0, it was possible numerically to retrieve the same CCF and bisector, proving that the simulation is returning coherent results. 

\section{Warping the CCFs to account for the limb-shift of spectral lines} \label{app:3}

On the Sun, the contrast of the granulation pattern decreases when going towards the limb \citep[][]{Sanchez-Cuberes-2003,Beckers-1980,Beckers-1978}, and consequently the amount of convective blueshift decreases as well. This limb-shift effect will affect the shape of the spectral lines, being "\emph{C}" shape on the stellar disc center a "\emph{D}" shape on the stellar limb. The variation of a few spectral line from the solar disc center to the limb have been measured on the Sun \citep[e.g.][]{Cavallini-1985b}, however these observations does not help us in modeling this limb-shift effect because we are working with CCFs. A CCF is a weighted average of nearly all the spectral lines in the visible, and because different spectral lines are affected differently by convection, the bisector of a CCF is different from the bisector of a given spectral line (see the left plot of Figure \ref{fig:app2-0}). 

\begin{figure*}
\begin{center}
\includegraphics[width=8cm]{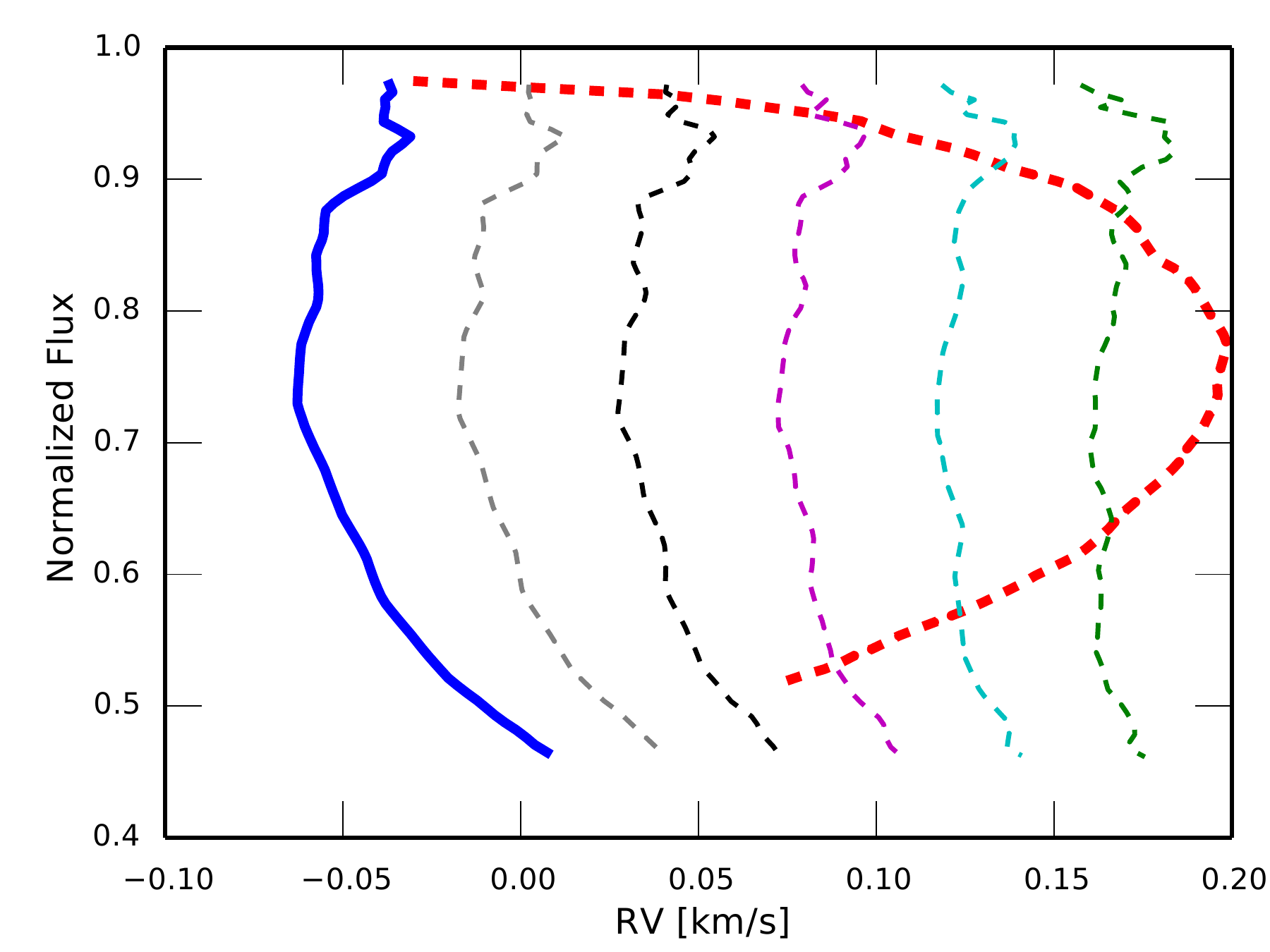}
\includegraphics[width=8cm]{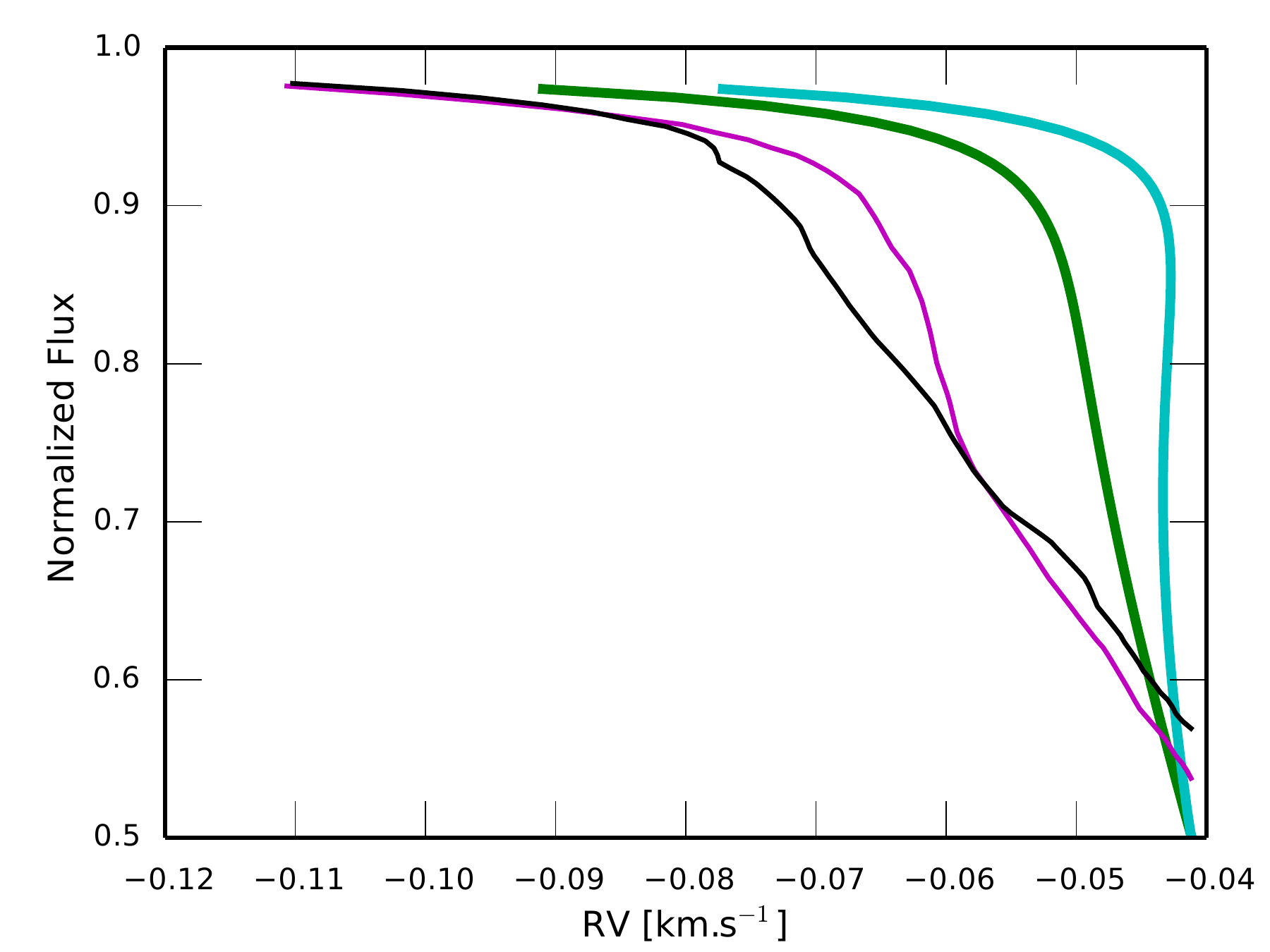}
\caption{\emph{Left:} Bisectors of the quiet photosphere CCF in blue (continuous line) and of a spot CCF in red (thick dashed line) as seen in Figure \ref{fig:2-2-0}. The thin dashed lines represent the variation of the bisector for different $\cos{\theta}$ angles as empirically accounted for in our model. From left to right, we can see the bisectors for $\cos{\theta} = 0.8$, 0.6, 0.4, 0.2 and 0. Note that for $\cos{\theta} = 1.0$, we use the quiet photosphere CCF bisector because it has been calculated from an FTS spectrum taken at the center of the solar disc. When approaching the limb, the bisector is redshifted and straighten to account for the reduction of the convective blueshift. \emph{Right:} From right to left, we compare the bisector of the integrated CCF obtained considering (cyan line) and not considering (green line) the limb-shift effect, in addition to the bisectors of the $\alpha$ Cen A (purple line) and 18 Scorpii (black line) CCFs observed with HARPS. The model not accounting for the limb-shift effect is a better match to the observations.}
\label{fig:app3-0}
\end{center}
\end{figure*}

To account for the limb-shift effect of spectral lines, we decided to warp the CCFs depending on the stellar disc position. The CCF for the quiet photosphere that we use throughout the paper have been extracted from a FTS spectrum taken at the disc center. We therefore have the correct CCF for the stellar disc center. On the limb, the granulation contrast decreases and therefore the amount of convective blueshift as well. The quiet photosphere CCF on the limb should therefore not include any convective blueshift, which implies that the CCF should be redshifted by $\sim350\,km\,s^{-1}$ and its bisector should be straight. For regions between the disc center and the limb, we interpolated linearly in $\cos{\theta}$ between the two bisectors of these extreme regions. The left plot of Figure \ref{fig:app3-0} shows the CCF bisectors for $\cos{\theta}= 1,0.8,0.6,0.4,0.2$ and 0. The CCF at any position on the disc is obtained by warping the quiet photosphere CCF taken at the disc center to reproduce the desired bisector.

To test if our empirical way of considering the limb-shift effect is realistic, we selected two solar twins observed with HARPS during their minimum of activity: $\alpha$ Cen A and 18 Scorpii. If the limb-shift effect is accounted for in a realistic way, the bisector of the integrated CCF estimated with our model without any active region should be similar to the CCF bisectors of these two stars. The result of this test is shown in the right plot of Figure \ref{fig:app3-0}. We can see that the bisector derived when considering the limb-shift effect is not matching the observations, and that the bisector obtained without this extra effect is a better match.

\begin{figure*}
\begin{center}
\includegraphics[width=8cm]{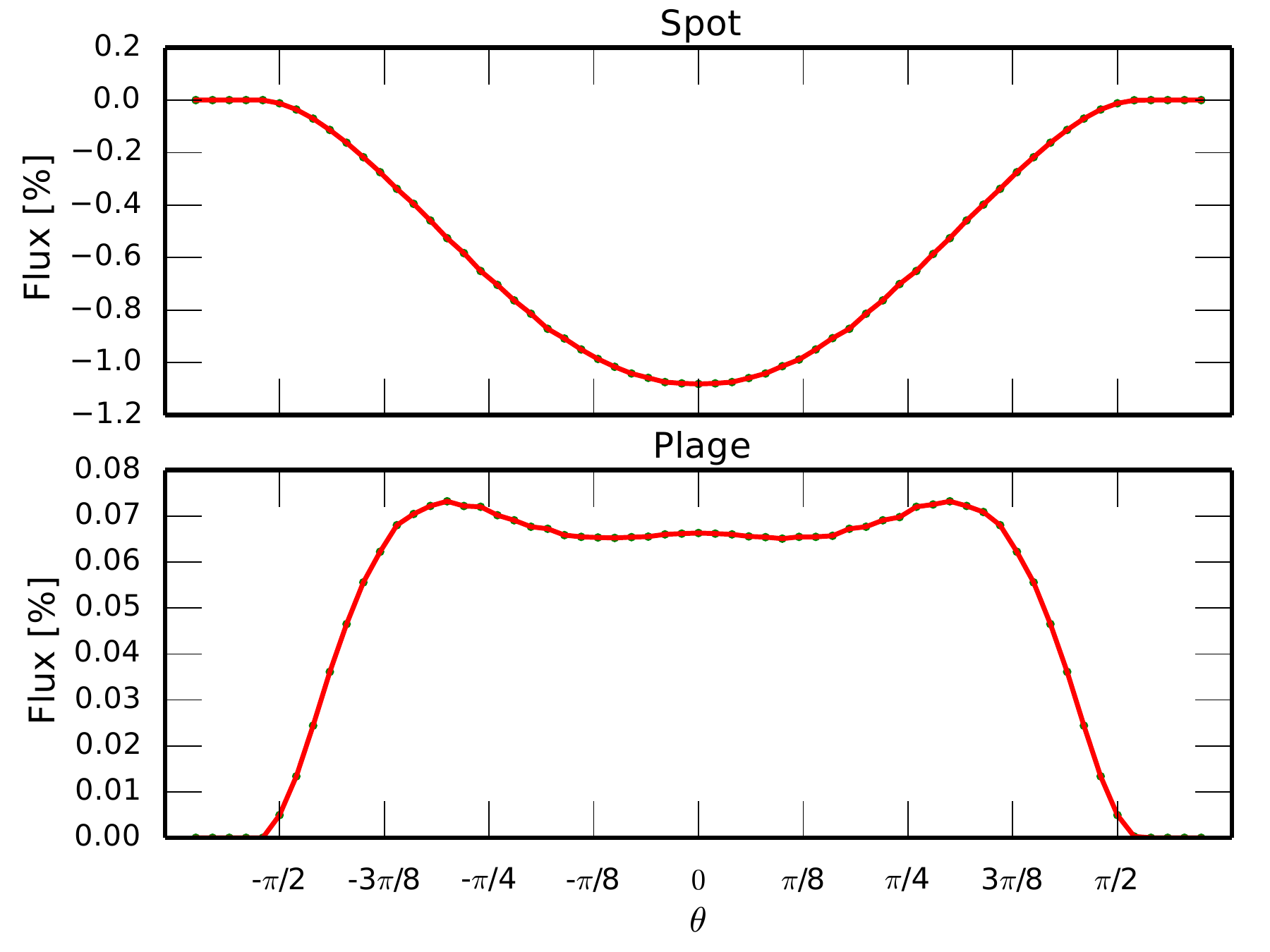}
\includegraphics[width=8cm]{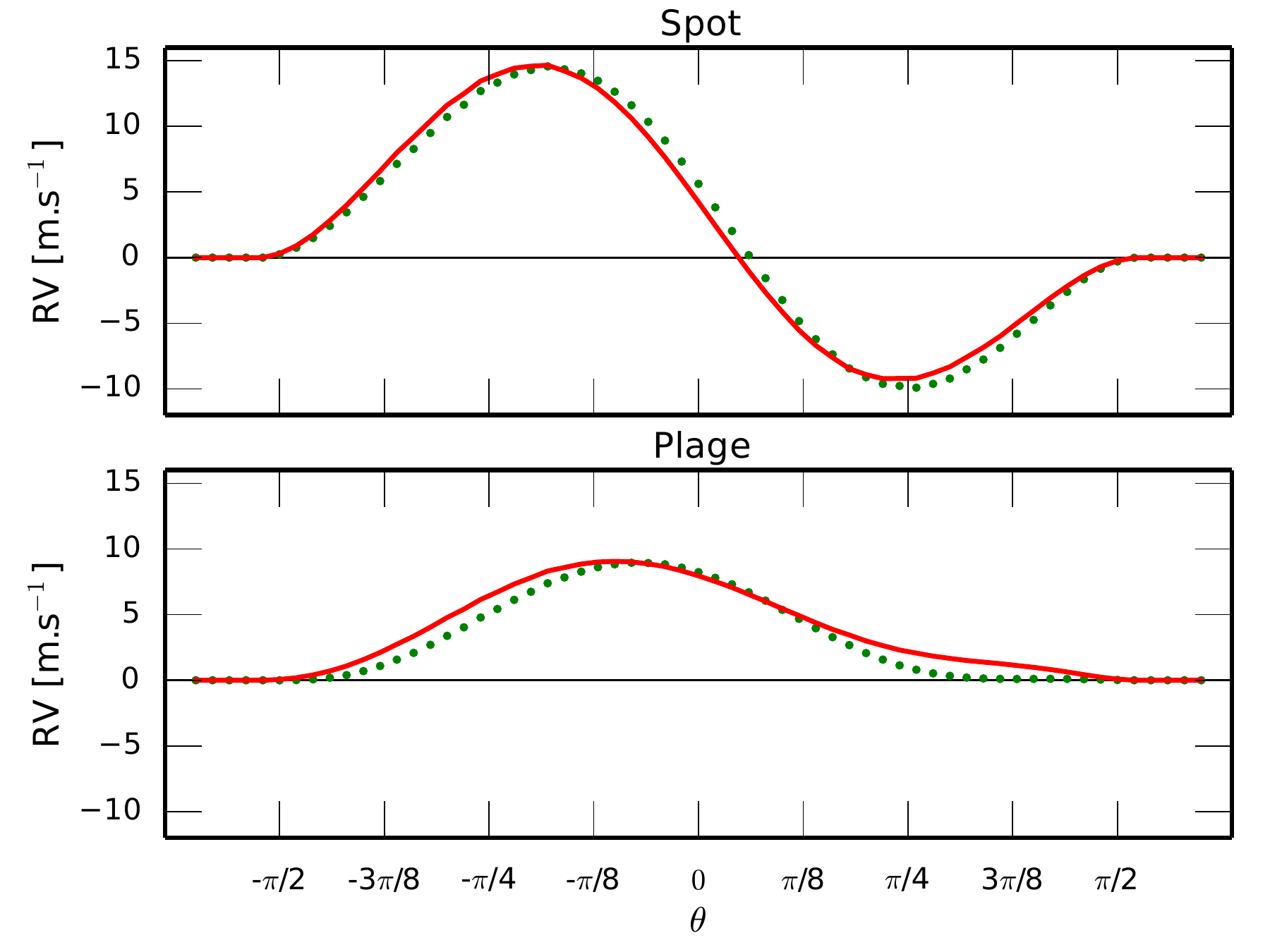}
\includegraphics[width=8cm]{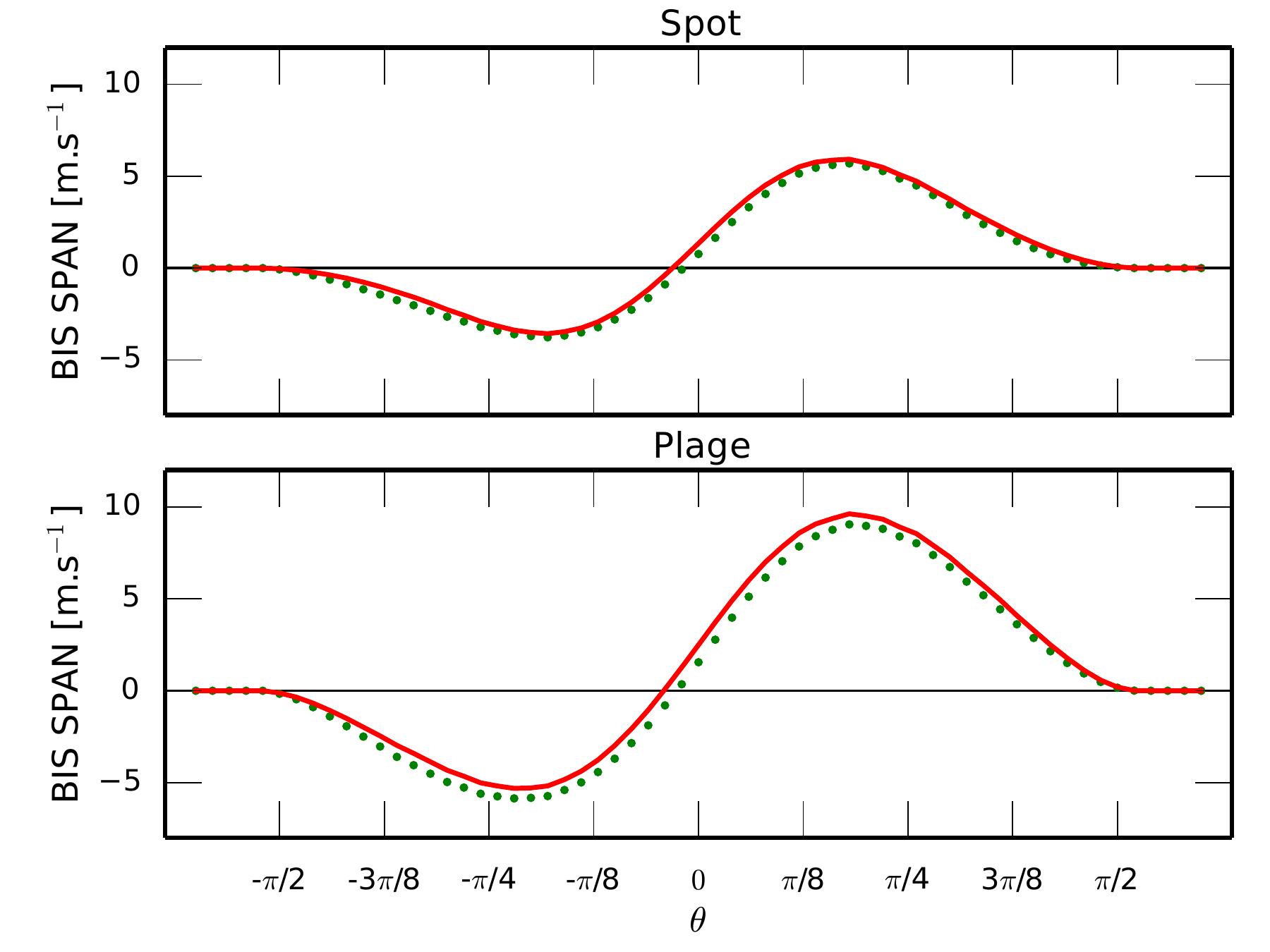}
\includegraphics[width=8cm]{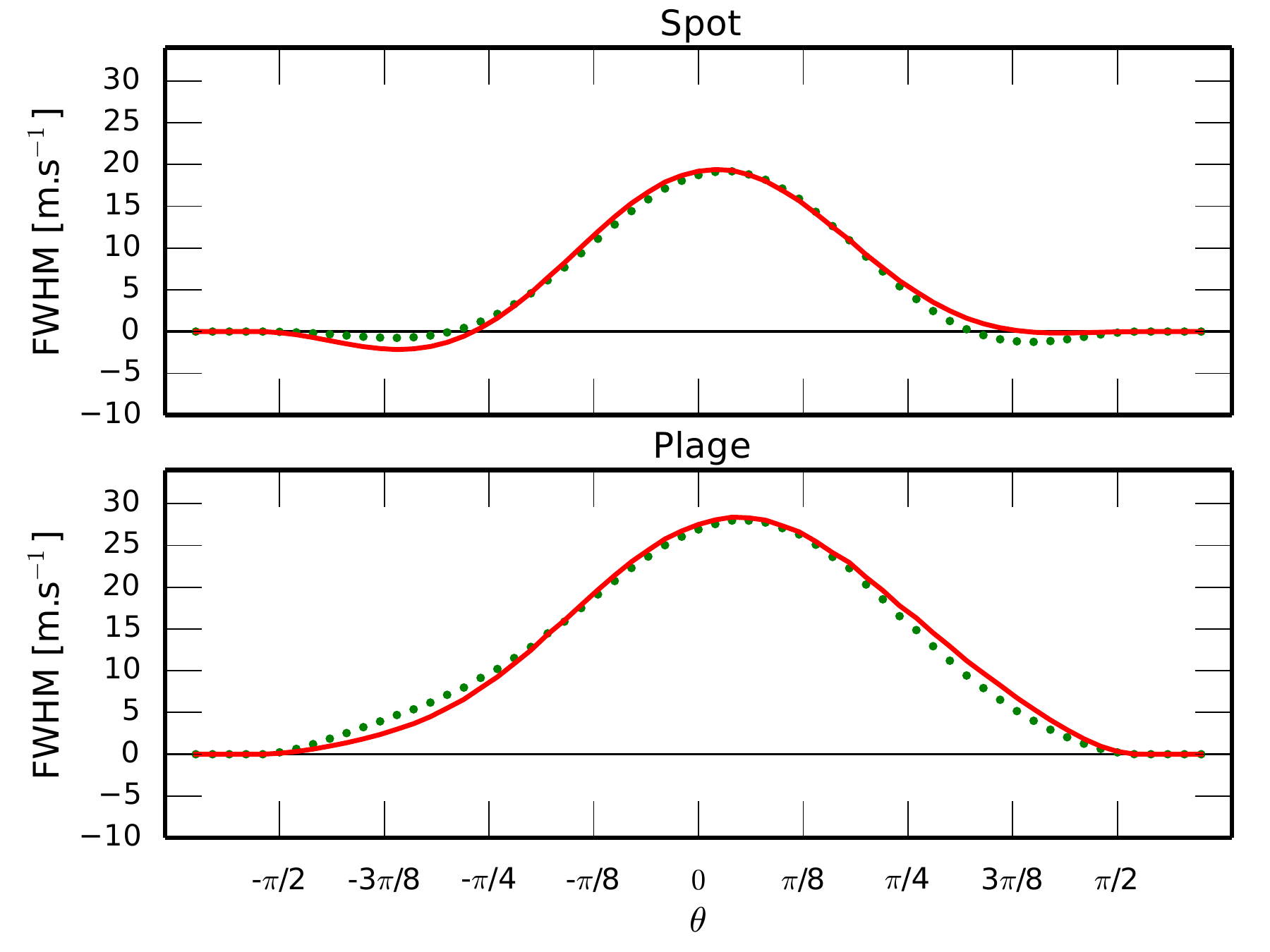}
\caption{Photometric, RV BIS SPAN, and FWHM variations induced by an equatorial spot or plage on an equator on star when considering the limb-shift effect (green dotted line) or not considering it (red continuous line). The size of the active region is 1\%. The contrast of the active region is 0.54 in the case of a spot ($663 K$ cooler than the effective temperature of the Sun, at 5293\,\AA.), and is given by Eq. \ref{eq:2-2-0} in the case of a plage. The active region is on the stellar disc center when $\theta =$ 0 and on the limb when $\theta = \pm\,\pi/2$.}
\label{fig:app3-1}
\end{center}
\end{figure*}

Following these results, we can conclude that our empirical way to include the limb-shift effect cannot reproduce the observations, and we decided not to include this effect at this stage. Further development to include this limb-shift effect are required using either solar observation at different $\cos{\theta}$ angles, or MHD simulations that can reproduce realistic bisector shape for different positions on the stellar disc. Nevertheless, comparing the photometric, RV, BIS SPAN and FWHM activity-induced variations derived with and without considering the limb-shift effect (Figure \ref{fig:app3-1}), we see that this effect, even if accounted for in a realistic way in the future, is not significantly modifying the activity-induced variations. This is because the limb-shift effect only influences the limb of the star that have a small weight in the disk-integrated CCF because of limb darkening.

\end{document}